\documentclass[aps,amsmath,amssymb,reprint,superscriptaddress,prd,nofootinbib,showkeys,showpacs]{revtex4-1}
\usepackage{graphicx, epsfig, bm, amsmath}
\usepackage{dcolumn}% Align table columns on decimal point
\usepackage{rotating}
\usepackage{relsize}
\usepackage{lmodern}
\usepackage{slantsc}

\usepackage{subfigure}

\ifx\pdfoutput\@undefined\usepackage[usenames,dvips]{color}
\else\usepackage[usenames,dvipsnames]{color}

\newcommand{\head}[2]{\multicolumn{1}{>{\centering\arraybackslash}p{#1}}{#2}}

\newcolumntype{C}[1]{>{\centering\arraybackslash}p{#1}}

\usepackage{array}

\newcommand{\ba}{\begin{eqnarray}}
\newcommand{\ea}{\end{eqnarray}}
\newcommand{\be}{\begin{equation}}
\newcommand{\ee}{\end{equation}}

\newcommand{\fsky}{f_{\text{sky}}}

\newcommand{\mC}{\bm{C}}

\newcommand{\mN}{\bm{N}}

\begin{document}

%%%%%%%%%%%%%%%%%%%%%%%%%%%%%%%%%%%%%%%%%%%%%%%%%

\newcommand{\book}{\cite{2013neco.book.....L}}
\newcommand{\bigrev}{\cite{2006PhR...429..307L}}
\newcommand{\wong}{\cite{2011ARNPS..61...69W}}
\newcommand{\smallrev}{\cite{2012arXiv1212.6154L}}

\newcommand{\lcdm}{$\Lambda$CDM\,}
\newcommand{\lcdmw}{$\Lambda$wCDM}
\newcommand{\pcnt}{\%}
\newcommand{\lcdmwa}{$\Lambda$waCDM}
\newcommand{\sa}{Sov. Astron. Lett.}
\newcommand{\jpb}{J. Phys. B.}
\newcommand{\natu}{Nature (London)}
\newcommand{\aas}{Bull. Am. Astron. Soc.}
\def\aap{A\&A}
\def\physrep{Phys. Rep.}
\def\apj{ApJ}
\def\apjs{ApJS}
\def\apjl{ApJL}
\def\mnras{MNRAS}
\def\aj{AJ}
\def\nat{Nature}
\def\aaps{A\&A Supp.}
\def\pra{Phys.Rev.A}         % Physical Review A: General Physics
\def\prb{Phys.Rev.B}         % Physical Review B: Solid State
\def\prc{Phys.Rev.C}         % Physical Review C
\def\prd{Phys.Rev.D}         % Physical Review D
\def\prl{Phys.Rev.Lett}      % Physical Review Letters
\def\araa{ARA\&A}       % Annual Review of Astron and Astrophys
\def\gca{GeCoA}         % Geochimica et Cosmochimica Acta
\def\pasp{PASP}              % Publications of the ASP
\def\pasj{PASJ}              % Publications of the ASJ
\def\apss{ApSS}
\def\jcap{JCAP}
\def\sovast{Soviet Astronomy}
\def\plb{Phys. Lett. B}

\newcommand{\az}[1]{\textcolor{blue}{#1}}

\newcommand{\sd}[1]{\textcolor{red}{[{\bf SD}: #1]}}

\preprint{APS/123-QED}
\title[]{Extreme data compression for the CMB}

\author{Alan Zablocki}
\affiliation{Department of Astronomy and Astrophysics, University of Chicago, Chicago, Illinois 60637, USA}
\affiliation{Kavli Institute for Cosmological Physics, University of Chicago, Chicago, Illinois 60637, USA}
\author{Scott Dodelson}
\affiliation{Fermilab Center for Particle Astrophysics, Fermi National Accelerator Laboratory, Batavia, Illinois 60510-0500, USA}
\affiliation{Department of Astronomy and Astrophysics, University of Chicago, Chicago, Illinois 60637, USA}
%\affiliation{Department of Astronomy and Astrophysics, University of Chicago, Chicago, IL 60637}
%\affiliation{Kavli Institute for Cosmological Physics, University of Chicago, Chicago, IL 60637}
\affiliation{Kavli Institute for Cosmological Physics, University of Chicago, Chicago, Illinois 60637, USA}

\date{\today}

\begin{abstract}
We apply the Karhunen-Lo\'eve methods to cosmic microwave background (CMB) data sets, and show that we can recover the input
cosmology and obtain the marginalized likelihoods in $\Lambda$ cold dark matter cosmologies in under a minute, much faster than Markov chain Monte Carlo
methods. This is achieved by forming a linear combination of the power spectra at each multipole $l$, and solving a system of simultaneous equations
such that the Fisher matrix is locally unchanged. Instead of carrying out a full likelihood evaluation over the whole parameter space, we need
evaluate the likelihood only for the parameter of interest, with the data compression effectively marginalizing over all other parameters. The weighting 
vectors contain insight about the physical effects of the parameters on the CMB anisotropy power spectrum $C_l$. The shape and amplitude of these vectors give
an intuitive feel for the physics of the CMB, the sensitivity of the observed spectrum to cosmological parameters, and the relative sensitivity of different
experiments to cosmological parameters. We test this method on exact theory $C_l$ as well as on a Wilkinson Microwave Anisotropy Probe (WMAP)-like CMB data
set generated from a random realization of a fiducial cosmology, comparing the compression results to those from a full likelihood analysis using CosmoMC. After showing that the method works, we apply
it to the temperature power spectrum from the WMAP seven-year data release, and discuss the successes and limitations of our method as applied to a real data set.
\end{abstract}

%\pacs{98.80.-k}% PACS, the Physics and Astronomy Classification Scheme.

%Use showkeys class option if keyword
                             %display desired
\maketitle
\section{Introduction}
\label{intro}
Modern astrophysical data sets are getting ever larger. This is driven in part by the increased size of the telescopes allowing large astronomical
surveys, as well as the increase in the detector number, their sensitivity, and the resolution. Future galaxy surveys like the Large Synoptic Survey
Telescope (LSST) and Euclid will observe on order $\sim10^9$ galaxies, while current cosmic microwave background (CMB) experiments such as Planck, the South Pole Telescope (SPT) and the Atacama 
Cosmology Telescope (ACT) already map
the microwave sky over more than $\sim10^7$ pixels. Data compression and sophisticated statistical methods applied to these extremely large data sets
have ushered us into the era of ``precision cosmology'', where the data is very well described by the simple six parameter $\Lambda$ cold dark matter (CDM)
model.

The large size of today's data sets often makes it impractical to carry out brute force likelihood calculations. This has therefore motivated a 
number of data compression methods to be developed for use in statistical analyses of galaxy redshift surveys \cite{1983ApJ...267..465D}
and CMB maps \cite{1993ApJ...417L...9S,1994ApJ...430L..85G}. A common approach is to compress the data quadratically into a number of power spectrum 
estimates; for galaxy redshift surveys, the compressed data set is a set of power spectrum estimates $P(k)$ and for CMB experiments, it is the anisotropy
power spectrum of fluctuations $C_l$. To obtain estimates of model parameters, one then performs a Bayesian likelihood analysis using Markov
chain Monte Carlo (MCMC) methods.

The Karhunen-Lo\`eve (KL) eigenvalue method was previously applied to both CMB maps \cite{1994PhRvL..72...13B} and
redshift surveys \cite{1996ApJ...465...34V}. The KL compression method 
can be generalized to two important examples for data sets with certain noise properties:
(i) the case where the mean is known and independent of model parameters and (ii) the case where the covariance is independent of model parameters
\cite{1997ApJ...480...22T}. Here we consider the second case, when the data vector is the power spectrum, $C_l$, itself.

This case was applied to galaxy spectra, where
the speedup in the likelihood computation was achieved using a set of orthonormal compression vectors \cite{2000MNRAS.317..965H,2001MNRAS.327..849R}
(akin to the Gram-Schmidt procedure, for which the order of vectors matters). The same procedure was also applied to mock CMB data for only three parameters, but 
it excluded experimental noise \cite{2002MNRAS.334..167G}. This covariance-independent case has been shown to occasionally produce multimodal likelihood peaks, 
in applications to planetary transit light curves \cite{2005MNRAS.362..460P} and gravitational wave data analysis \cite{2011MNRAS.413L..66G}, 
though there are ways to mitigate these problems, albeit at an increase in computation time by as much as a factor of 20. 

More recently, minimizing the computational cost of an exact CMB likelihood and power spectrum estimation using linear 
compression was investigated in \cite{2015ApJS..221....5G} using Wilkinson Microwave Anisotropy Probe (WMAP) data as an example, while in \cite{2013PhRvD..88d3522W} the authors looked at
efficiently summarizing CMB data using two shift parameters and the physical baryon density $\Omega_b h^2$ to obtain dark energy constraints. In \cite{2003ApJ...596..725C}, the authors showed
that a nonlinear transformation of cosmological parameters can also serve as a form of data compression, which yields a set of normal parameters with a 
Gaussian likelihood distribution, although in that case there is no reduction in the number of parameters.

In this work we create the weighting vectors according to the prescription found in \cite{1997ApJ...480...22T}. Instead of creating a set of
orthonormal vectors we create a linear combination of all the data, such that the resulting {\it mode} holds the most information on the parameter
of interest, with the data compression automatically marginalizing over all other parameters. We carry out this procedure for six $\Lambda$CDM parameters,
although we have tested our methods on extensions to $\Lambda$CDM, e.g., by including the tensor-to-scalar ratio $r$ parameter.

In contrast to work carried out in \cite{2000MNRAS.317..965H,2001MNRAS.327..849R,2005MNRAS.362..460P,2011MNRAS.413L..66G} our method uses only
one mode, offering a significant speedup in obtaining the marginalized likelihoods, and it does not depend on the order of the parameters. We note that
the choice of parametrization will matter when investigating models with known or unknown degeneracies.

The paper is organized as follows: in Sec. \ref{formal} we introduce the extreme compression (EC) method and describe its implementation on CMB spectra.
In Sec. \ref{implement} we implement the compression for a single parameter and describe the marginalization procedure for the whole parameter space.
In Sec. \ref{wmaplikeexp}, we derive the compression vectors and discuss their physical characteristics as applied to the CMB. We then test our method
on two mock data sets, including experimental noise and compare against results obtained using MCMC. As a further test, we analyze the
WMAP seven-year CMB specturm in Sec. \ref{results} and conclude in Sec. \ref{conclude}.
\section{Developing the formalism}
\label{formal}
In this section we briefly review some special cases of data compression presented in \cite{1997ApJ...480...22T}. We then develop the case where the
covariance of the data is assumed to be known and independent of the model parameters, and apply this method to the CMB power spectrum.
\subsection{Compressing the Fisher information matrix}
\label{fish_comp}
The log-likelihood $\mathcal{L}$ for a Gaussian probability distribution can be written as
\begin{equation}
 -2 \mathcal{L}=n \mbox{ln}\ 2\pi + \mbox{ln det } \bm {\textbf{Cov}} + (\textbf{x} - \bm \mu)^{t} \bm {\textbf{Cov}}^{-1}  (\textbf{x} - \bm \mu) \label{1},
\end{equation}
where the covariance matrix is $\bm {\textbf{Cov}} = \langle (\textbf{x} - \bm \mu)^{t}(\textbf{x} - \bm \mu) \rangle$ and $\bm {\mu}$ is the mean
$\langle \textbf{x} \rangle$.
The Fisher information matrix is defined as
\begin{equation}
 \textbf{F}_{ij}=-\left\langle \frac{\partial^2 \mathcal{L}}{\partial \theta_{i}\partial \theta_{j}}\right \rangle = -\langle \mathcal{L}_{,ij}\rangle, \label{3}
\end{equation}
and is a measure of the curvature of the likelihood around the maximum likelihood point $\theta_{\text{ML}}$. Working through some matrix algebra it can be shown that the Fisher matrix can be written as
\begin{equation}
 \textbf{F}_{ij} =  \frac{1}{2} \mbox{ Tr} [\textbf{A}_i \textbf{A}_j + \bm {\textbf{Cov}}^{-1} \textbf{M}_{ij} ], \label{4}
\end{equation}
where $\textbf{A}_i= \bm {\textbf{Cov}}^{-1} \bm {\textbf{Cov}}_{,i} = (\mbox{ln }\bm {\textbf{Cov}})_{,i}$ and $\textbf{M}_{ij}=\langle\textbf{D}_{,ij}\rangle=\bm{\mu}^{t}_{,i}\bm{\mu}_{,j}+\bm{\mu}^{t}_{,j}\bm{\mu}_{,i}$\cite{1996ApJ...465...34V}.

We can perform a linear compression on our data set $\textbf{x}$ with
\begin{equation}
 \bm{y} = \textbf{Bx} \label{5},
\end{equation}
where $\textbf{B}$ is the compression matrix of size $ n' \times n$ and $\bm {y}$ is the resulting data set of
dimension $n'$. It can be shown  \cite{1997ApJ...480...22T} that for $n=n'$ and $\textbf{B}$ invertible, the new Fisher matrix after the linear
compression, $\bm \tilde{\textbf{F}}_{ij}$, is given by
\begin{equation}
 \tilde{\textbf{F}}_{ij} =\frac{1}{2} \mbox{ Tr} [\textbf{B}^{-t} (\textbf{A}_i \textbf{A}_j + 
 \bm {\textbf{Cov}}^{-1} \textbf{M}_{ij}) \textbf{B}^{t} ] = \textbf{F}_{ij} \label{6}.
\end{equation}
The Fisher matrix is thus unchanged. For $n' < n$, the matrix $\textbf{B}$ is not invertible and each row of $\textbf{B}$ specifies one
number in the new data set. For the simplest case where only one linear combination of the data is selected so that $\textbf{B}$ has just one 
row, $\textbf{B} = \textbf{b}^{t}$ 
 the diagonal entries of the Fisher matrix are
\begin{equation}
 \tilde{\text{F}}_{ii} =\frac{1}{2} \left( \frac{\textbf{b}^{t}  
\bm {\textbf{Cov}}_{,i} \textbf{b} }{\textbf{b}^{t} \bm {\textbf{Cov}}\ \textbf{b}} 
\right)^2 + \frac{\left( \textbf{b}^{t} \bm {\mu}_{,i} \right)^{2} }{\left( \textbf{b}^{t} \bm {\textbf{Cov}}\ \textbf{b} \right)} \label{7}.
\end{equation}

How can we use this result to estimate the value of some parameter $\theta_{i}$ and the error $\Delta\theta_i$ associated with it? We wish 
to define $ \textbf{b}^t$ such that the compressed data set carries as much information about parameter $\theta_i$ as possible. That is, we aim to
minimize the error on $\theta_i$. To do so, we maximize the element of the Fisher matrix $\tilde{\text{F}}_{ii}$.  The solution in general is nonlinear in $\textbf{b}$. Inspection of Eq.~(\ref{7}) shows
that the Fisher matrix now consists of two terms, one of which depends on the derivative of the covariance $\bm {\textbf{Cov}}_{,i}$ and another that depends
on the derivative of the mean $\bm {\mu}_{,i}$. Assuming that the CMB covariance matrix is weakly dependent on the parameters, even though
this assumption is not quite correct at low multipoles, yields an interesting result. In that case, 
the Fisher matrix is just
\begin{equation}
 \tilde{\text{F}}_{ii} = \frac{\left( \textbf{b}^{t} \bm {\mu}_{,i} \right)^{2} }{\left( \textbf{b}^{t} \bm {\textbf{Cov}}\ \textbf{b} \right)} \label{8}.
\end{equation}
Maximizing this leads to the solution $\textbf{b}=\bm{\textbf{Cov}}^{-1} \bm{\mu}^{}_{,i}$.
Our compressed data set, $y=\textbf{b}^t\textbf{x}$, now consists of just one number $y_{i}$,
\begin{equation}
 y_{i}=\bm{\mu}^{t}_{,i} \bm{\textbf{Cov}}^{-1} \textbf{x} \label{9}.
\end{equation}
In this case the compressed Fisher matrix is given by
\begin{equation}
 \text{F}_{ii}=\bm{\mu}^{t}_{,i} \bm{\textbf{Cov}}^{-1} \bm{\mu}^{}_{,i} \label{9b}.
\end{equation}

\subsection{Applying data compression to the CMB power spectrum}
\label{compress_CMB}
The CMB temperature anisotropies form a scalar 2D field on the sky and are often expanded in spherical harmonics
\begin{equation}
\frac{\Delta T}{T}(\theta,\phi)=\sum_{l}\sum_{m} a_{lm}Y_{lm}(\theta,\phi), \label{9d}
\end{equation}
where $\Delta T$ is the temperature variation from the mean, \emph{l} is the multipole, $Y_{lm}(\theta,\phi)$ 
is the spherical harmonic function of degree \emph{l} and order $m$, and $a_{lm}$ are the expansion coefficients or multipole moments.
The variance $\delta_{ll'}\delta_{mm'}C^{}_{l}=\langle a_{lm}^{*} a_{l'm'}^{}\rangle$, where $\delta_{ll'}$ is the Kronecker delta function, contains all the statistical information.
\begin{figure}[t!]
\vspace{-0.6cm}
\centering
\includegraphics[width=0.37\textwidth, angle=-90]{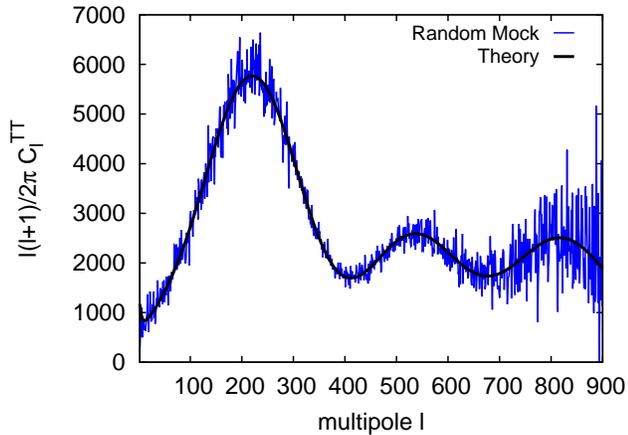} % WMAP up to 900  linear  
\caption[]{Temperature power spectrum obtained from a random realization (solid blue) and the theory power spectrum computed with the 
Boltzmann code CAMB \cite{Lewis:1999bs} in solid black. We ignore the effect of lensing on the CMB.}\label{clsthrn}
\end{figure}
Here we use the temperature power spectrum so that the data vector is
\begin{equation}
 \textbf{x} = \frac{1}{2l+1} \sum_{m=-l}^{m=l} |a^{T}_{lm}|^{2}, \label{11}
\end{equation}
such that $\langle \textbf{x}  \rangle = \bm {\mu} = C_{l}$. We are therefore carrying out a quadratic precompression \cite{1997ApJ...480...22T}.
In Fig.~\ref{clsthrn} we compare the theory temperature power spectrum with that of a random realization for a WMAP-like experiment. The compressed data set for a given parameter $\theta_i$ is a single linear combination of the $C_l$'s:
\begin{equation}
y_i=\sum_{l} {\frac{\partial C^{}_{l}}{\partial\theta_{i}}} {\text{Cov}}_{}^{-1} (C_{l}^{},C_{l}^{})\ \frac{1}{2l+1} \sum_{m=-l}^{m=l} |a^{}_{lm}|^{2}. \label{12}
\end{equation}

The measurement of the angular power spectrum $C_{l}$ has characteristic uncertainty due to finite beam size and a 
limit on the number of modes we observe on the sky known as cosmic variance, with the variance at each multipole given by
\begin{equation}
 {\text{Cov}}_{}(C^{}_{l},C^{}_{l})=\frac{2}{(2l+1)f_{\text{sky}}}(C^{}_{l} + N^{}_l)^{2}  \label{eqn:cmbnoise2},
\end{equation}
where $f_{\text{sky}}$ is the fraction of the sky covered by the experiment. For maps made with Gaussian beams the noise term $N_l$ has
the form \cite{1995PhRvD..52.4307K}
\begin{equation}
 N^{}_l = (\sigma \theta)^2 e^{l(l+1)\, \theta^2 / 8\text{ln}2 } \label{14},
\end{equation}
where $\sigma$ and $\theta$ are the sensitivity ($\Delta T/T$) and angular resolution in radians respectively.

The expected value $\langle y_{i} \rangle$ is then
\begin{equation}
\langle y_i \rangle=\sum_{l} {\frac{\partial C^{}_{l}}{\partial\theta_{i}}} {\text{Cov}}_{}^{-1}(C_{l}^{},C_{l}^{})\ C^{}_{l}  \label{15},
\end{equation}
and $\langle y_{i} \rangle$ carries all the information contained in the data on $\theta_i$.
 We can define the coefficients $\alpha^{i}_{l}$ to be
\begin{equation}
{\alpha}^{i}_{l} = {\frac{{\partial C}^{}_{l}}{{\partial\theta}_{i}^{}}}  {\text{Cov}}_{}^{-1} (C_{l}^{},C_{l}^{})  \label{16},
\end{equation}
so that
\begin{equation}
 \langle y_i \rangle=\sum_{l} {\alpha}^{i}_{l} \ {C}_{l}^{}  \label{17}.
\end{equation}
For a given parameter $\theta_i$, the coefficients $\alpha^i_l$ describe the combination of multipoles that carry the information about $\theta_i$.

The variance of $\langle y_{i} \rangle$ is
\begin{equation}
\sigma_{\langle y_{i} \rangle}^2=%\langle y_{i}^2 \rangle - \langle y_{i} \rangle^2\nonumber\\
%&=&
\langle y_{i}^2 \rangle - \sum_{l,l'} {\alpha}^{i}_{l} {\alpha}^{i}_{l'} {C}^{}_{l} C_{l'}.
%\nonumber\\
%&=&\sum_{l,l'} {\alpha}^{i}_{l} {\alpha}^{i}_{l'} \left(\frac{\sum_{m,m'} \langle a_{lm} a^{*}_{lm} a_{l'm'} a^{*}_{l'm'}\rangle}{(2l+1)(2l'+1)}-{C}^{}_{l} C_{l'}\right) \label{18}\nonumber\\
\end{equation}
Since the $a_{lm}$ are Gaussian fields, the resulting four-point functions are easily evaluated and 
\begin{equation}
\sigma_{\langle y_{i} \rangle}^2=\sum_{l=1} {\alpha}^{i}_{l} \ {\text{Cov}}_{}(C^{}_{l},C^{}_{l}) \ \alpha_{l}^{i} \label{21}.
\end{equation}
Using the expected value and variance of $\langle y_{i} \rangle$ we can rewrite the compressed Fisher matrix given by Eq.~(\ref{9b}) as
\begin{equation}
\text{F}_{ii}^{y} = \left (\frac{d y}{d \theta_i} \right )^2 \frac{1}{\sigma_{\langle y_{i} \rangle}^2}\label{21b}.
\end{equation}
We can compare the error bars obtained from the extremely compressed Fisher matrix above to the error bar obtained with Eq.~(\ref{9b}), which is identical
to the Fisher information matrix for the CMB as
\begin{equation}
{\text{F}_{\textit{ij}}^{\text{CMB}}} = \sum_{l} {\frac{{\partial C}^{}_{l}}{{\partial\theta}_{i}^{\ }}} {\text{Cov}}_{}^{-1} (C_{l}^{},C_{l}^{}) {\frac{{\partial C}^{}_{l}}{{\partial\theta}_{j}^{\ }}} \label{9c}.
\end{equation}
\section{Implementation}
\label{implement}
\subsection{One parameter example}
\begin{figure}[t]
\vspace*{-1cm}
\centering
\includegraphics[width=0.25\textwidth, angle=-90]{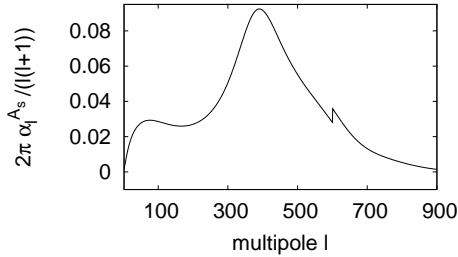}
\caption{\label{fig:wmap7_TT_1_alphas} The compression vector on the scalar power spectrum amplitude $A_s$. The discontinuity at $l\sim600$ is due to a drop in the WMAP experimental noise.} %(we use the same noise as the WMAP likelihood code).}
\end{figure}
\begin{figure}[h]
   \vspace{0.4cm}
\centering
\includegraphics[width=0.16\textwidth, angle=-90]{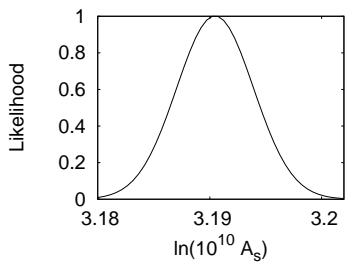}
\vspace*{5mm}
\caption[]{Unmarginalized likelihood for the log power of the primordial curvature perturbations. The data used here is the exact theory $C_l$.}\label{clsthrn334}
\end{figure}

Using the prescription in the previous section we are now able to compress the CMB temperature power spectrum into just a handful of numbers. To illustrate the
procedure we first choose a simple one parameter example focusing on the scalar power spectrum normalization parameter $A_s$. 
Using Eq.~(\ref{16}) and choosing a fiducial point at which to compute the derivative of $C_l$ with respect to ln$(10^{10} A_s)$, we obtain the weighting
vector on $A_s$, which is plotted in Fig.~\ref{fig:wmap7_TT_1_alphas}.
In general, we expect the weights to start with a small amplitude at low $l$,
where cosmic variance is high, then to increase until the experimental noise starts to dominate. For WMAP, this starts at $l\sim900$, with the weights
decreasing to zero between an $l$ of 900--1200. A simple test of this compression is to use the theory $C_l$'s as the {\it data vector}, and with WMAP-like noise, compute the likelihood for $A_s$. This is depicted in Fig.~\ref{clsthrn334}.
The one curve there is actually three curves, (i) the likelihood computed using a single mode $y_{A_s}$:
\begin{equation}
-2\text{ln}\text{L} = \frac{(y_{A_s} - \bar y_{A_s})^2}{2 \sigma_{\langle y_{A_s} \rangle}^2},
\end{equation}
(ii) the likelihood using the full set of $C_l$'s, and (iii) the Fisher (Gaussian) approximation with the variance obtained from Eq.~(\ref{21b}). All three
approaches give the same answer, showing that in this simple case, the compression works well.
 \subsection{Two parameter model example}
In the previous section we showed how to compress a data set and obtain the likelihood for a single parameter. As can be seen in Fig.~\ref{clsthrn334}, the likelihood
is quite narrow, and the error on ln$(10^{10} A_s)$ is very small. In this section, we will show how to compress accounting for a second parameter,  obtaining
marginalized distributions very quickly. 

Each compressed data set $y_i$, by design, carries all the information on the parameter of interest $\theta_i$. However, it will also have some
sensitivity to the other parameter, a sensitivity that we would like to remove. We now show with this simple two-dimensional example how
to remove the unwanted sensitivity, essentially marginalizing over the remaining parameter.

We begin by forming a linear combination of $y_{1}$ and $y_{2}$ for the first parameter as
\begin{equation}
 y^{'}_{1} = c_{1}y_{1} +c_{2}y_{2},
\end{equation}
with $ y^{}_{1}  = \sum_{l} \alpha^{1}_{l} C_{l}^{}$ and $ y^{}_{2}  = \sum_{l} \alpha^{2}_{l} C_{l}^{}$, where $c_1$ and $c_2$ are chosen by the requirement that $y'_1$ does not depend on $\theta_2$.  For this to be independent of $\theta_{2}$ we require that the derivative of $y^{'}_{1}$ with respect to $\theta_{2}$ vanishes.
We then obtain
\begin{equation}
{\frac{\partial y_{1}^{'}}{{\partial\theta}_{2}^{ \ }}} = c_1 \bigg [ \sum_{l} \alpha^{1}_{l}
\ {\frac{\partial C_{l}^{}}{{\partial\theta}_{2}^{}}}\bigg ] + c_2 \bigg [ \sum_{l} \alpha^{2}_{l}
\ {\frac{\partial C_{l}}{{\partial\theta}_{2}^{}}} \bigg ] = 0.
\end{equation}
The quantities in square brackets are just the Fisher matrix elements so that the equation for $ y^{'}_{1}$ is
\begin{equation}
{\frac{\partial y_{1}^{'}}{{\partial\theta}_{2}^{ \ }}}=0 = c_{1}\text{F}_{12} + c_{2}\text{F}_{22}.\label{sol1}
\end{equation}
This fixes the ratio of the two coefficients, and $c_1$ can be set to unity, so that the new, marginalized vector $y'_1$ is
\begin{equation}
 y^{'}_{1}  = \sum_{l} \alpha^{'1}_{l} \ C_{l}^{}\label{aaa}
\end{equation}
with
\begin{equation}
\alpha^{'1}_{l} = \alpha^{1}_{l} -\frac{\text{F}_{12}}{\text{F}_{22}}\alpha^{2}_{l}.\label{32}
\end{equation}
Repeating the procedure for the second parameter yields the weighting vector 
\begin{equation}
\alpha^{'2}_{l} = \alpha^{2}_{l} -\frac{\text{F}_{12}}{\text{F}_{11}}\alpha^{1}_{l}.\label{32b}
\end{equation}
We note that in two dimensions, this particular example is equivalent to the common approach of creating an orthonormal basis
using the Gram-Schmidt process in quantum mechanics. More specifically, the dot product (defined by $\textbf{b}^t \bm{\textbf{Cov}}\ \textbf{b}$) is only zero for the combinations of 
$\bm \alpha^{2'} \bm{\textbf{Cov}}\ \bm \alpha^1$ and $\bm \alpha^{1'} \bm{\textbf{Cov}}\ \bm \alpha^2$ with $\bm \alpha^{2'} \bm{\textbf{Cov}}\ \bm \alpha^{1'} \ne 0$.

\begin{figure}[t!]
\centering
\includegraphics[width=0.16\textwidth, angle=-90]{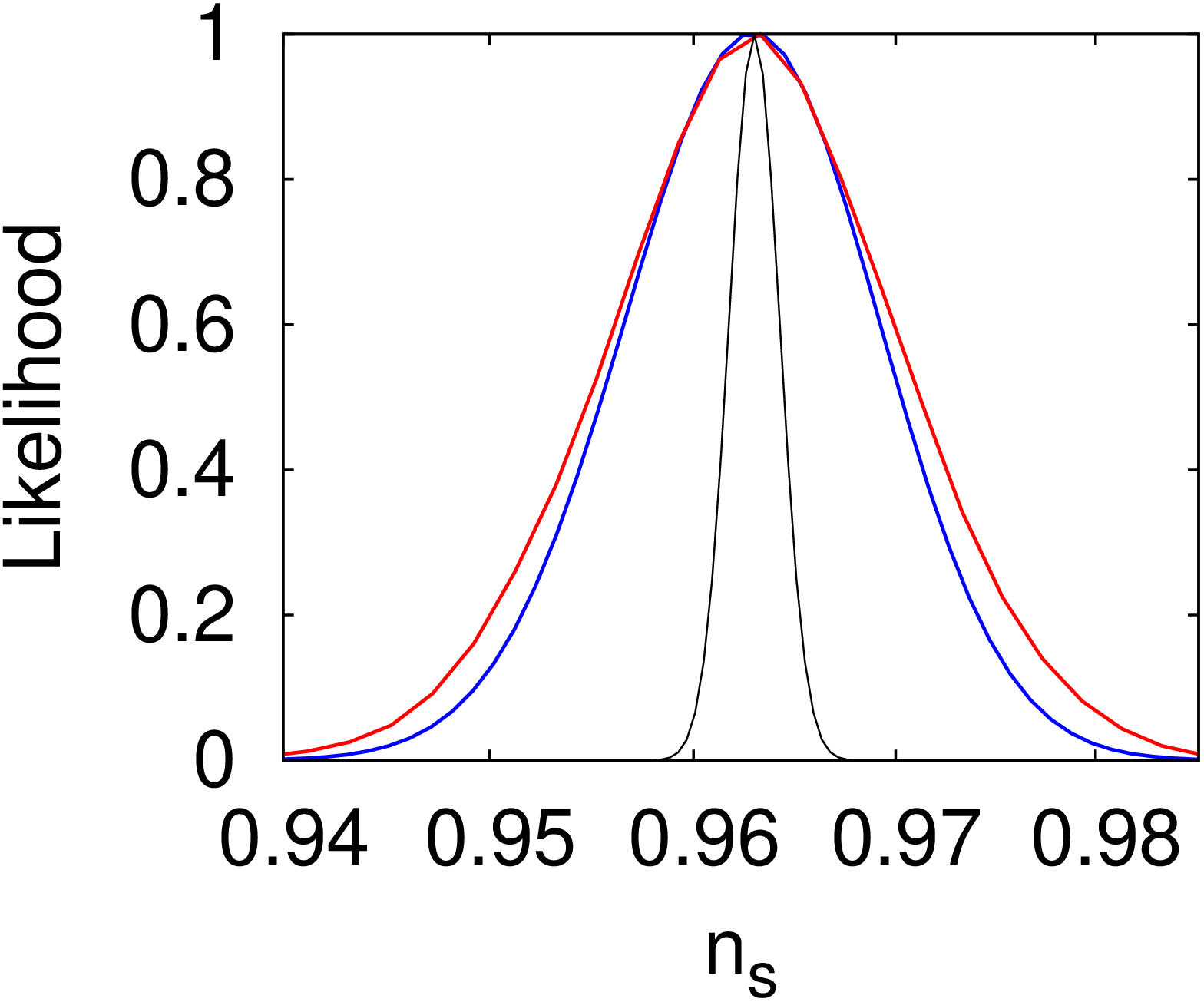}
\includegraphics[width=0.16\textwidth, angle=-90]{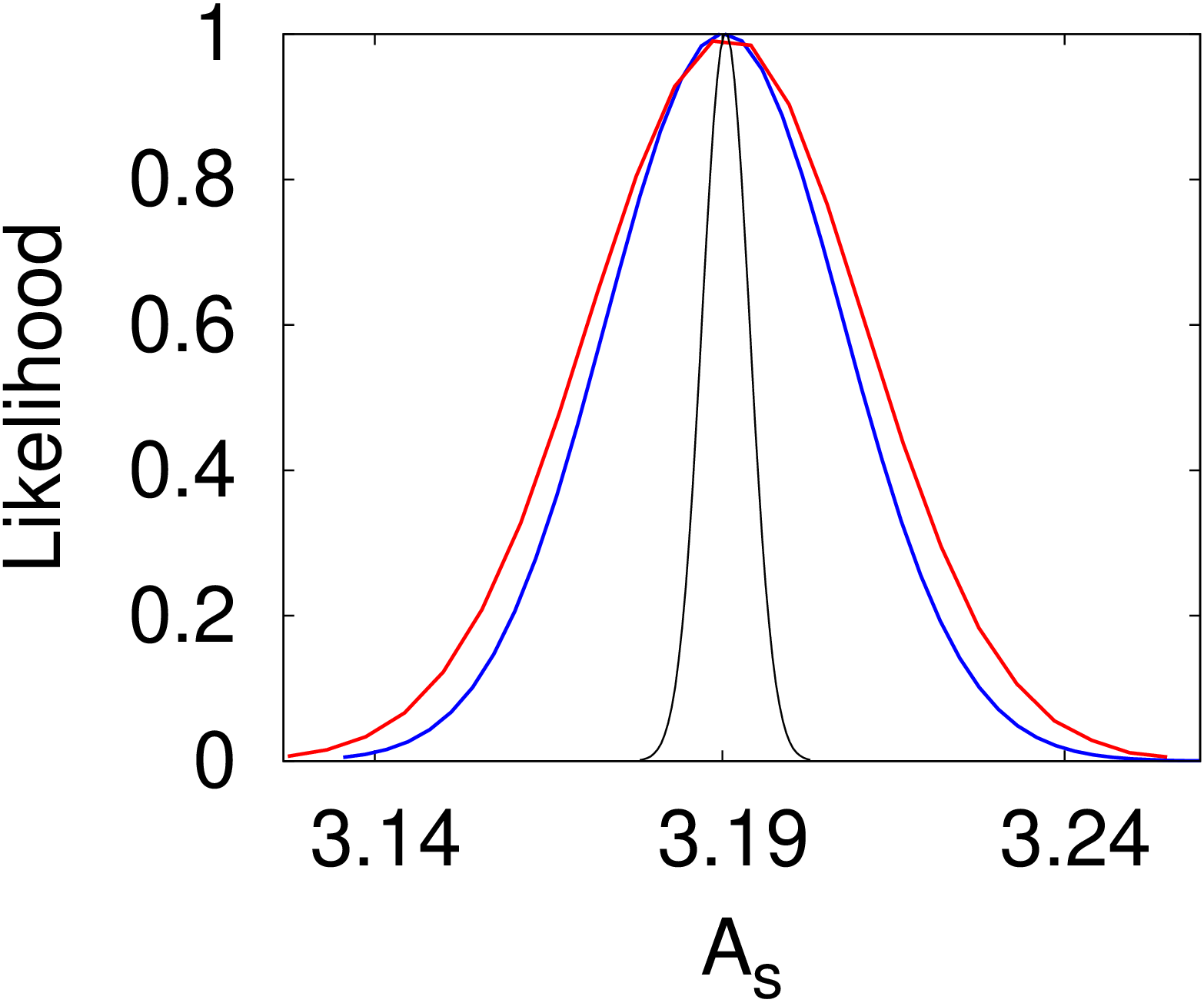}
\vspace*{5mm}
\caption[]{Marginalized likelihoods on the spectral index $n_s$ and the scalar power spectrum amplitude $A_s$ using extreme compression (solid blue) 
and the exact result from MCMC (solid red) for the two parameter toy model.
Marginalization is achieved using the solutions for $\alpha^{'1}_{l}$ and $\alpha^{'2}_{l}$ and Eqs.~(\ref{32}) and (\ref{32b}).
The unmarginalized case is shown in solid black for reference. The data used is the exact theory $C_l$.}
\label{clsthrn22}
\end{figure}

As an example, consider the compressed data set for $n_s$ and $A_s$. All the information about each parameter%parameters
is contained in a single $\chi^2$; e.g., 
\begin{equation}
\chi^2_{n_s} = \frac{(y^{'}_{n_s} - \bar y^{'}_{n_s})^2}{2 \sigma_{\langle y^{'}_{n_s} \rangle}^2}
\end{equation}
is a function of $n_s$ only. With information on the other parameter removed, we
need explore only one dimension to get the marginalized posterior. This is why the method is much faster than spanning the
full two-dimensional likelihood space. If we sample each dimension 20 times, the full likelihood is obtained with
only $2\times 20=40$ samples instead of $20^2=400$. And of course, as the parameter space gets larger, the difference 
becomes much more pronounced. In Fig.~\ref{clsthrn22}, we show the $n_s$ and $A_s$ marginalized likelihoods for
exact theory $C_l$ from the full likelihood and the compression given by Eqs.~(\ref{32}) and (\ref{32b}).
\subsection{Generalizing to higher dimensions}
\label{get_res_6}
Based on the results of the previous section we now present the general problem for $n$ parameters along with the solutions.
The most general linear combination of all the data in a model with $n$ parameters can be written as
\begin{equation}
y^{'}_{1} = c_{1}y_{1} + c_{2}y_{2} + \dots + c_{n}y_{n},\label{44}
\end{equation}
such that the compressed mode $y^{'}_{1}$ carries all the information on the first parameter $\theta_1$, with information on all other parameters removed.
To obtain the extreme compressed $\theta_1$ mode, $y^{'}_{1}$, we must solve the matrix problem
\begin{equation}
 \begin{pmatrix}
  \text{F}_{22} & \text{F}_{23} & \cdots & \text{F}_{2n} \\
  \text{F}_{32} & \text{F}_{33} & \cdots & \text{F}_{3n} \\
  \vdots  & \vdots  & \ddots & \vdots  \\
  \text{F}_{n2} & \text{F}_{n3} & \cdots & \text{F}_{nn} 
  \end{pmatrix}
  \begin{pmatrix}
         c_2 \\
         c_3 \\
         \vdots\\
         c_n
        \end{pmatrix}
  =\begin{pmatrix}
         -\text{F}_{12} \\
         -\text{F}_{13} \\
          \vdots  \\
         -\text{F}_{1n}
        \end{pmatrix}.
  \end{equation}
This yields $n-1$ unique constants on the $n-1$ coefficients $c_i (i>1)$ and $c_1$ can be set to unity. The same procedure holds for all other modes: for mode $i=\alpha$, the coefficients are determined by the general equation
\begin{equation}
\text{F}'_{\alpha,ij}c_j = -\text{F}_{\alpha i},
\end{equation}
where $\text{F}'_{\alpha}$ is the Fisher matrix with row and column $\alpha$ removed.

In the next section we calculate the weighting vectors for a WMAP-like experiment, and apply the compression method to mock WMAP data sets.

\section{Tests on a WMAP-like experiment}
\label{wmaplikeexp}
 
We now apply this formalism to obtain marginalized likelihoods from synthetic data from a WMAP-like experiment (mock data sets with WMAP noise) to 
see how well we can recover the parameters using extreme compression. We use the same parametrization as CosmoMC, with $100*\theta_{\text{MC}}$, an 
approximation for $r_{s}(z_{\star}) / D_A(z_{\star})$, the angular scale of the sound horizon at last scattering, replacing $\Omega_{\Lambda}$ or $H_0$ due to a known geometric degeneracy in the CMB (see Appendix \ref{bad_par}).
The fiducial cosmology assumed is: $\omega_c=\Omega_c h^2=0.1109$, $\omega_b=\Omega_b h^2 = 0.02258$, $100*\theta_s=1.039485$, $n_s=0.963$, ln$(10^{10} A_s) = 3.1904$ and $\tau=0.088$.
We first obtain the posterior distributions assuming that the data vector is the exact theory $C_l$, and then test on a more realistic mock data set using a random realization
of the fiducial cosmology.

 \subsection{WMAP weighting vectors}
      \begin{figure*}[t!]
   \includegraphics[width=0.22\textwidth, angle=-90]{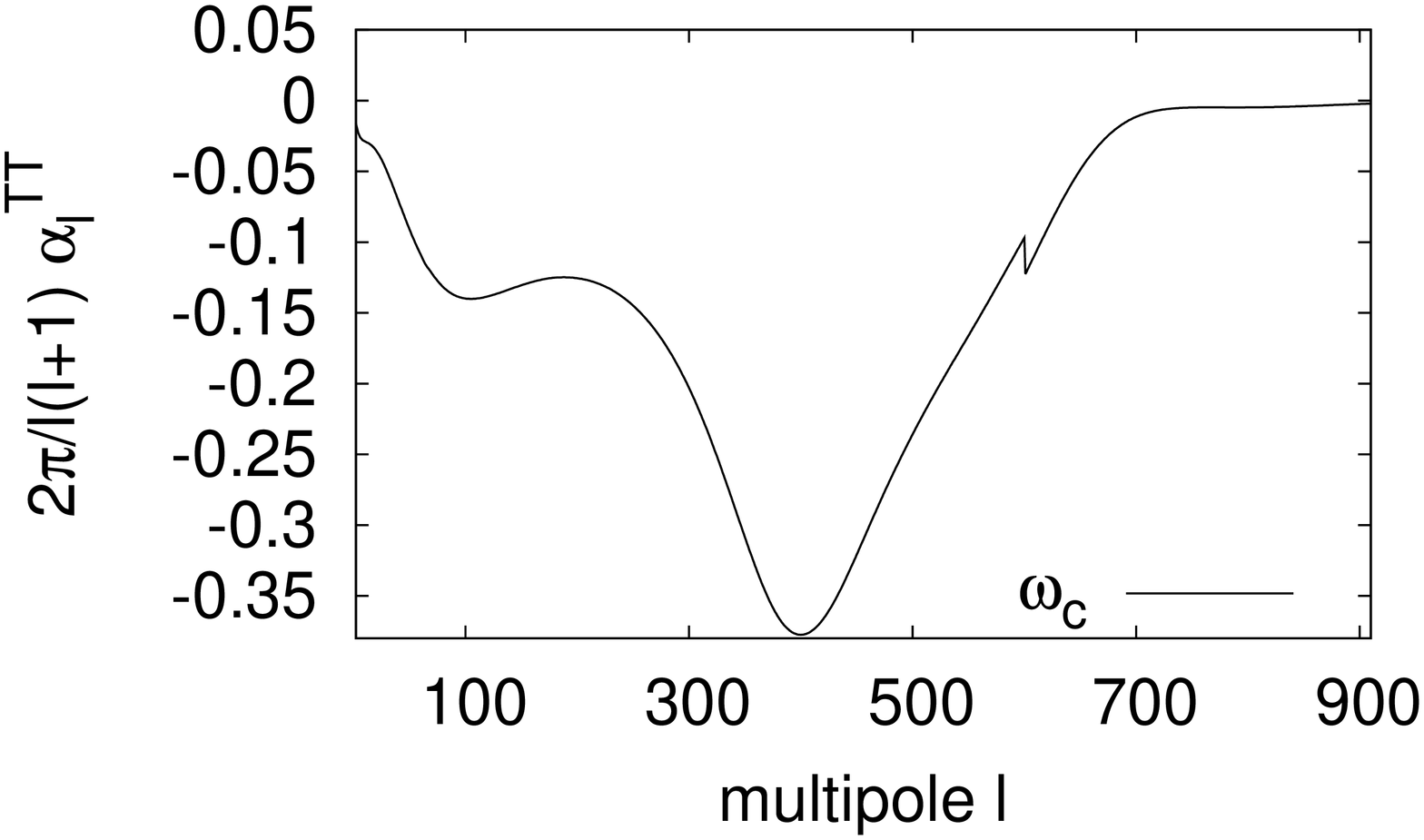}
  %  \vspace*{-5mm}
   \includegraphics[width=0.22\textwidth, angle=-90]{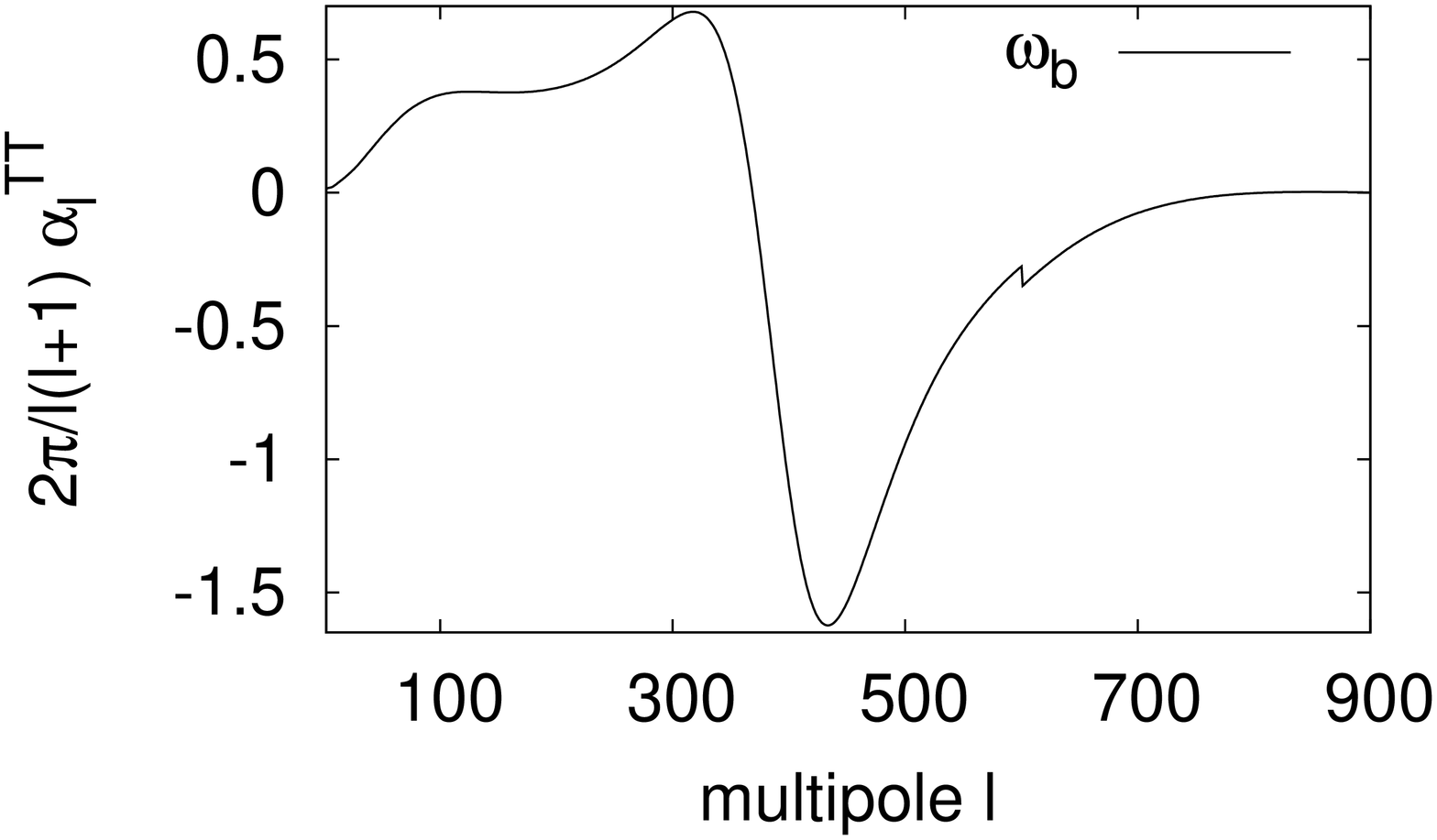}
  %  \vspace*{-5mm}
   \includegraphics[width=0.22\textwidth, angle=-90]{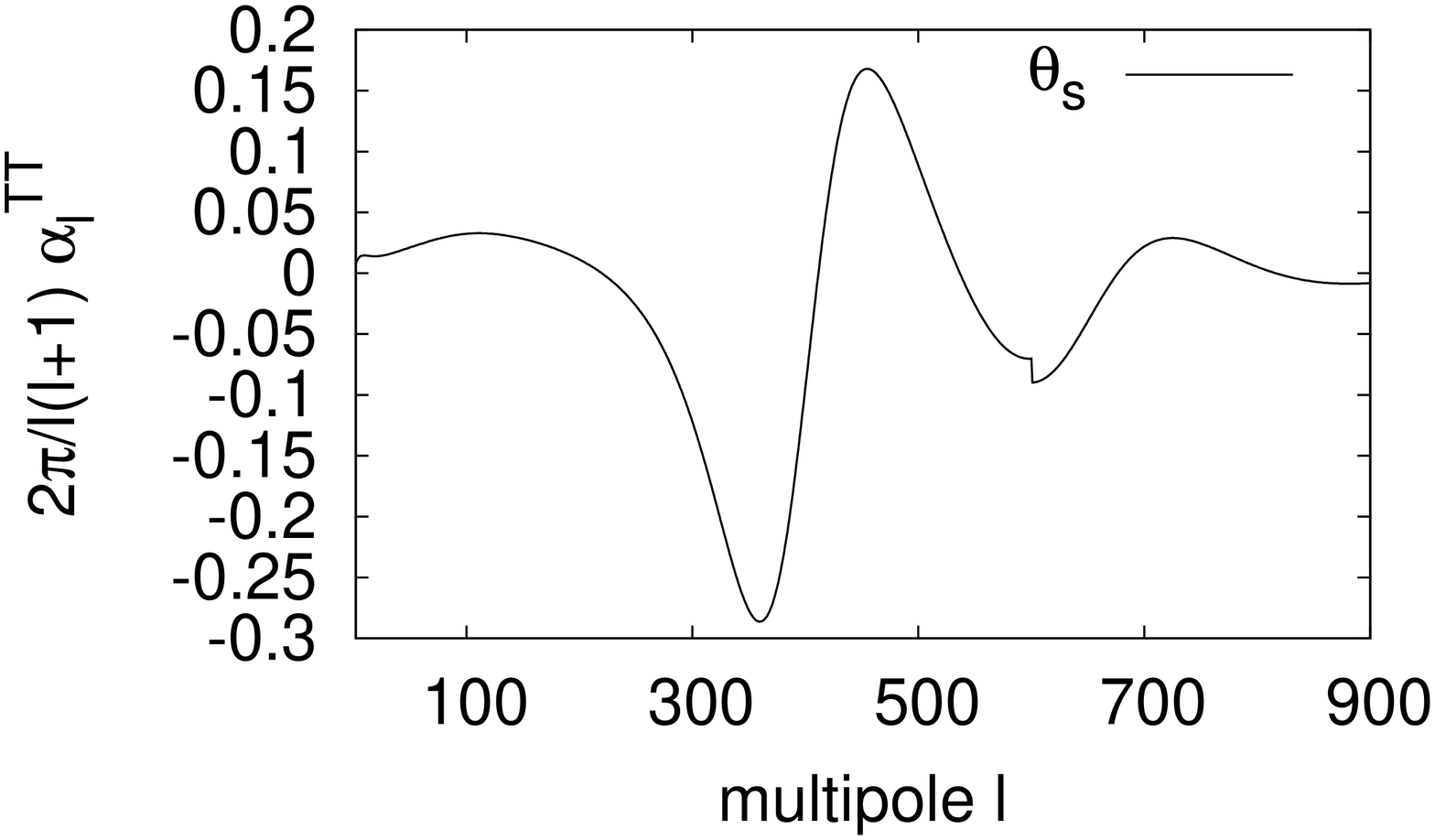}
   \includegraphics[width=0.22\textwidth, angle=-90]{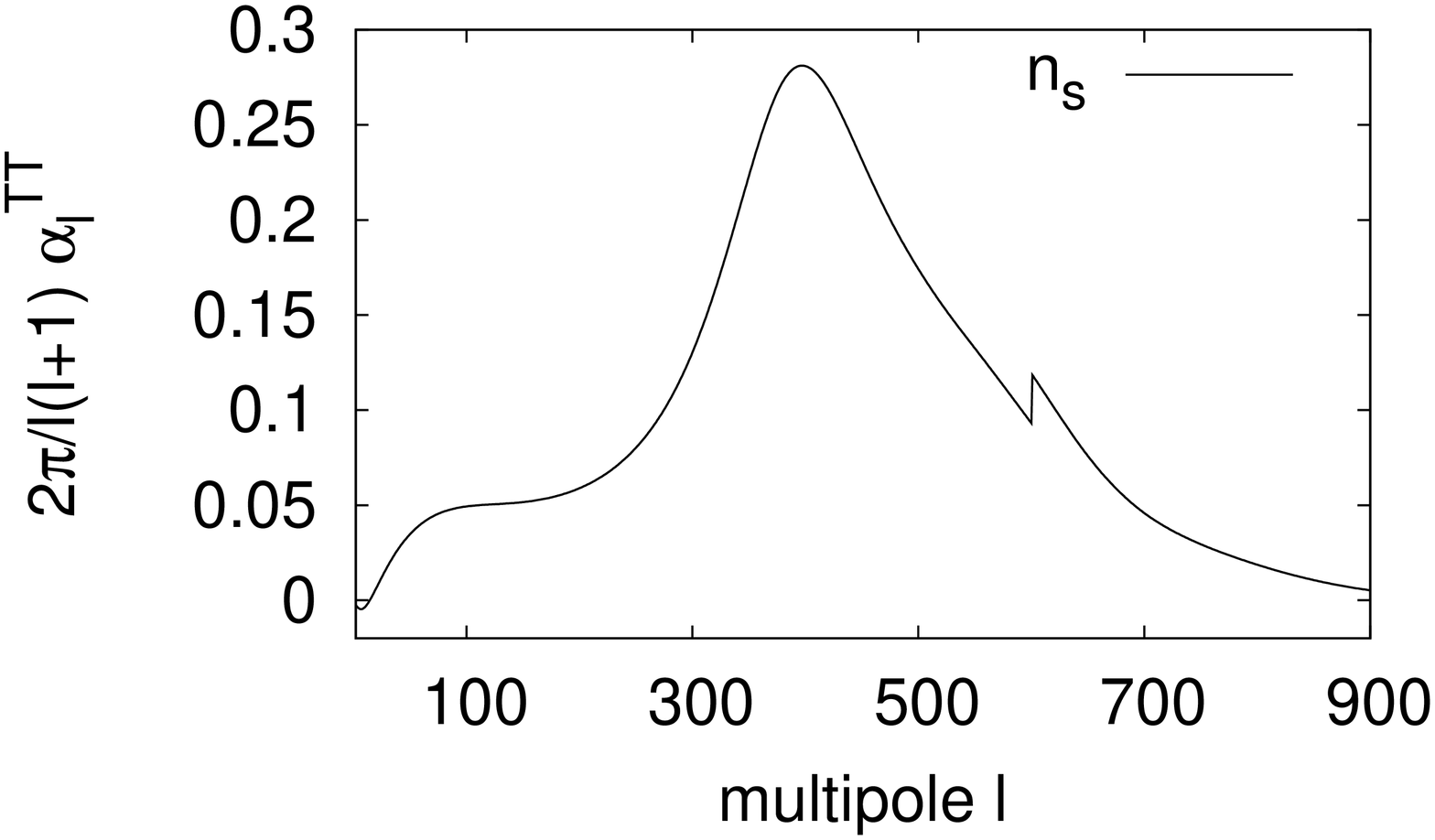}
   \includegraphics[width=0.22\textwidth, angle=-90]{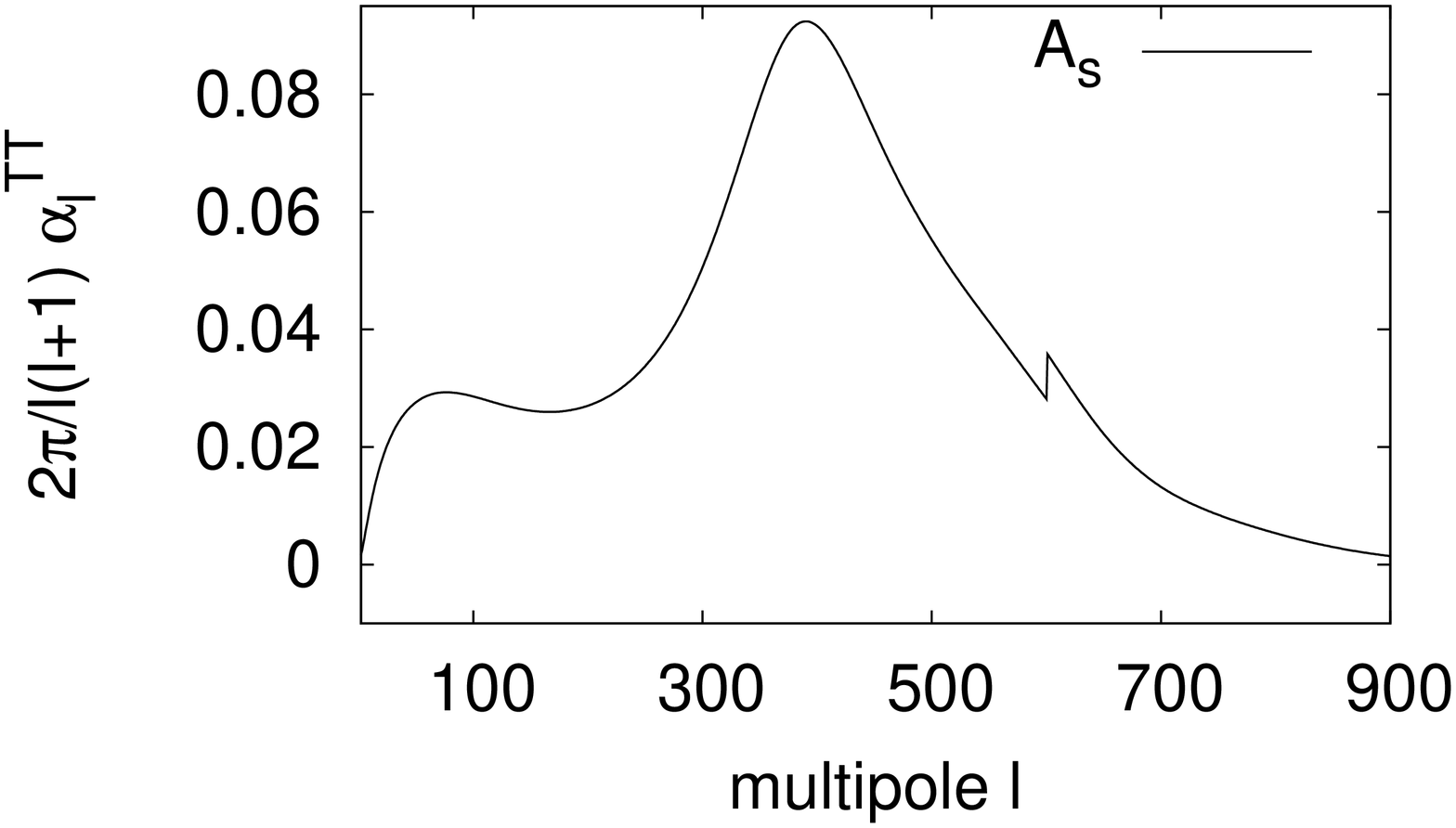}
   \includegraphics[width=0.22\textwidth, angle=-90]{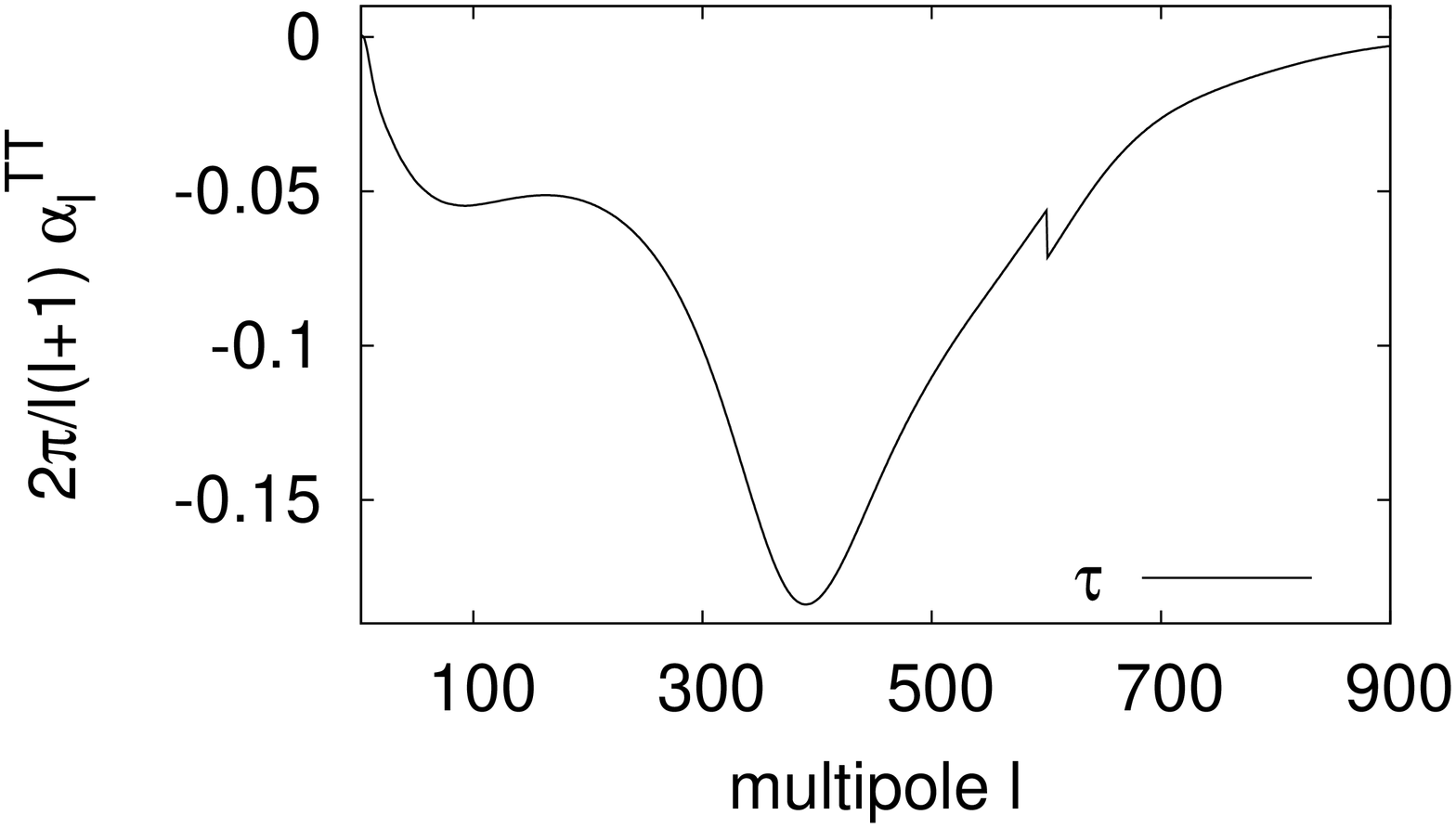}
  \caption{\label{fig:wmap7_TT_6_alphas} The six $\Lambda$CDM weighting vectors $\alpha^{i}_l$ for a CMB experiment with WMAP noise and sky coverage. Each vector is used to compress the temperature power 
 spectrum $C^{TT}_{l}$, into a single number $y_{i}$ that carries all the information  on each parameter $\theta_i$. A general feature
 of these vectors, is that their amplitudes are small at low-$l$, where cosmic variance is large [Eq.~(\ref{eqn:cmbnoise2})], and at high-$l$, where
 experimental noise dominates. The weights go down to zero between $l=900$ and $l=1200$. All six vectors reach their maximum amplitude between $l$ of 330-440. The jump at $l\sim600$ is due to a 
 discontinuity in WMAP noise.}
  \end{figure*}

In Sec. \ref{compress_CMB}, we showed that to achieve locally lossless compression of our CMB data set we need to compute the covariance of the data (where
 data is the spectrum $C_l$) and the derivative of the data with respect to the cosmological parameters in the $\Lambda$CDM model. To calculate the weighting vectors
 for the CMB power spectrum, we obtain the six derivatives of the power spectrum with respect to the parameter vector
 $\Theta = \{\omega_c,\omega_b,100 \theta_s,n_s,A_s,\tau\}$. We use a double sided derivative formula with a step size of 3\% (we use 0.5\% for the derivative
 with respect to $\theta_s$). 
  \begin{figure*}[t!]
\includegraphics[width=0.22\textwidth, angle=-90]{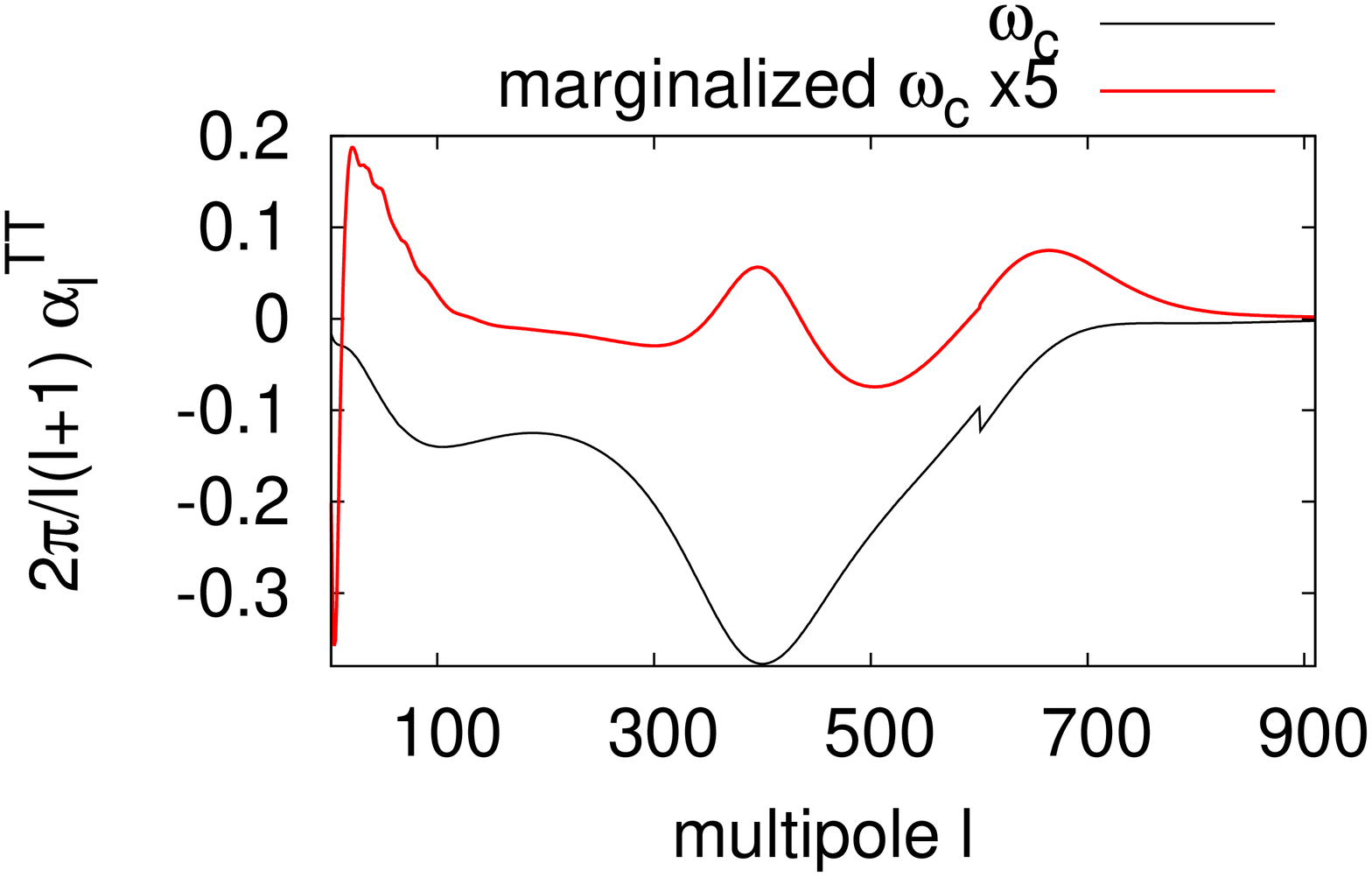}
\vspace*{-1mm}
\includegraphics[width=0.22\textwidth, angle=-90]{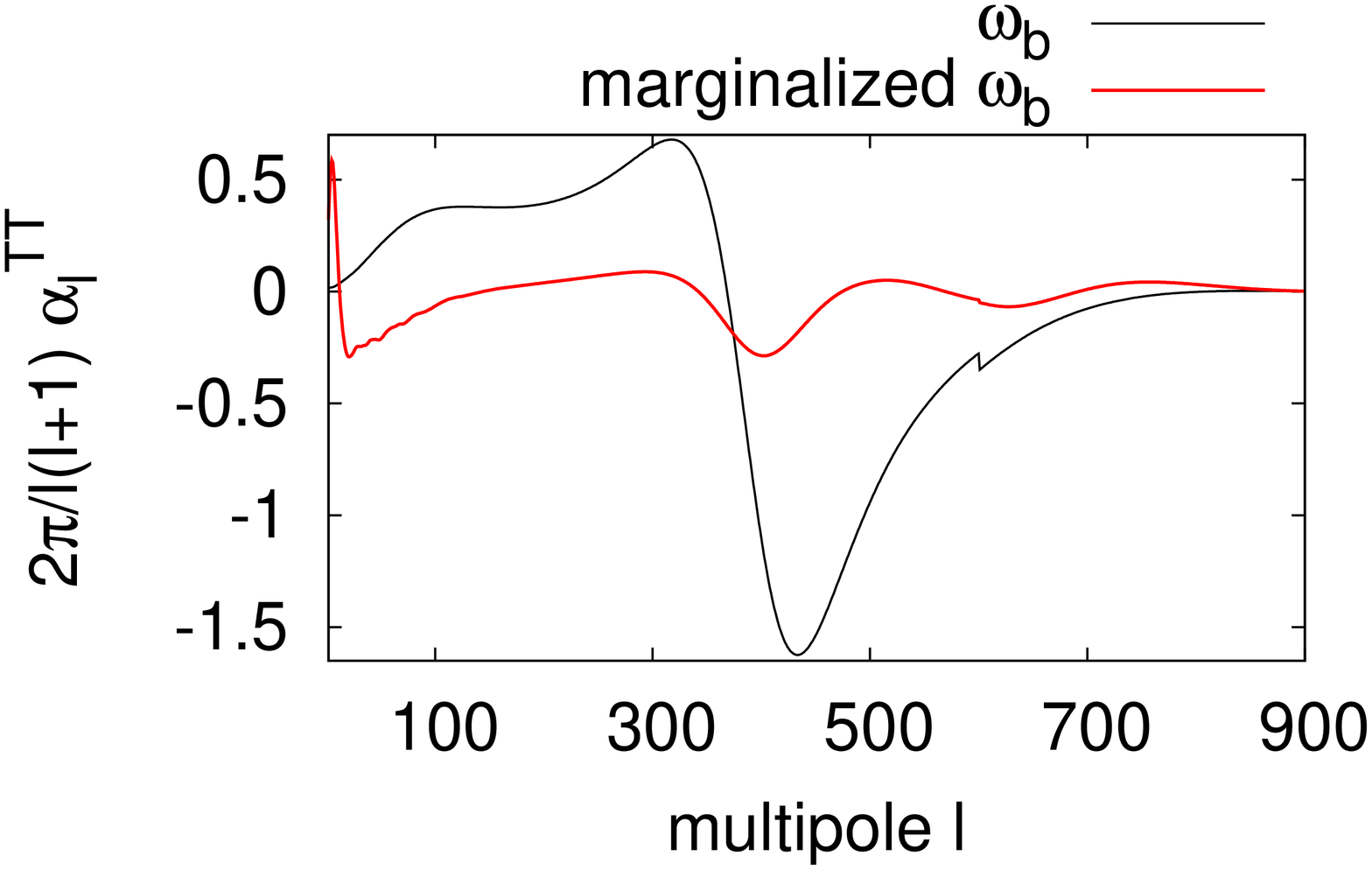}
\vspace*{-1mm}
\includegraphics[width=0.22\textwidth, angle=-90]{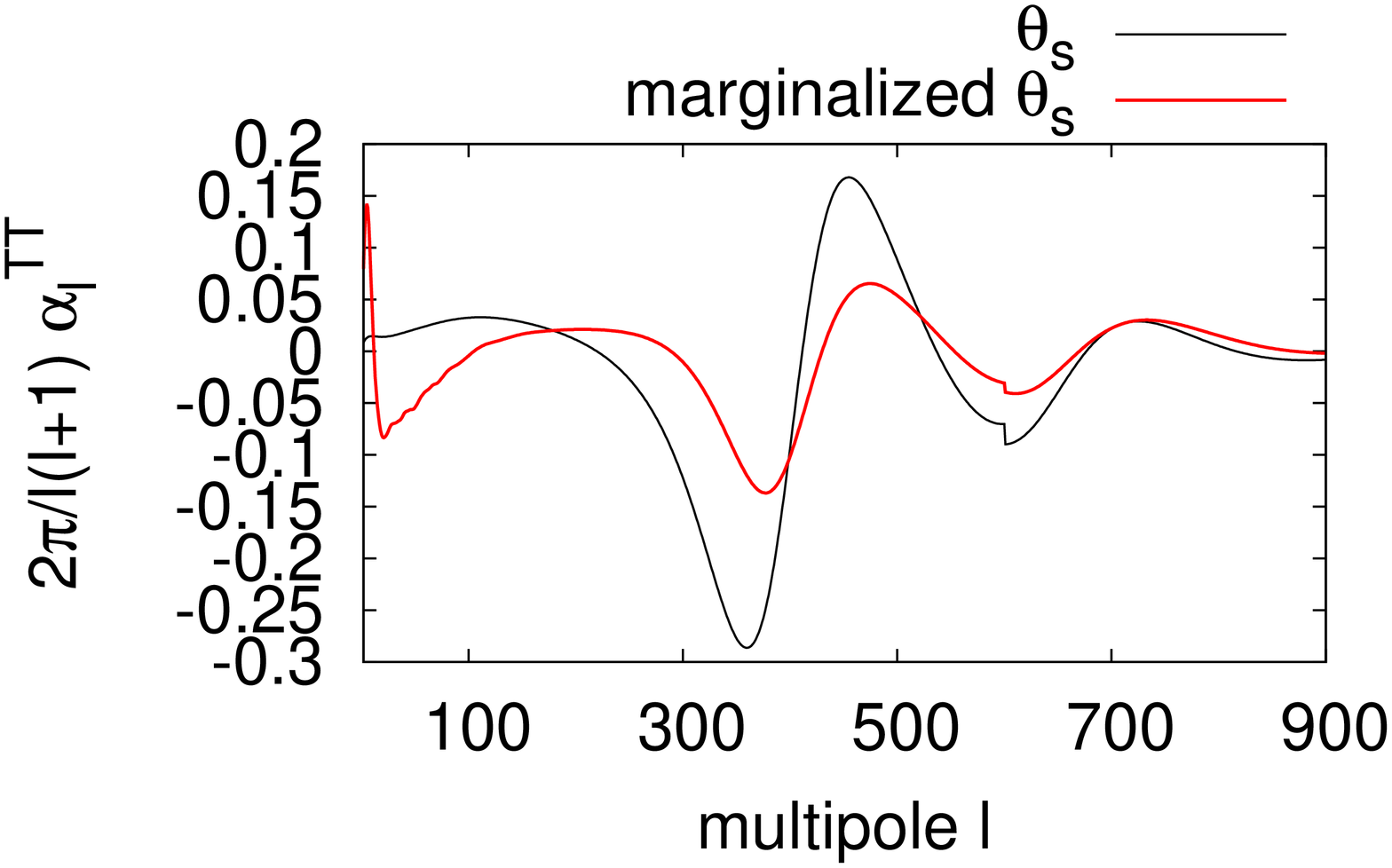}
\includegraphics[width=0.22\textwidth, angle=-90]{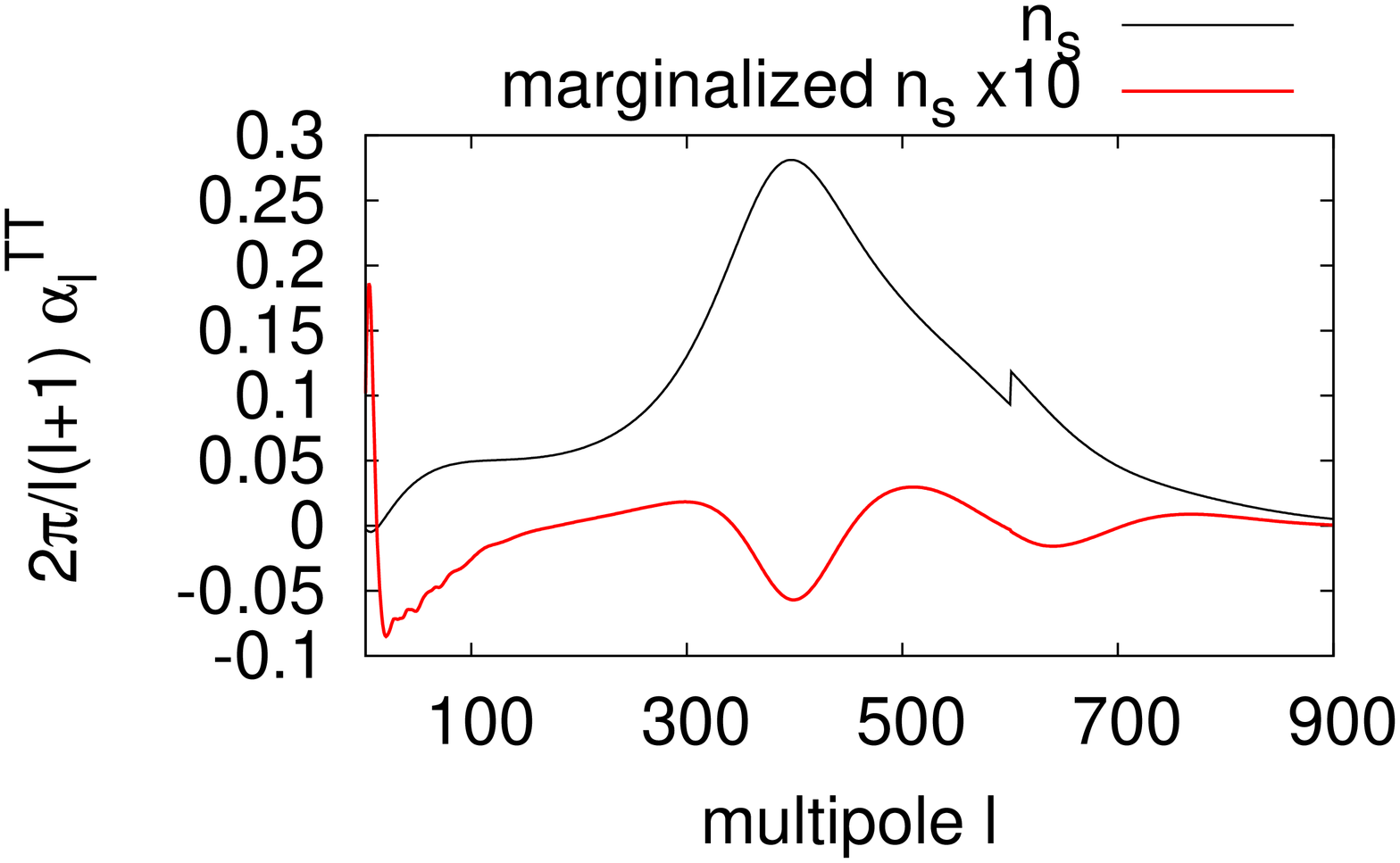}
\includegraphics[width=0.22\textwidth, angle=-90]{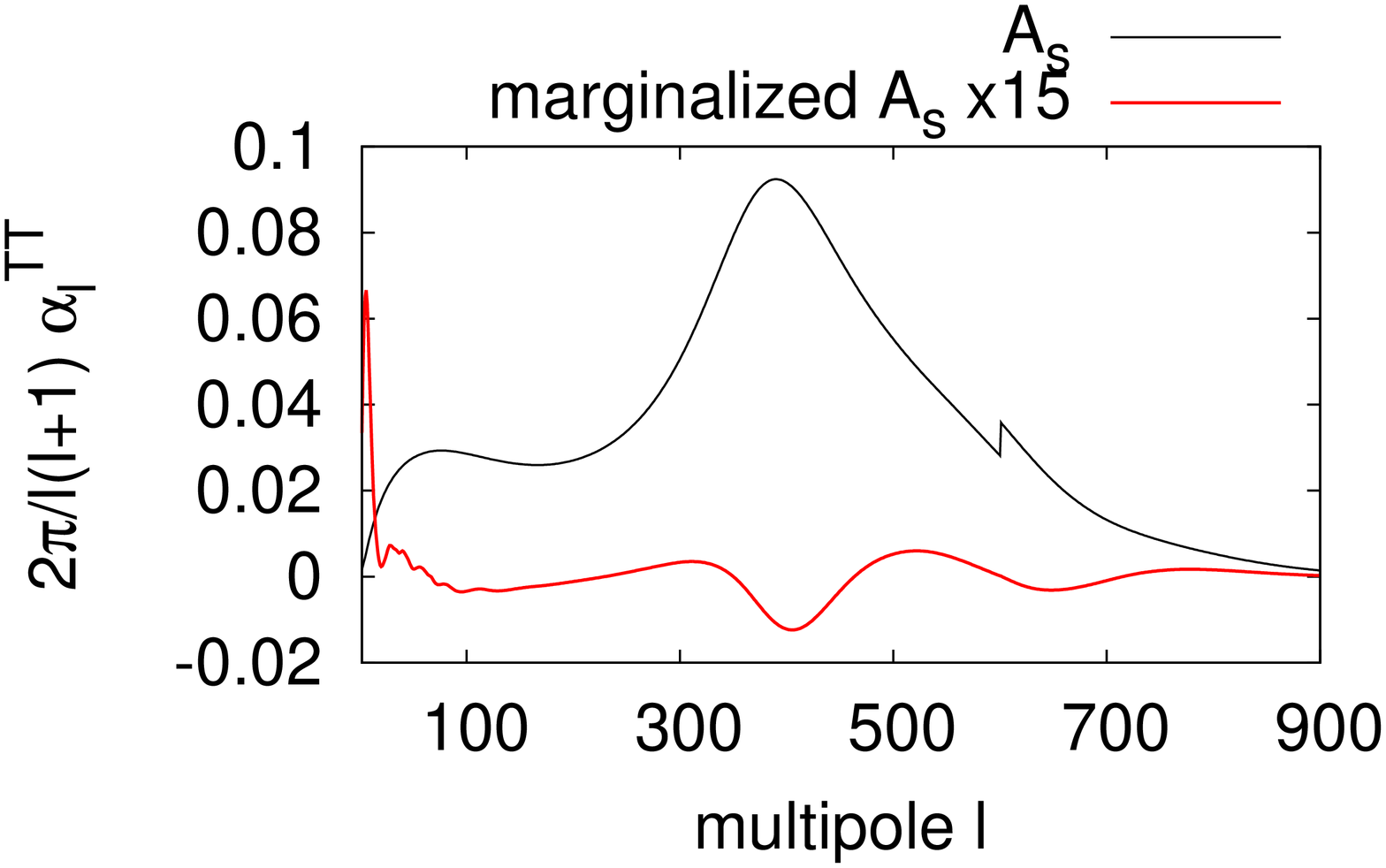}
\includegraphics[width=0.22\textwidth, angle=-90]{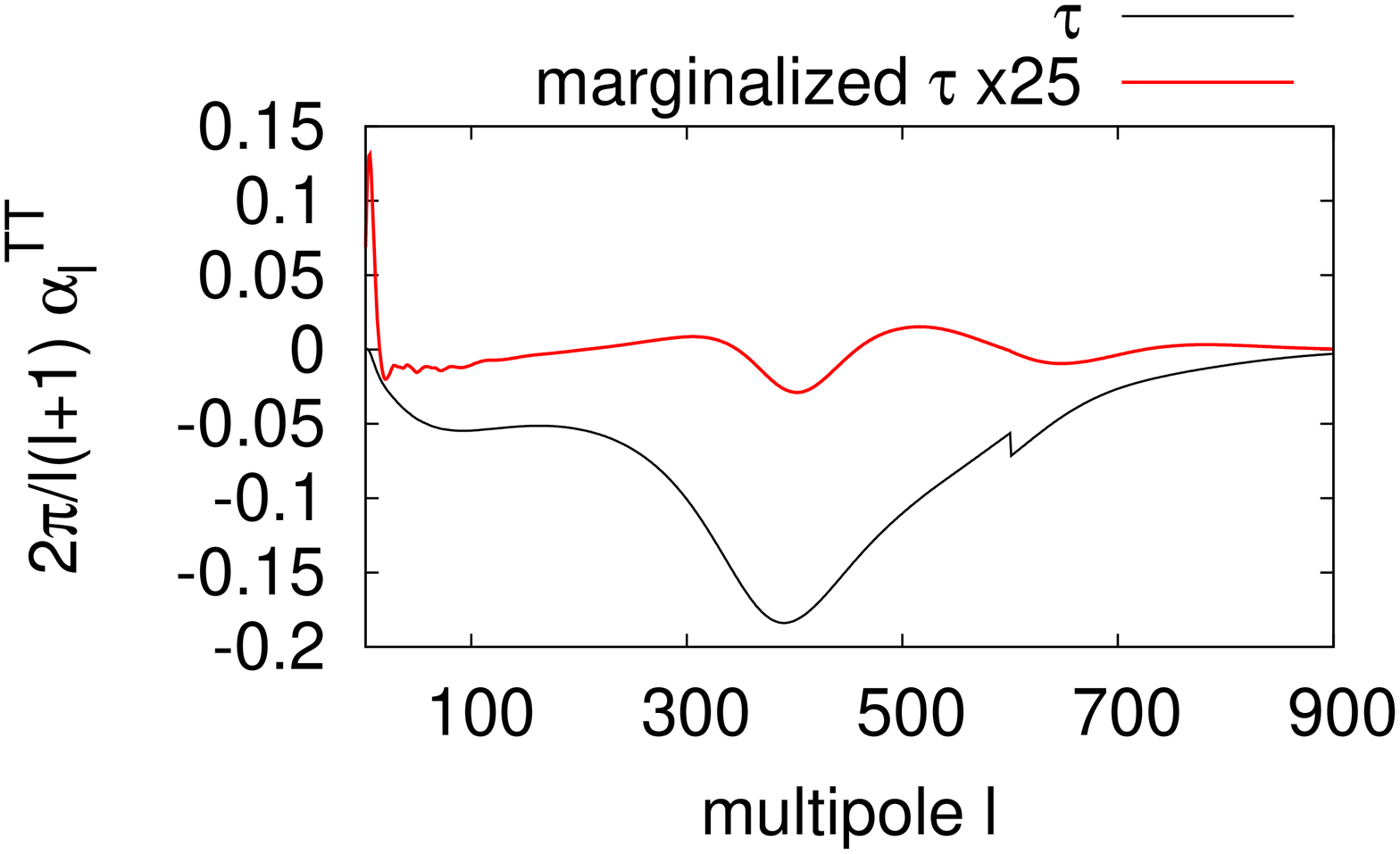}
\caption{\label{fig:wmap7_TT_6d_alphas} Comparison of the compression vectors $\alpha^{i}_l$ for WMAP before (solid black lines) and after marginalization
(solid red lines) using the prescription in Sec. \ref{get_res_6}. Some of the marginalized vectors have been multiplied by a scale factor to ease the
comparison. Note the apparent decrease in the amplitude in each vector, once we take out the information on all the other parameters.}
\end{figure*}
 \begin{figure*}[t!]
\includegraphics[width=0.22\textwidth, angle=-90]{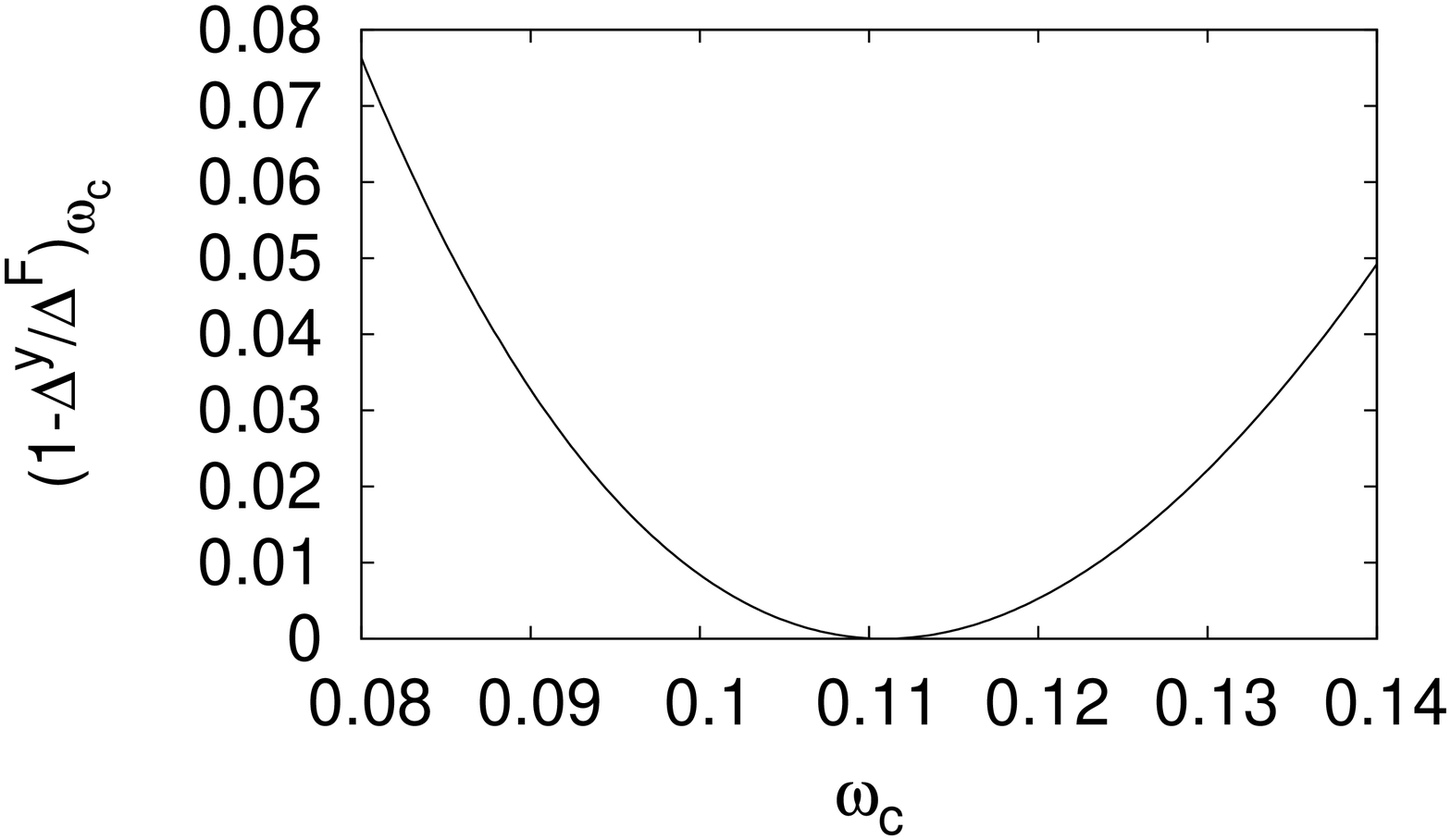}
  \vspace*{-1mm}
\includegraphics[width=0.22\textwidth, angle=-90]{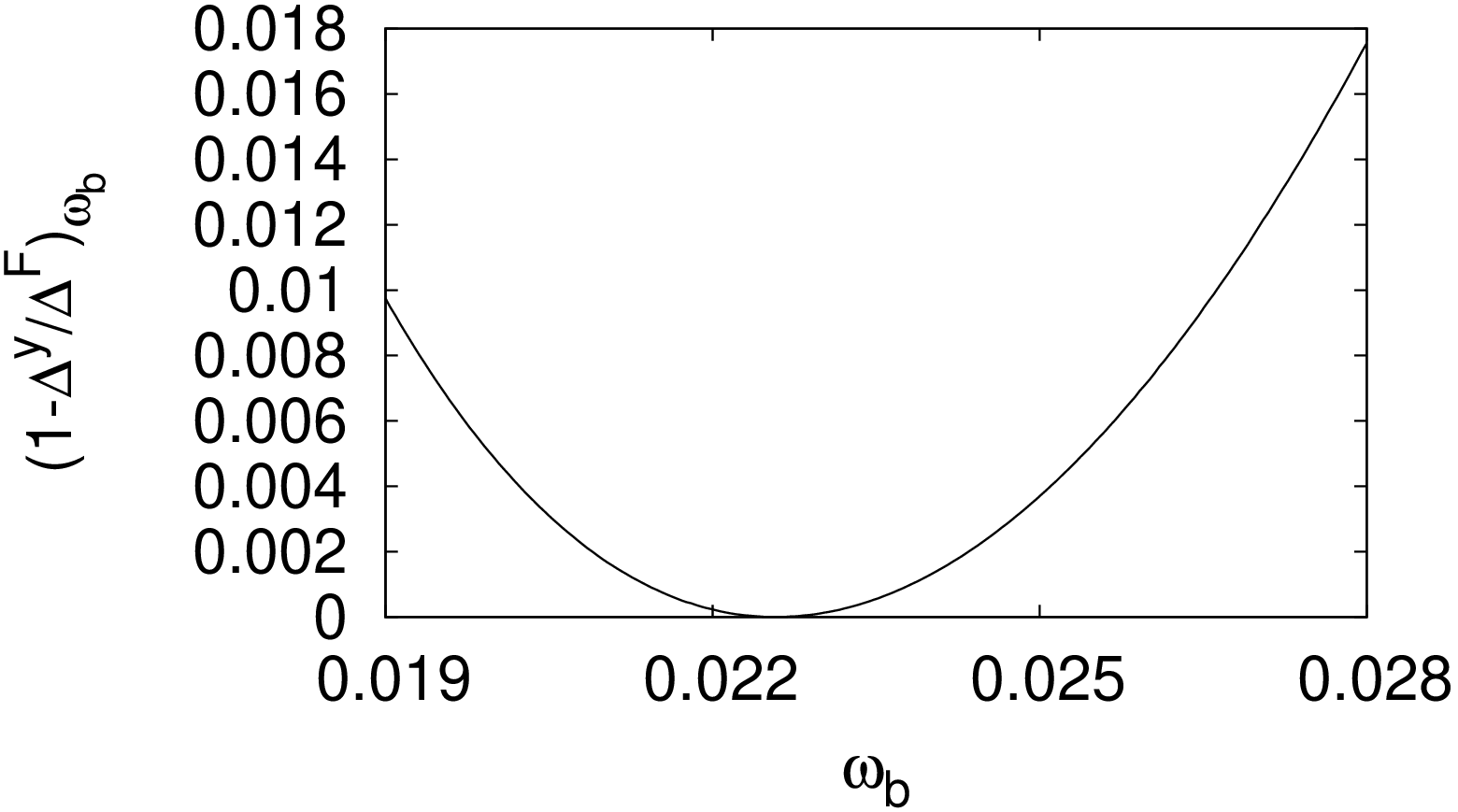}
 \vspace*{-1mm}
\includegraphics[width=0.22\textwidth, angle=-90]{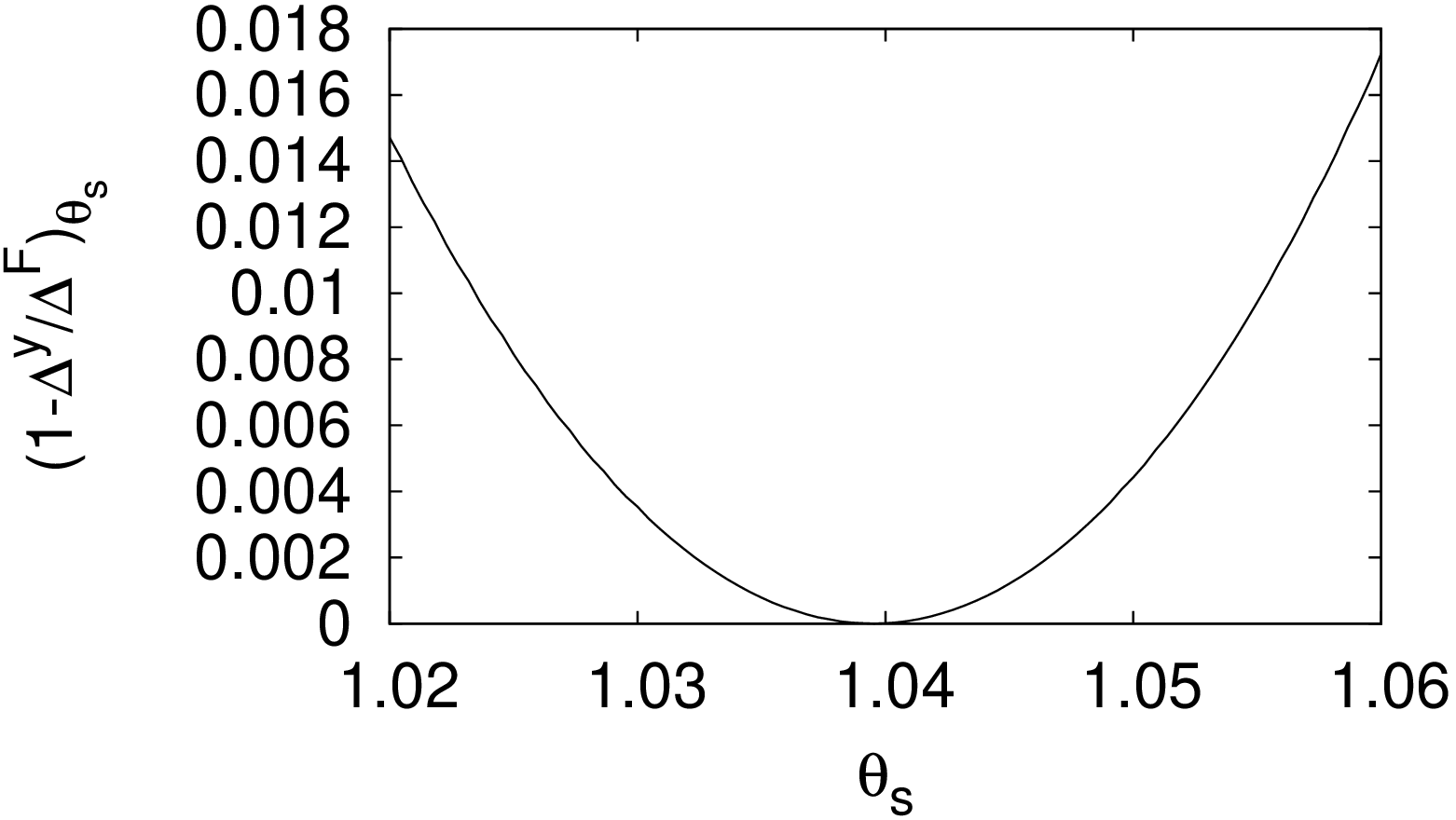}
\includegraphics[width=0.22\textwidth, angle=-90]{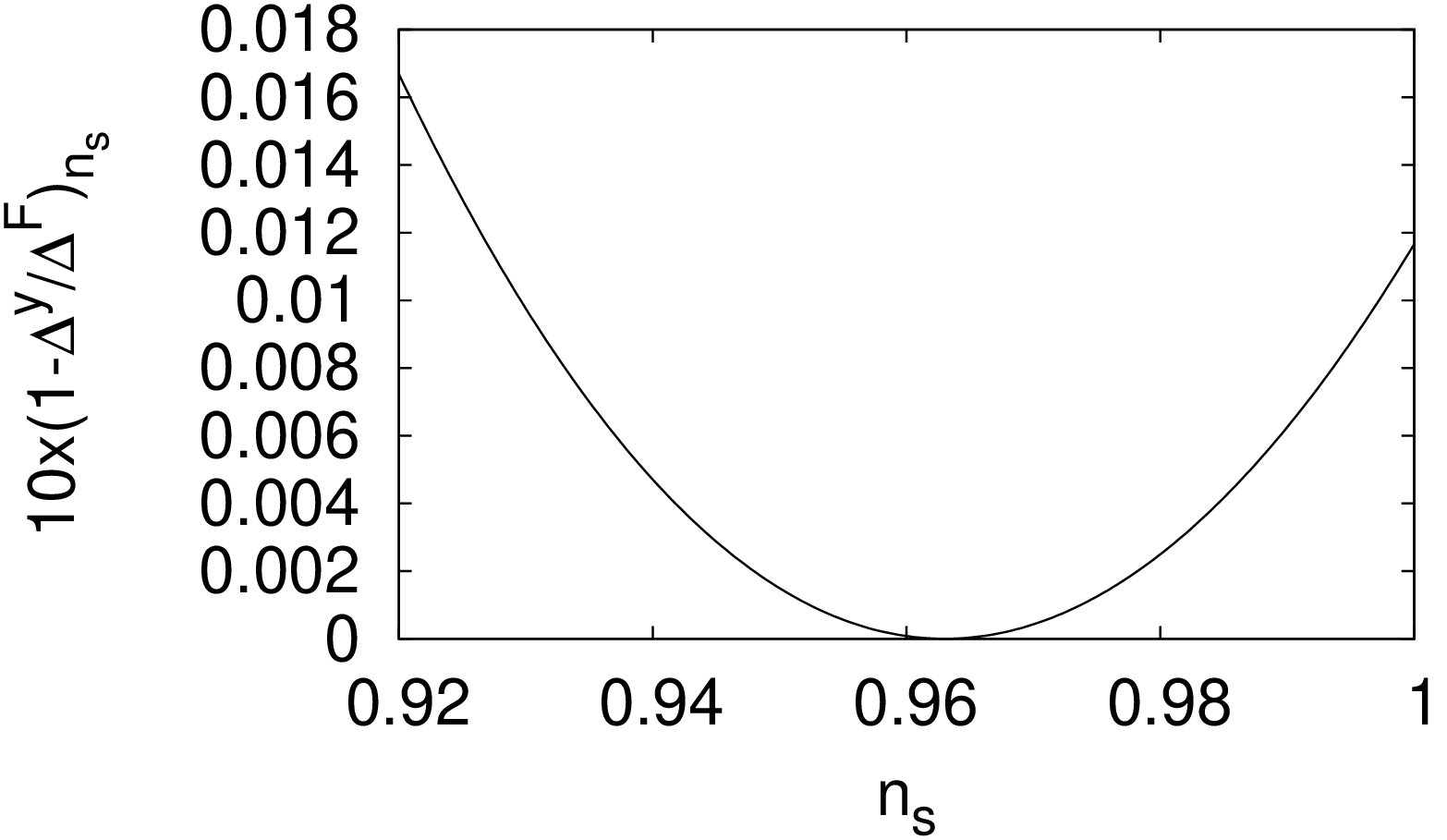}
\includegraphics[width=0.22\textwidth, angle=-90]{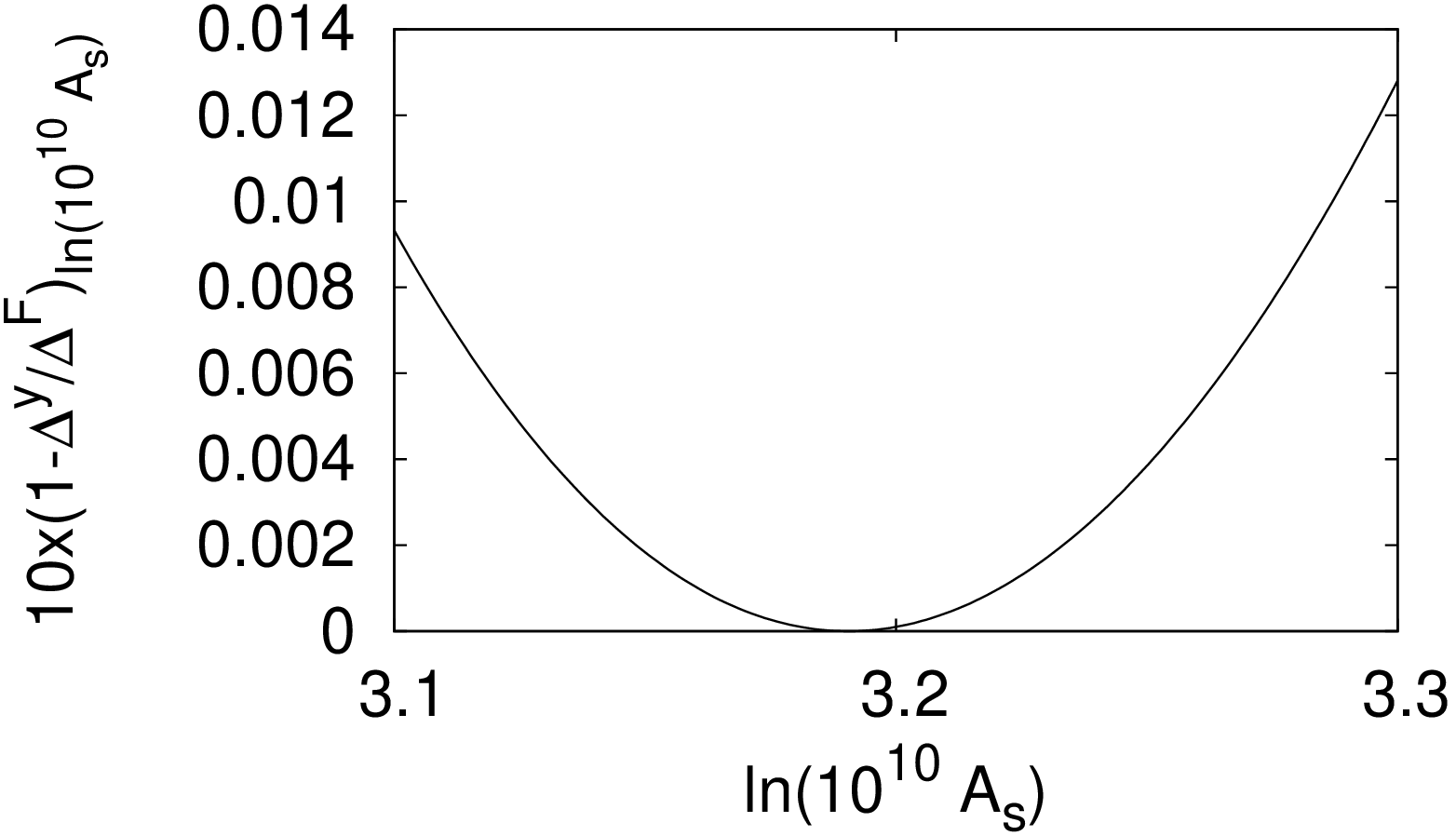}
\includegraphics[width=0.22\textwidth, angle=-90]{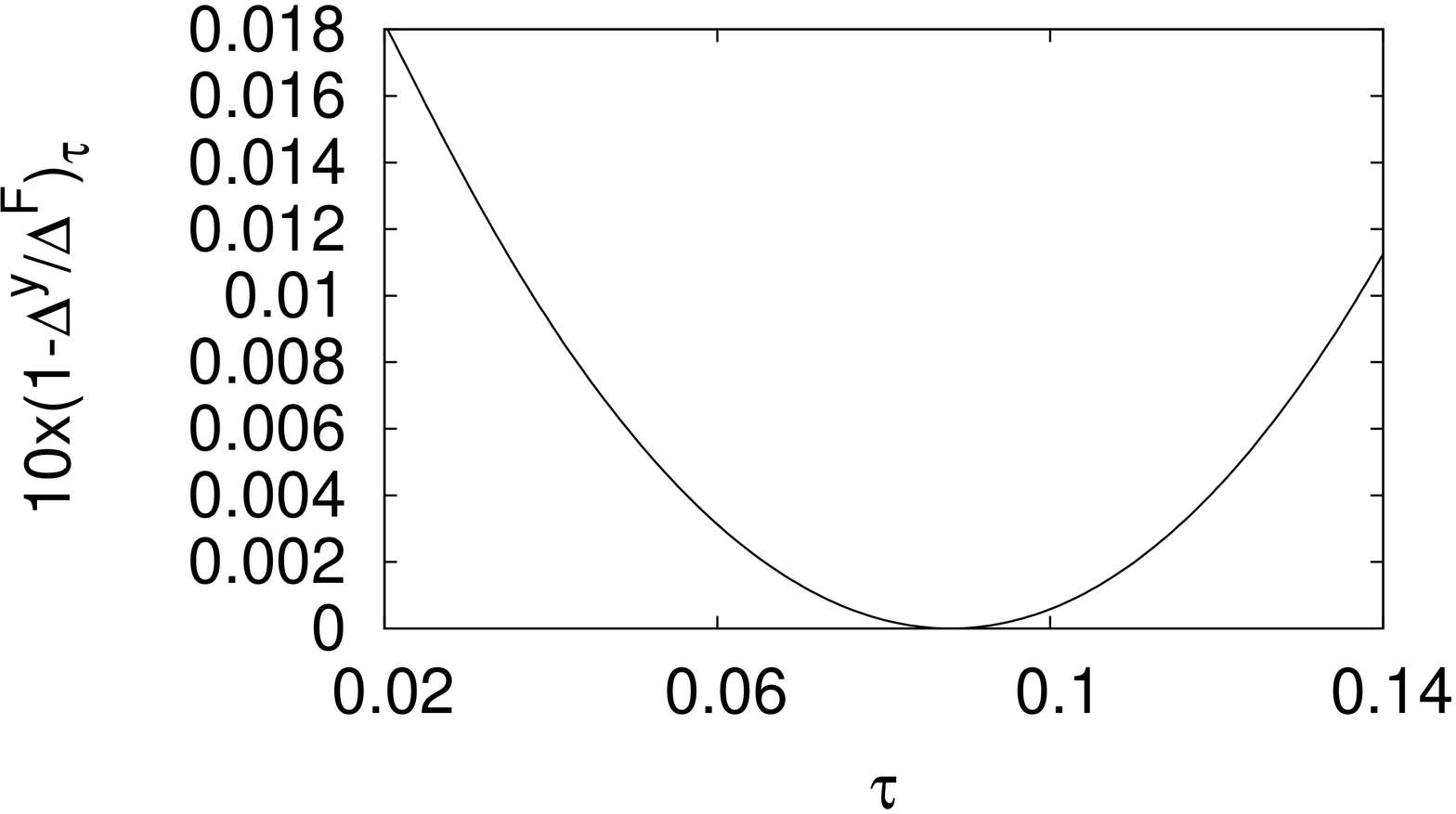}
  \caption{\label{fig:wmap7_TT_6_error} How well can we recover the Fisher matrix using the new compressed data set? 
  We show that the compressed Fisher matrix (Sec. \ref{fish_comp}) is the same as the full matrix in Eq.~(\ref{9c}).
  Shown is the ratio between the error from the compressed Fisher matrix and the Fisher matrix using all
  the data as a function of the value of the parameter assumed when computing the coefficients $\alpha^{i}_{l}$.
  By construction the Fisher matrix is unchanged at the fiducial value of a parameter (and in this case the maximum likelihood point $\theta_{\text{ML}}$). Since the Fisher matrix
  remains the same, the compression is locally lossless. Some plots have been scaled since the ratios are very small. When computing the $C_l$ derivative with respect to $\theta_s$, we keep
  $\omega_c$ and $\omega_b$ constant ($\Omega_{\Lambda}$ and $H_0$ are changed to keep a flat universe). In general, when computing the derivatives with respect to $\omega_c$
  and $\omega_b$, we hold $\theta_s$ constant. However in the plots on $\omega_c$ and $\omega_b$ above, we do not keep $\theta_s$ constant.}
  \end{figure*} 
   \begin{figure}[t!]
  \centering
\includegraphics[width=0.16\textwidth, angle=-90]{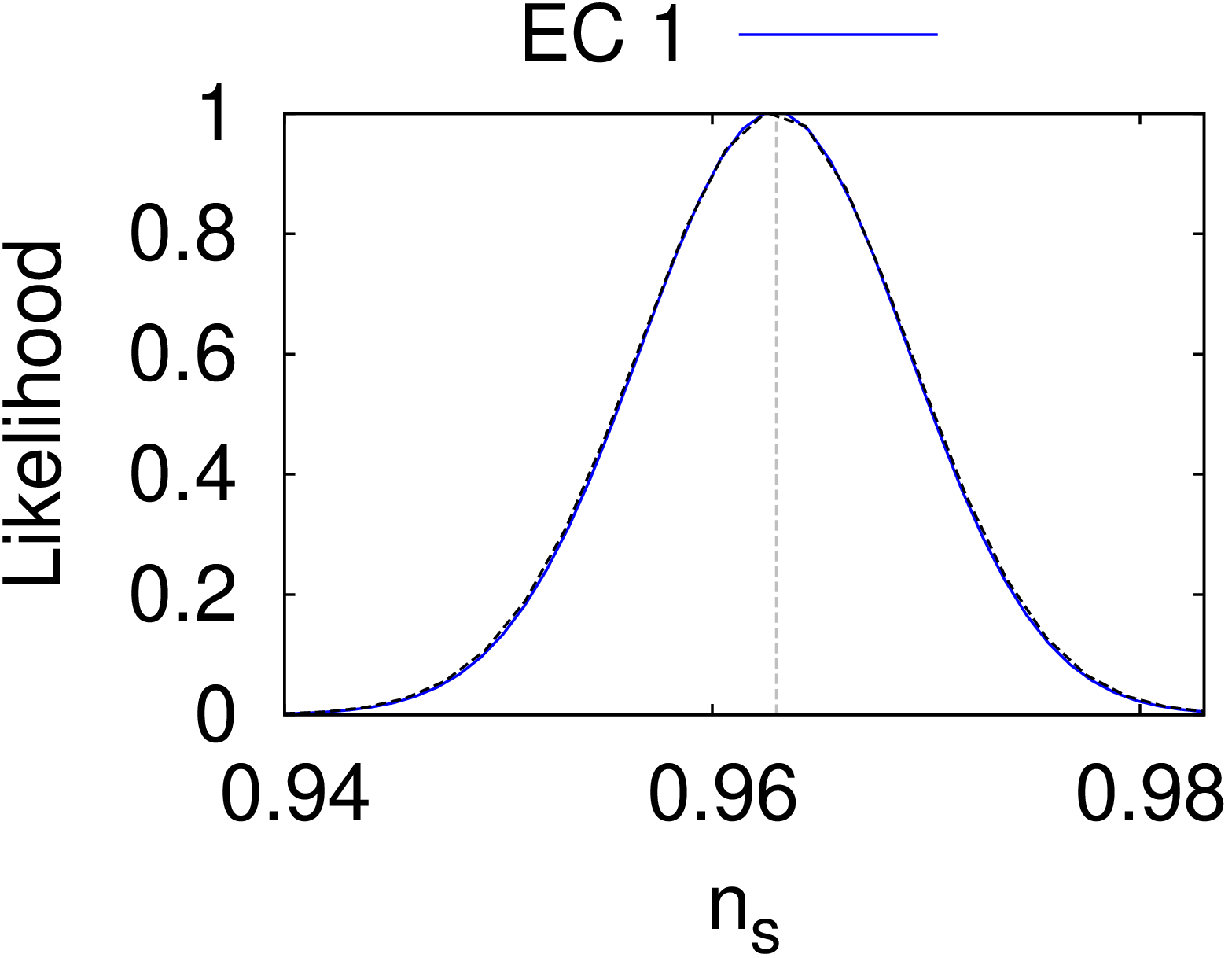}
\includegraphics[width=0.16\textwidth, angle=-90]{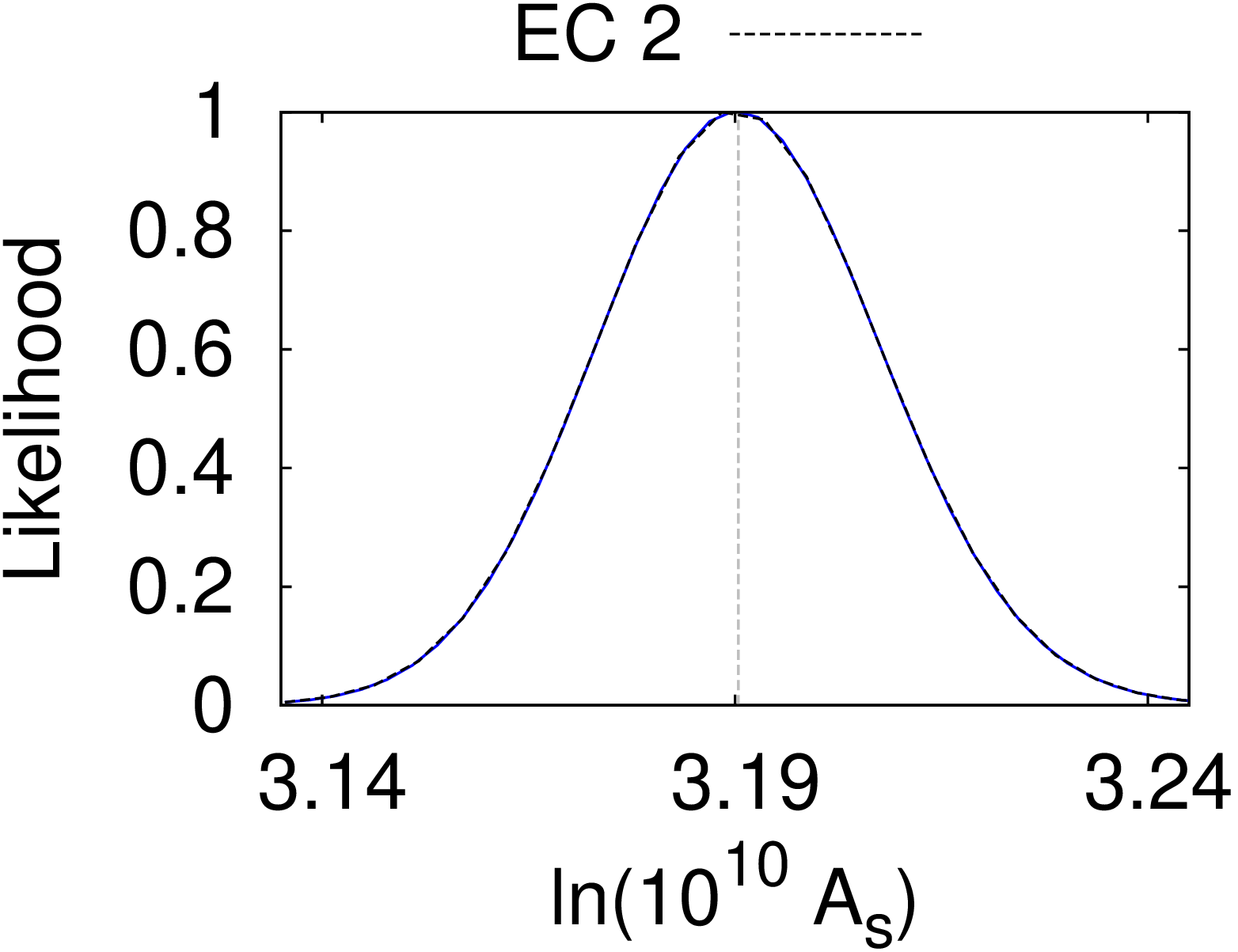}
\vspace*{5mm}
\caption[]{We plot the recovered likelihoods for the case where the weighting vectors are computed with a different fiducial cosmology, and note
the excellent agreement between the two compressions shown in dashed black and solid blue.}
\label{clsthrn22ccc}
\end{figure}
   \begin{figure*}[t!]
\centering
  \includegraphics[width=0.22\textwidth, angle=-90]{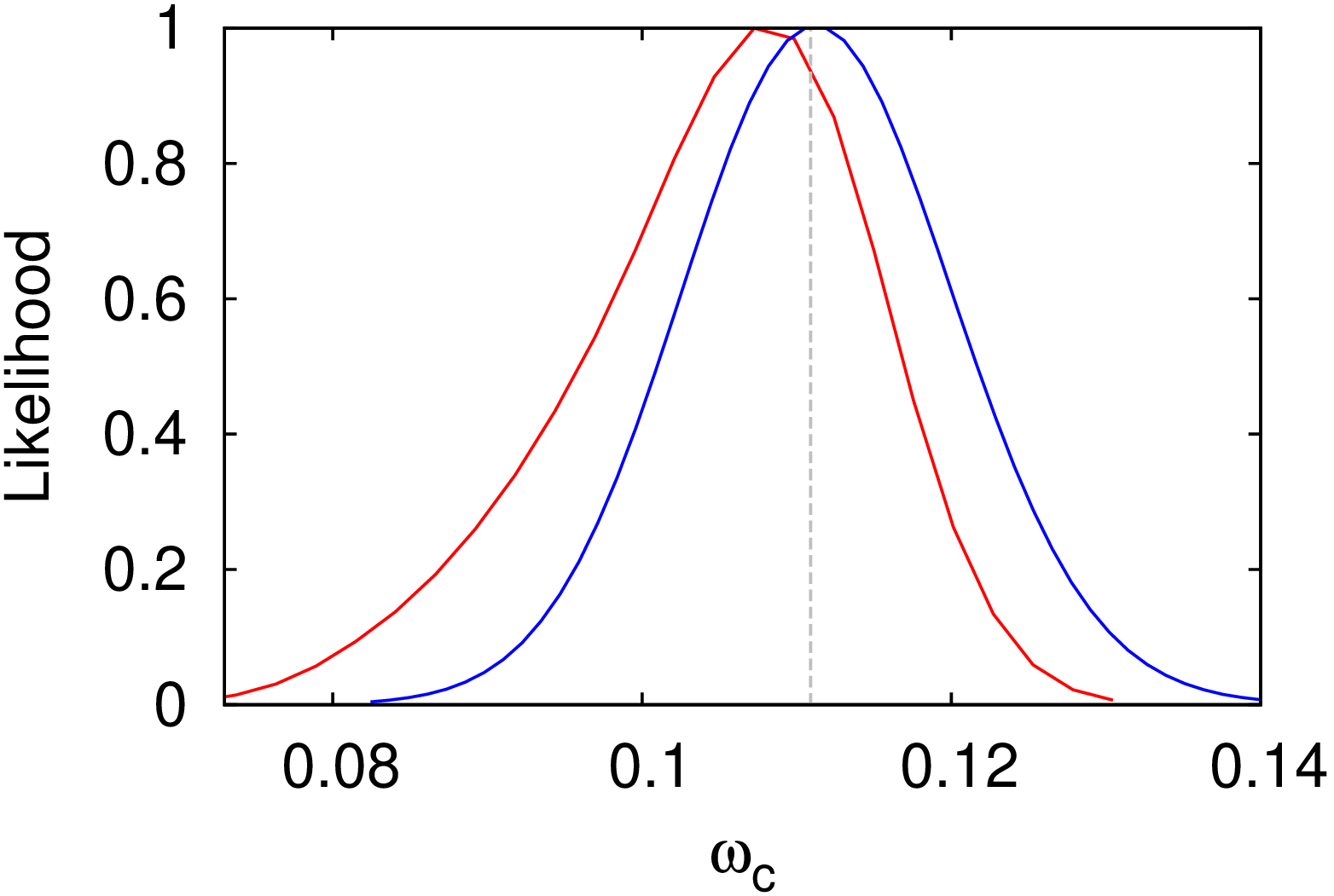}
   \vspace*{3mm}
     \includegraphics[width=0.22\textwidth, angle=-90]{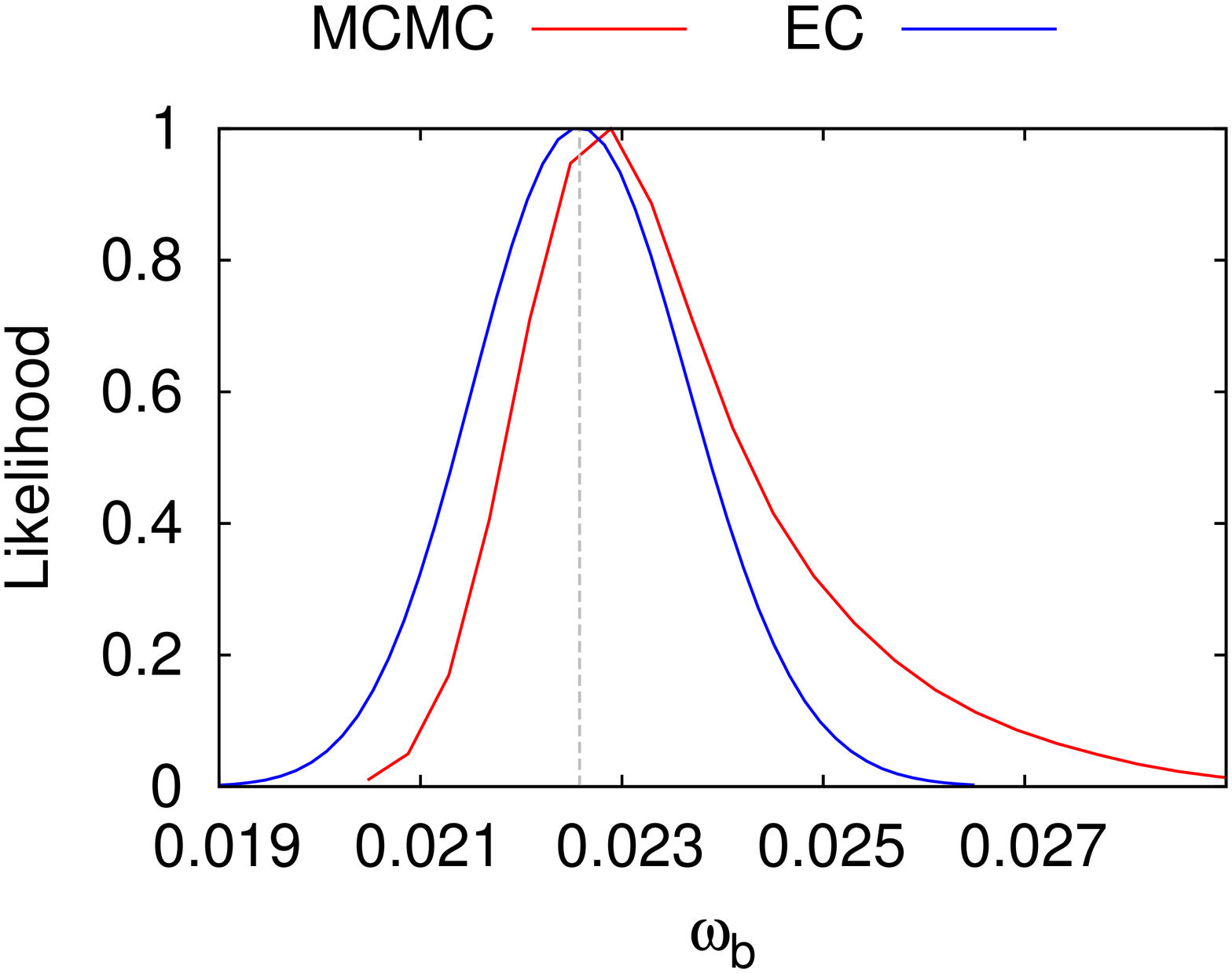}
       \includegraphics[width=0.22\textwidth, angle=-90]{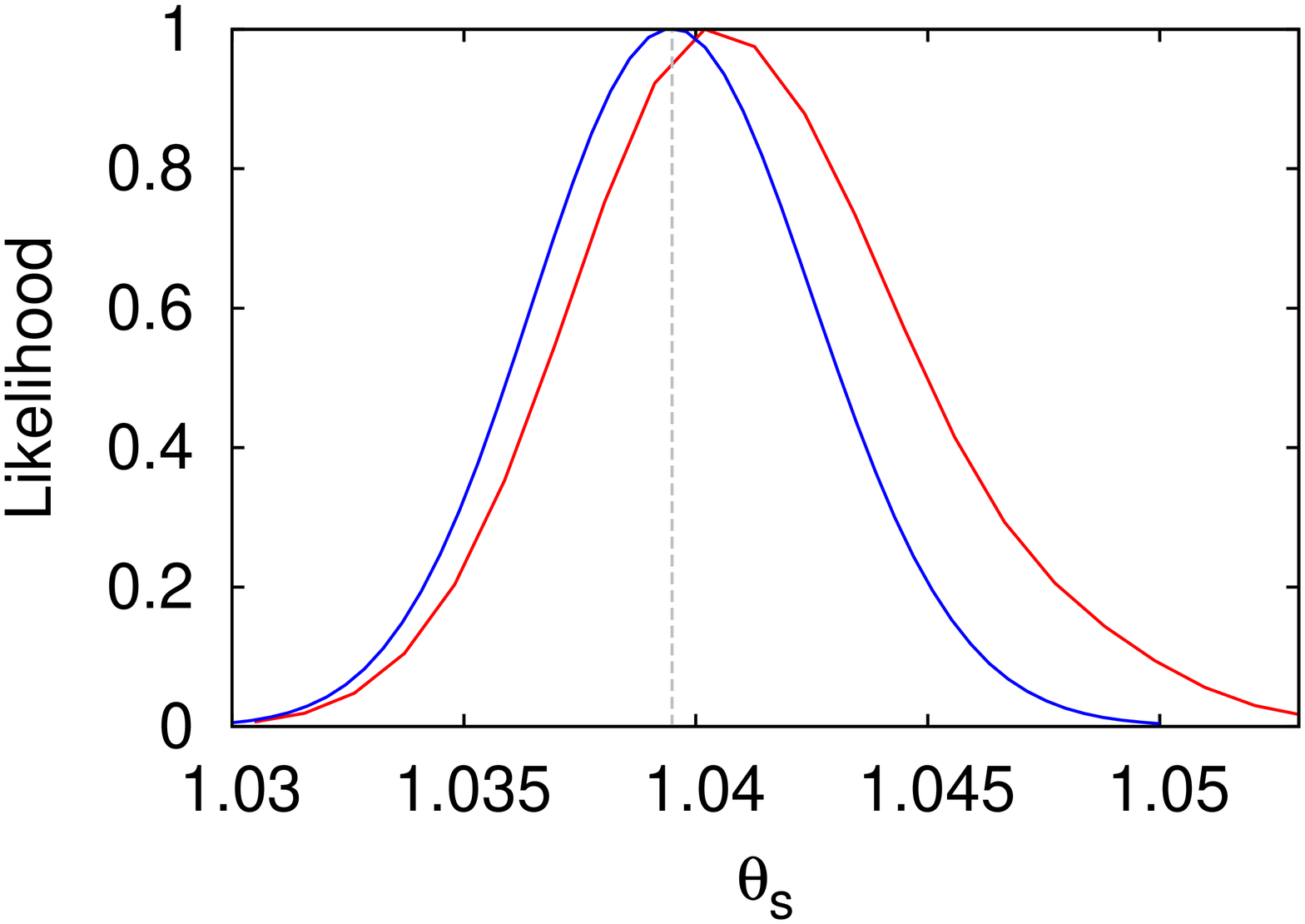}
    \includegraphics[width=0.22\textwidth, angle=-90]{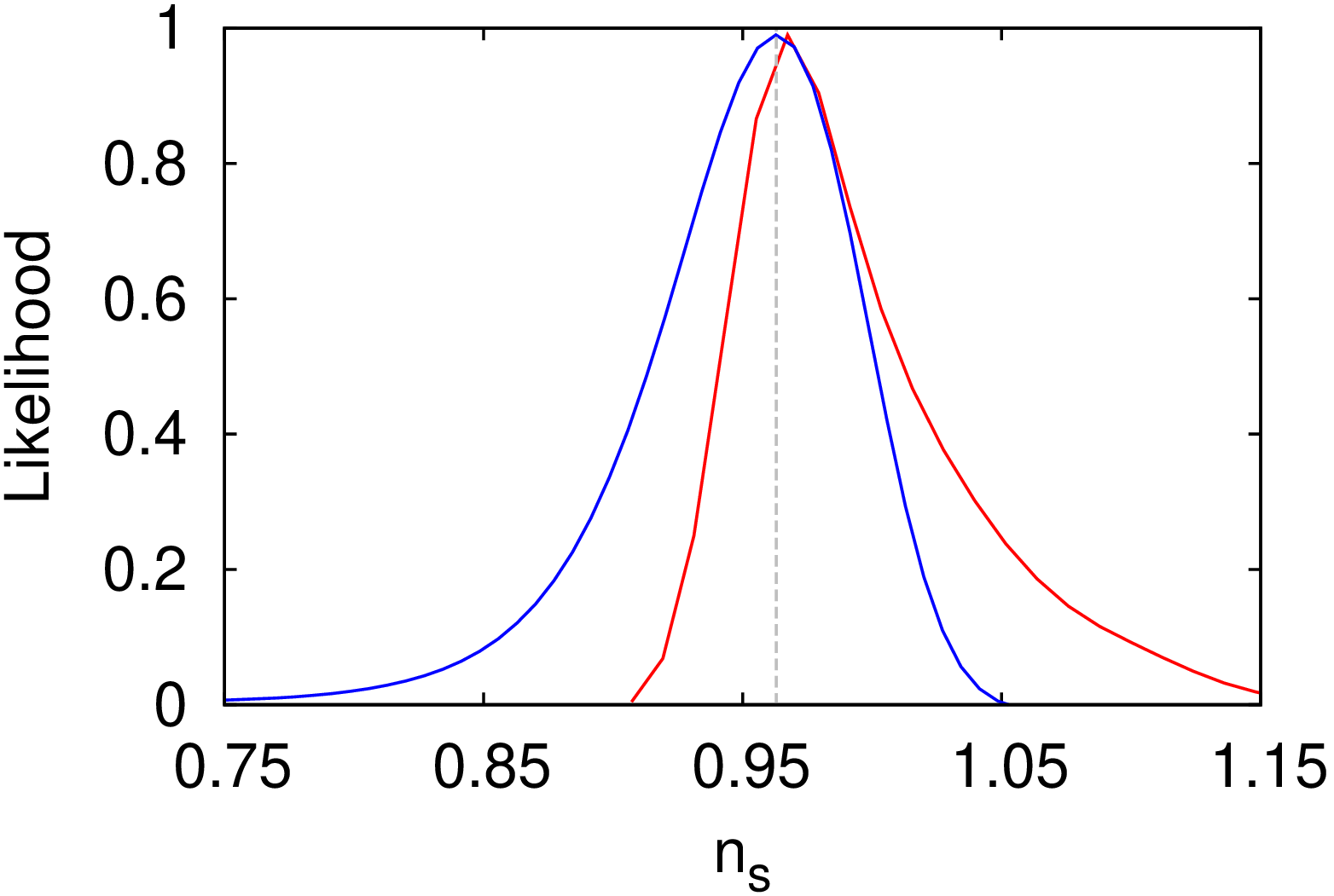}
      \includegraphics[width=0.22\textwidth, angle=-90]{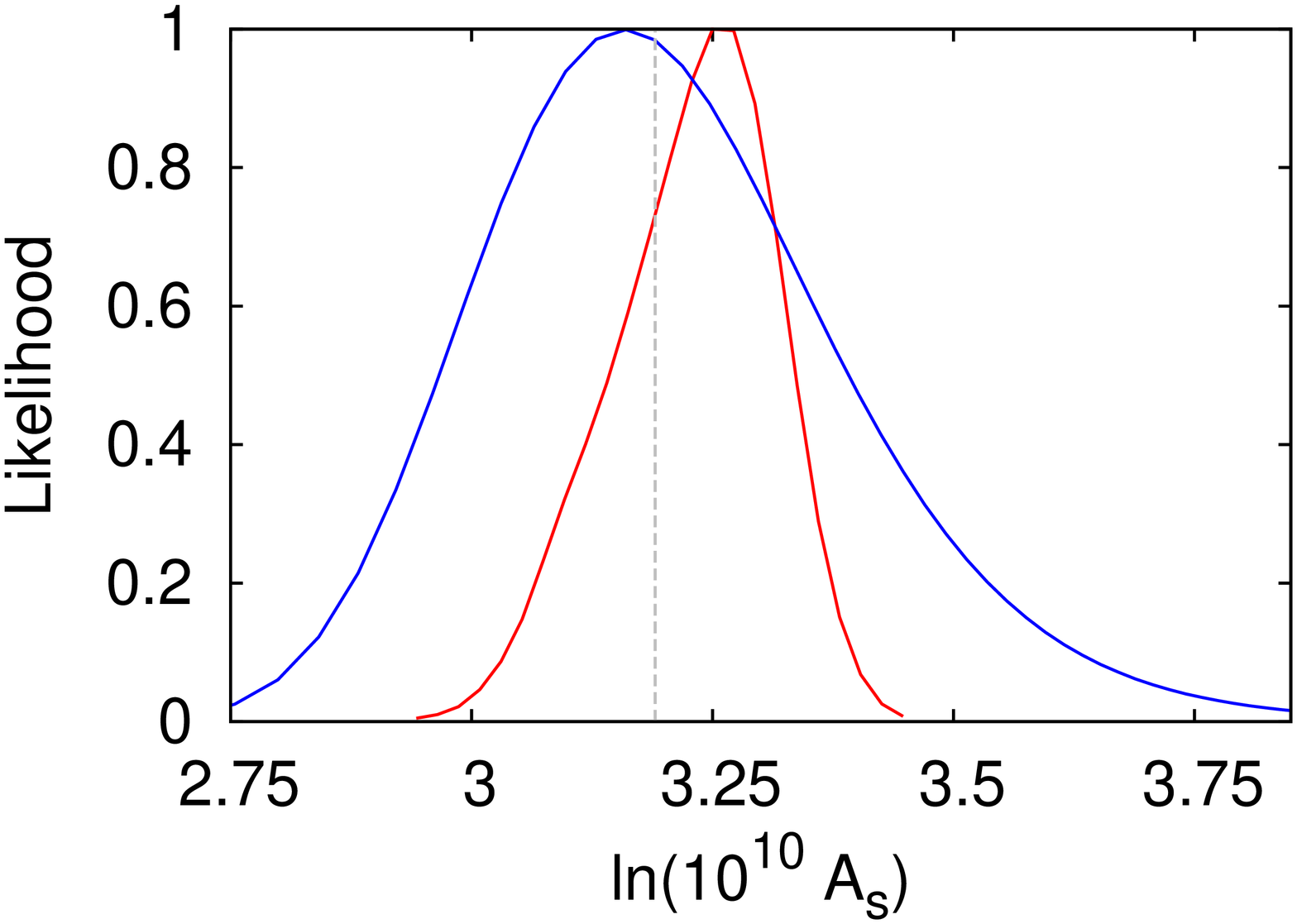}
        \includegraphics[width=0.22\textwidth, angle=-90]{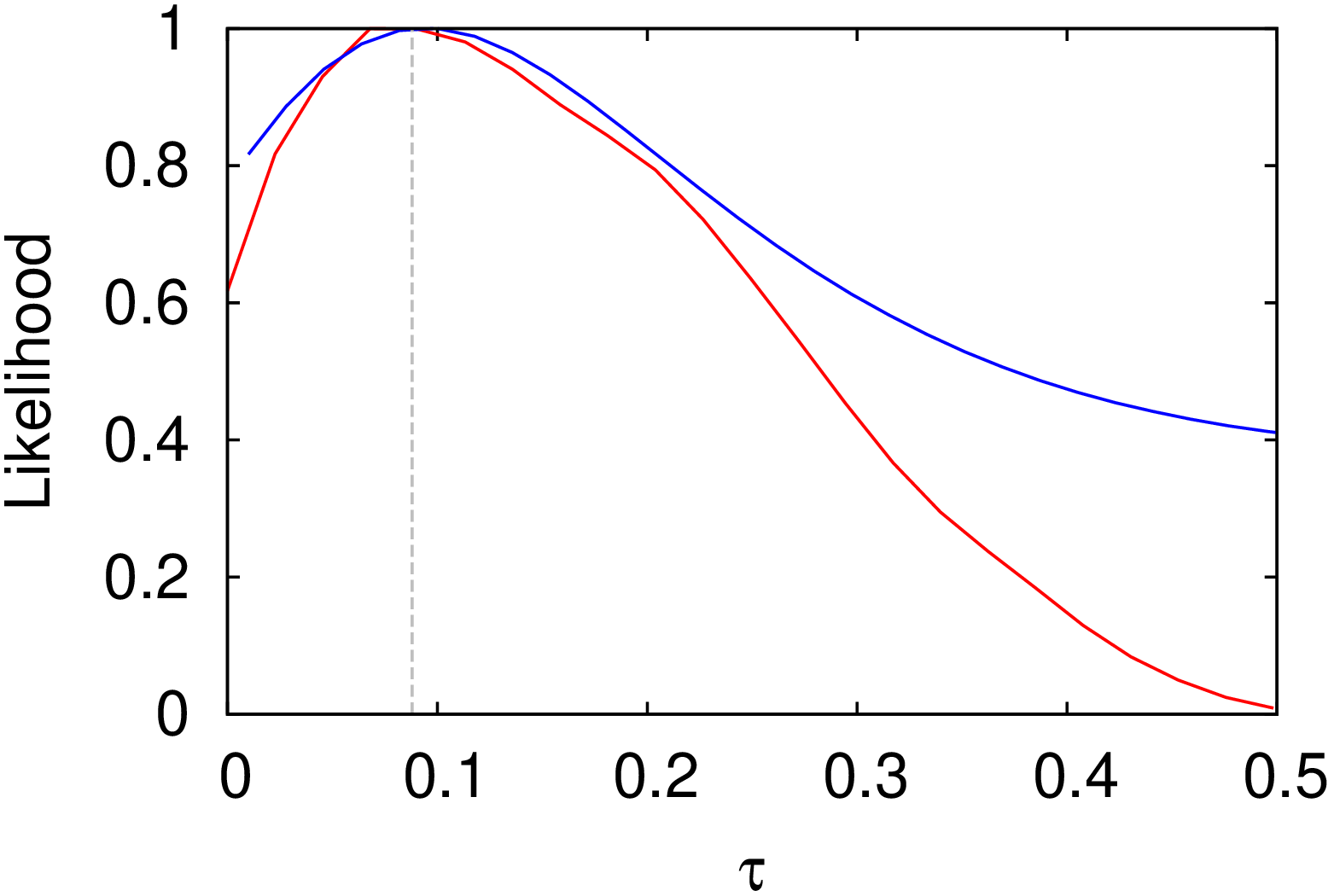}
   \vspace*{6mm}
  \caption{\label{fig:WMAPMCMCmarglike_TT_th} Comparison between the recovered WMAP posteriors from a single compression vector in blue and the
  marginalized likelihoods from a MCMC analysis in red. The data used here is the exact theory temperature power spectrum
  $C_{l}$, with the fiducial cosmology set to WMAP seven-year best-fit parameter values (gray vertical lines). The constraints on the optical 
  depth $\tau$ from the temperature spectrum $C_{l}$ alone are weak, which is reflected in the wide likelihood distribution, although
  even in this case the recovered likelihood peaks at the fiducial value of $\tau=0.088$. Since the parameter combination $A_s e^{-2\tau}$ determines
  the overall amplitude of the observed CMB anisotropy, the recovered value of $\text{ln}(10^{10}A_s)$, the log power of the primordial
  curvature perturbations is slightly biased. Here, we reach the hard limit in the sampler inserted for the redshift of reionization $z_{\text{re}}=40$, which corresponds to $\tau \sim 0.6$. This signals
  that the temperature data alone does not constrain the full six parameter $\Lambda$CDM model very well. }
\end{figure*} 
  \begin{figure*}[htp]
     \includegraphics[width=0.22\textwidth, angle=-90]{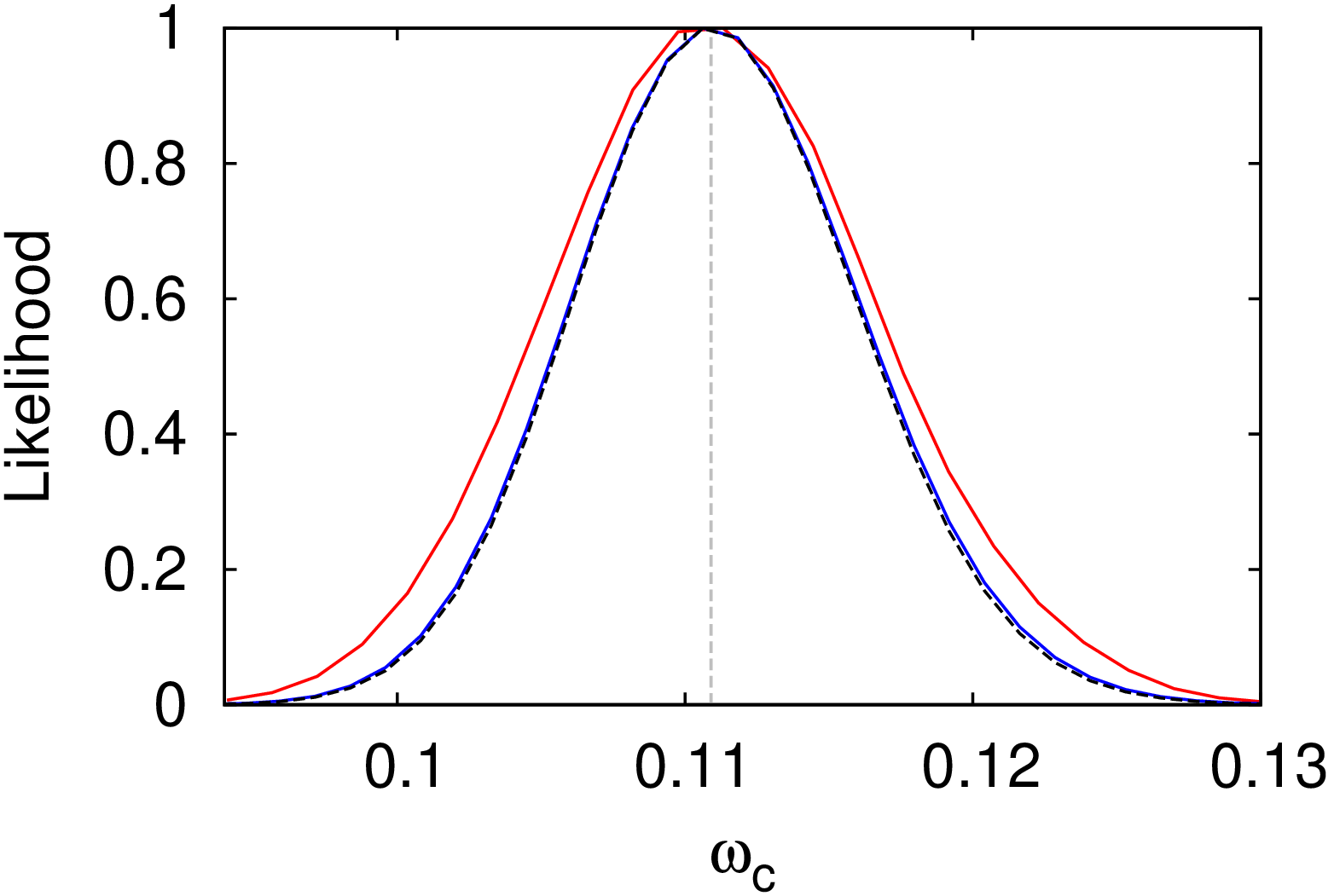}
     \vspace*{3mm}
  \includegraphics[width=0.22\textwidth, angle=-90]{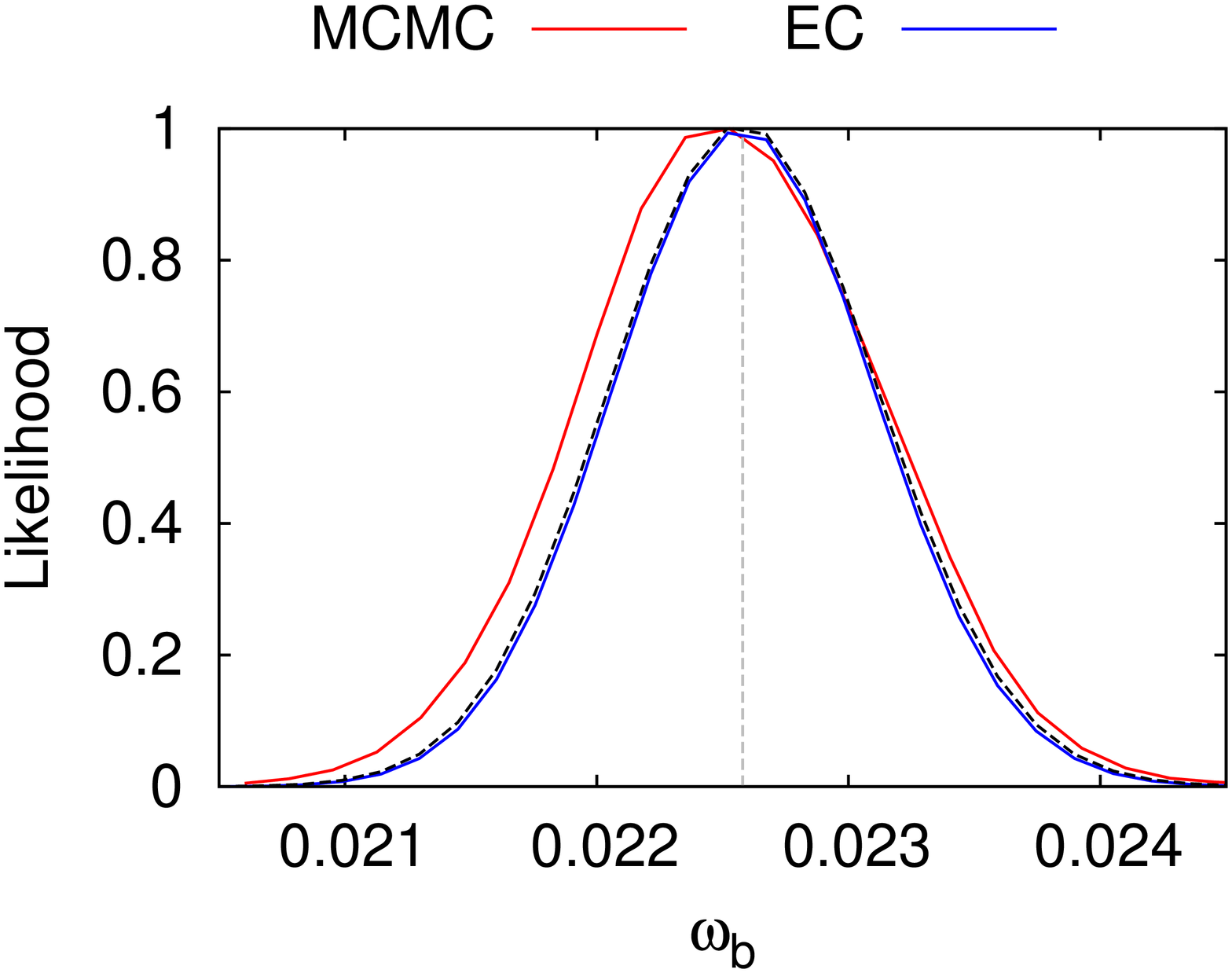}
  \includegraphics[width=0.22\textwidth, angle=-90]{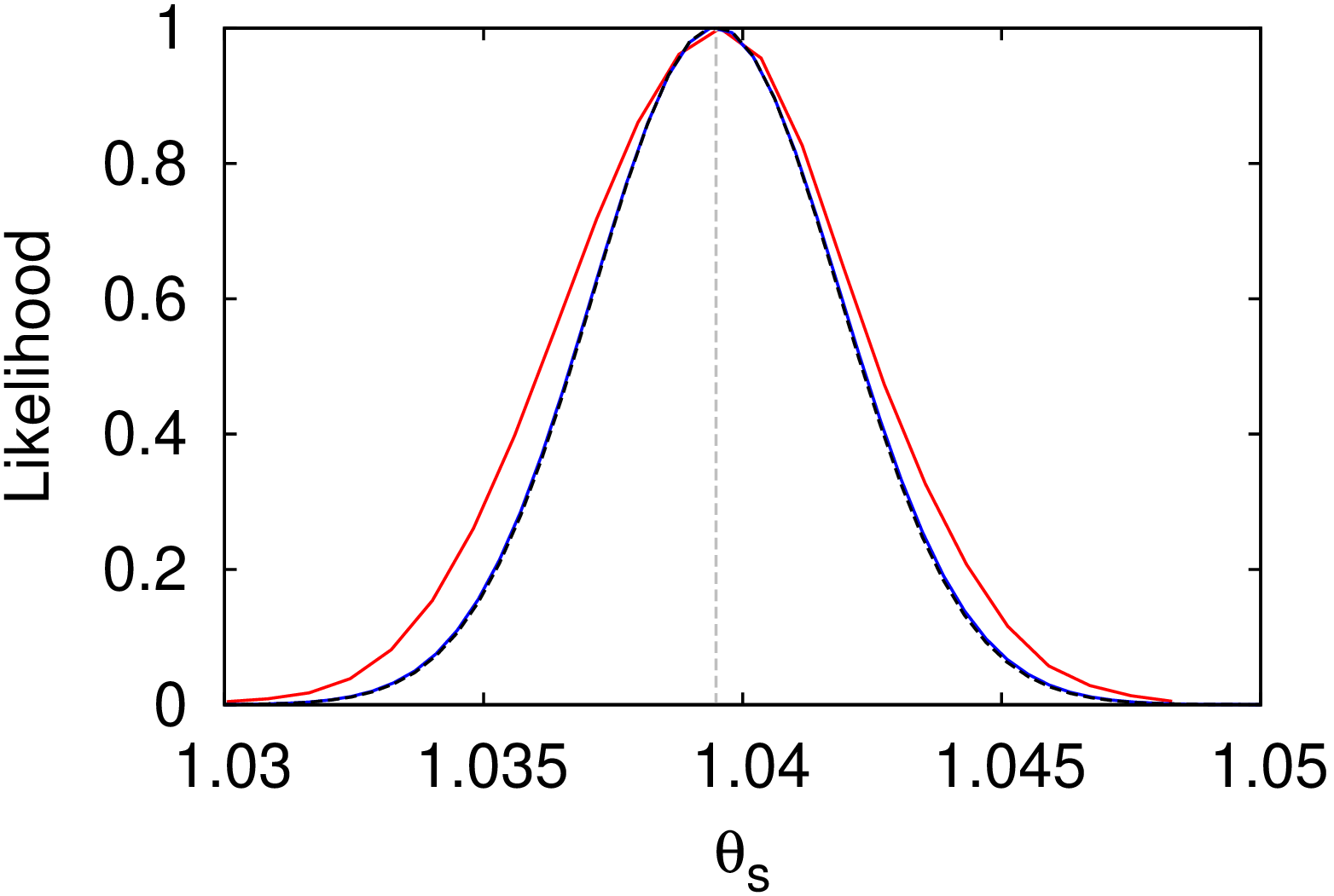}
  \includegraphics[width=0.22\textwidth, angle=-90]{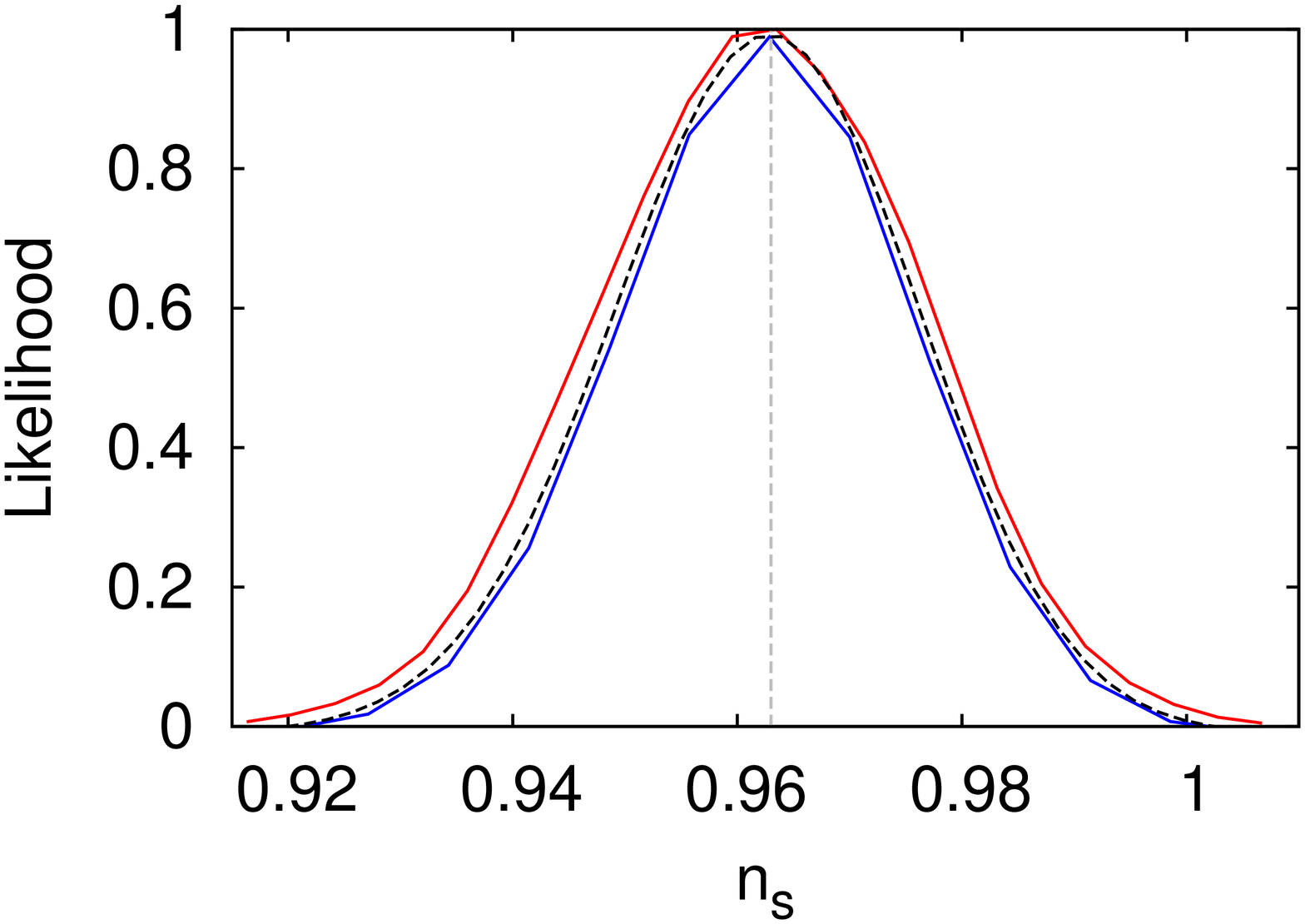}
  \includegraphics[width=0.22\textwidth, angle=-90]{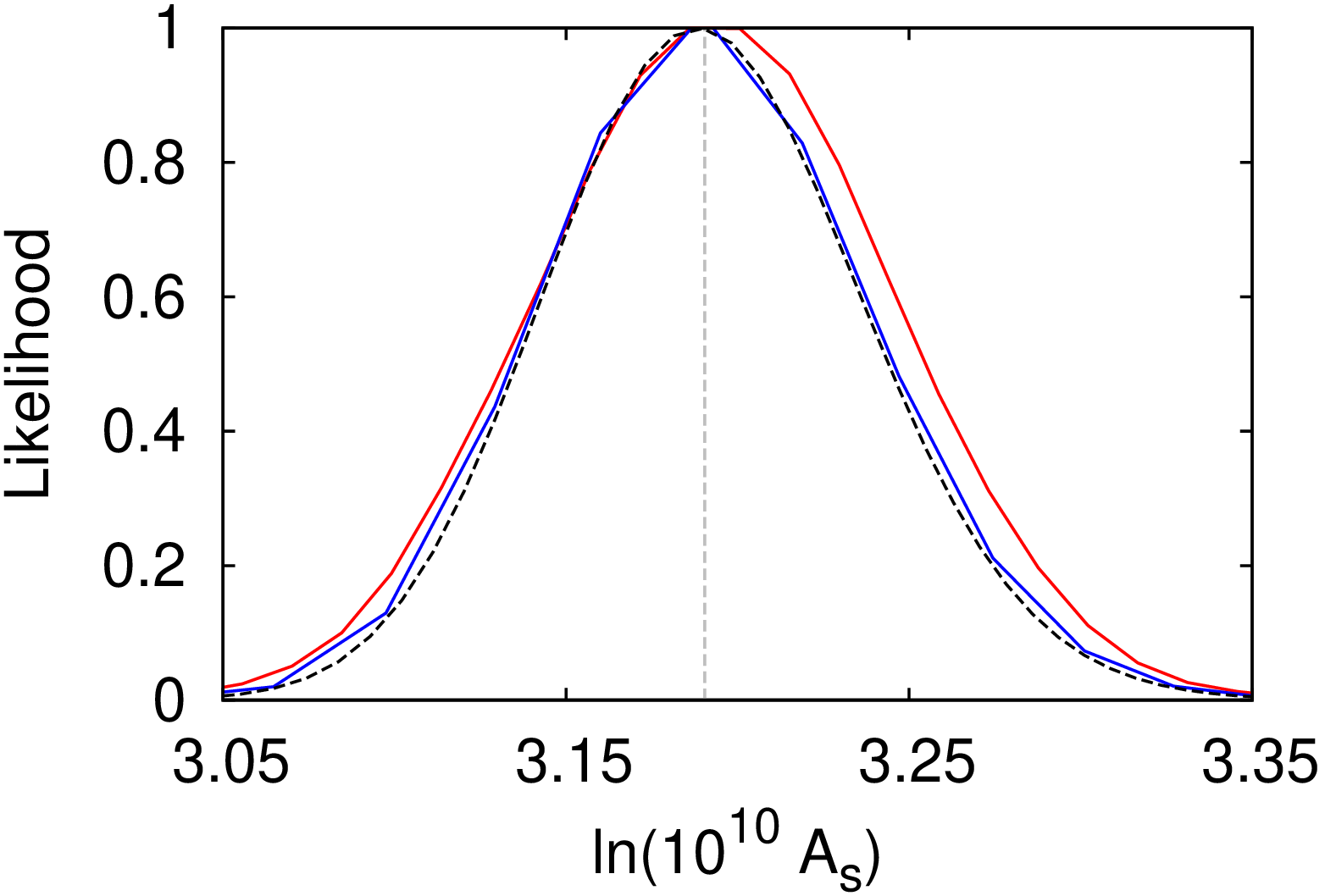}
   \vspace*{6mm}
  \caption{\label{fig:WMAPMCMCmarglike_TT_th_5D} Same as Fig.~\ref{fig:WMAPMCMCmarglike_TT_th} but with $\tau$ held fixed. The posterior distributions
  from the MCMC (solid red lines) agree well with the distributions from the EC analysis (solid blue lines). We also plot the likelihood obtained with 
  weighting vectors which are computed with a different fiducial cosmology (dashed black lines), and note the excellent agreement between the
  two compressions. To compute a second set of weighting vectors we use the following set of
  parameters: $\omega_c=0.12$, $\omega_b=0.0235$, $100*\theta_s=1.0485995$, $n_s=0.98$, ln$(10^{10} A_s) = 3.258$, and $\tau=0.085$.}
  \end{figure*}
  \begin{figure*}[t!]
  \includegraphics[width=0.22\textwidth, angle=-90]{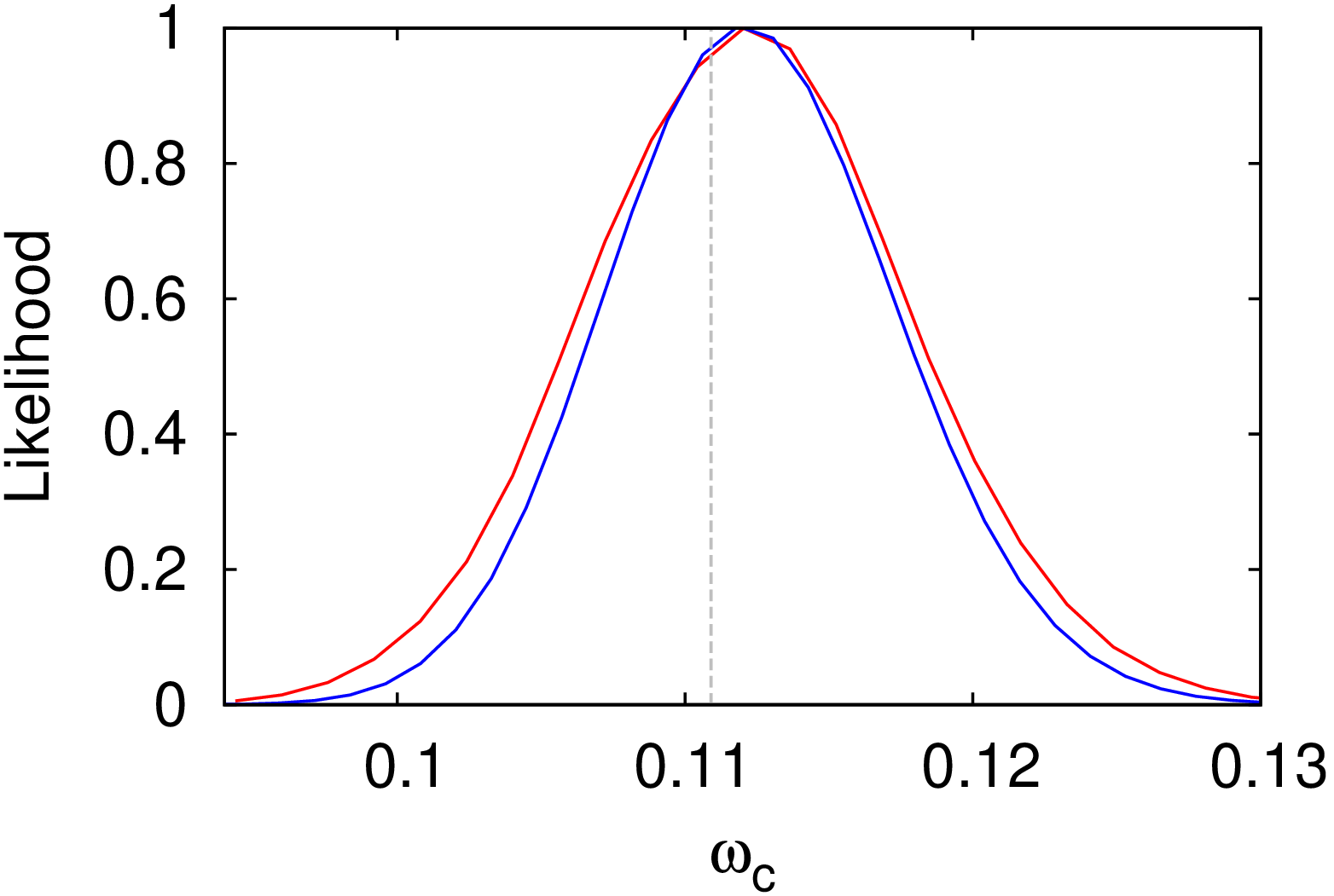}
       \vspace*{3mm}
  \includegraphics[width=0.22\textwidth, angle=-90]{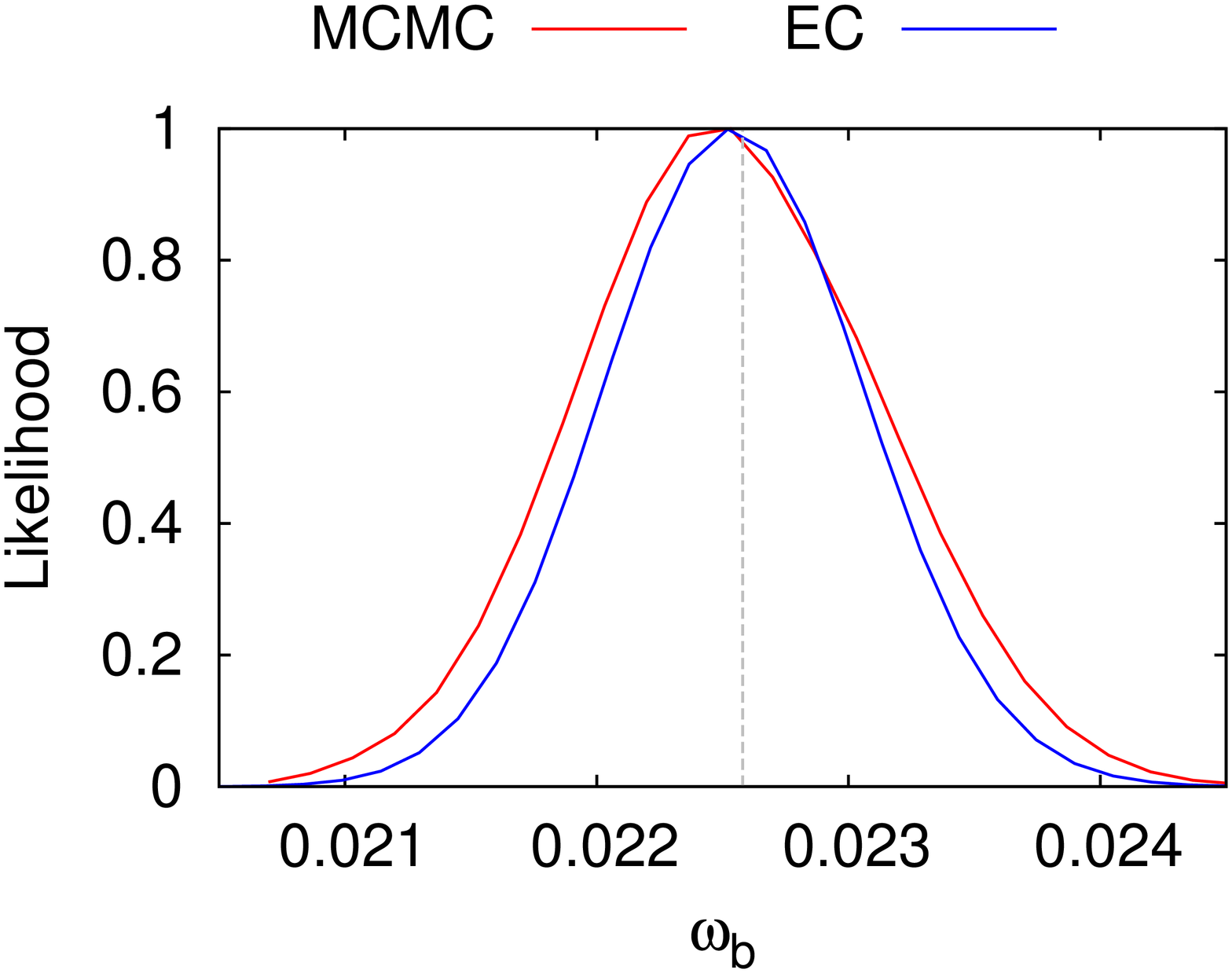}
  \includegraphics[width=0.22\textwidth, angle=-90]{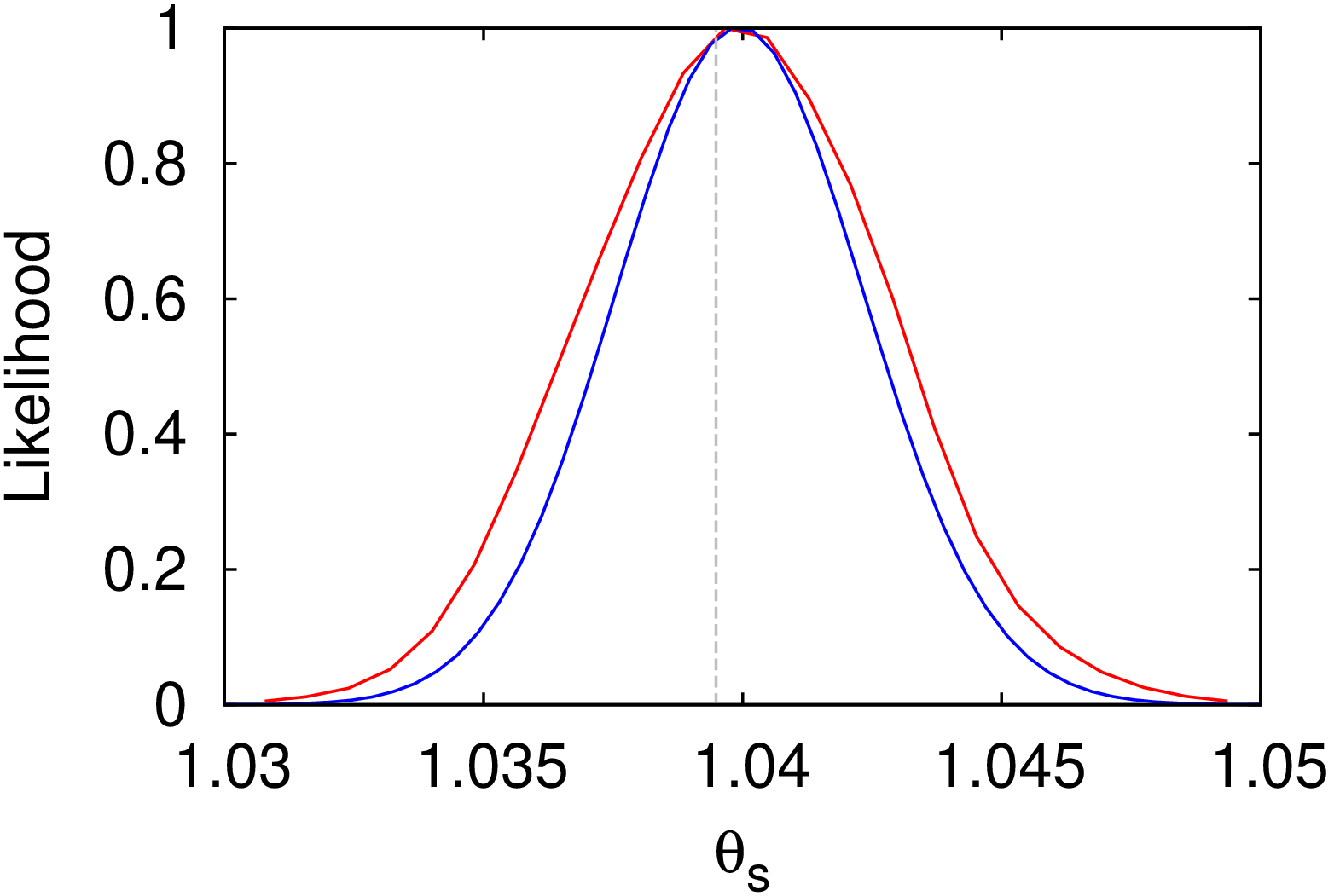}
  \includegraphics[width=0.22\textwidth, angle=-90]{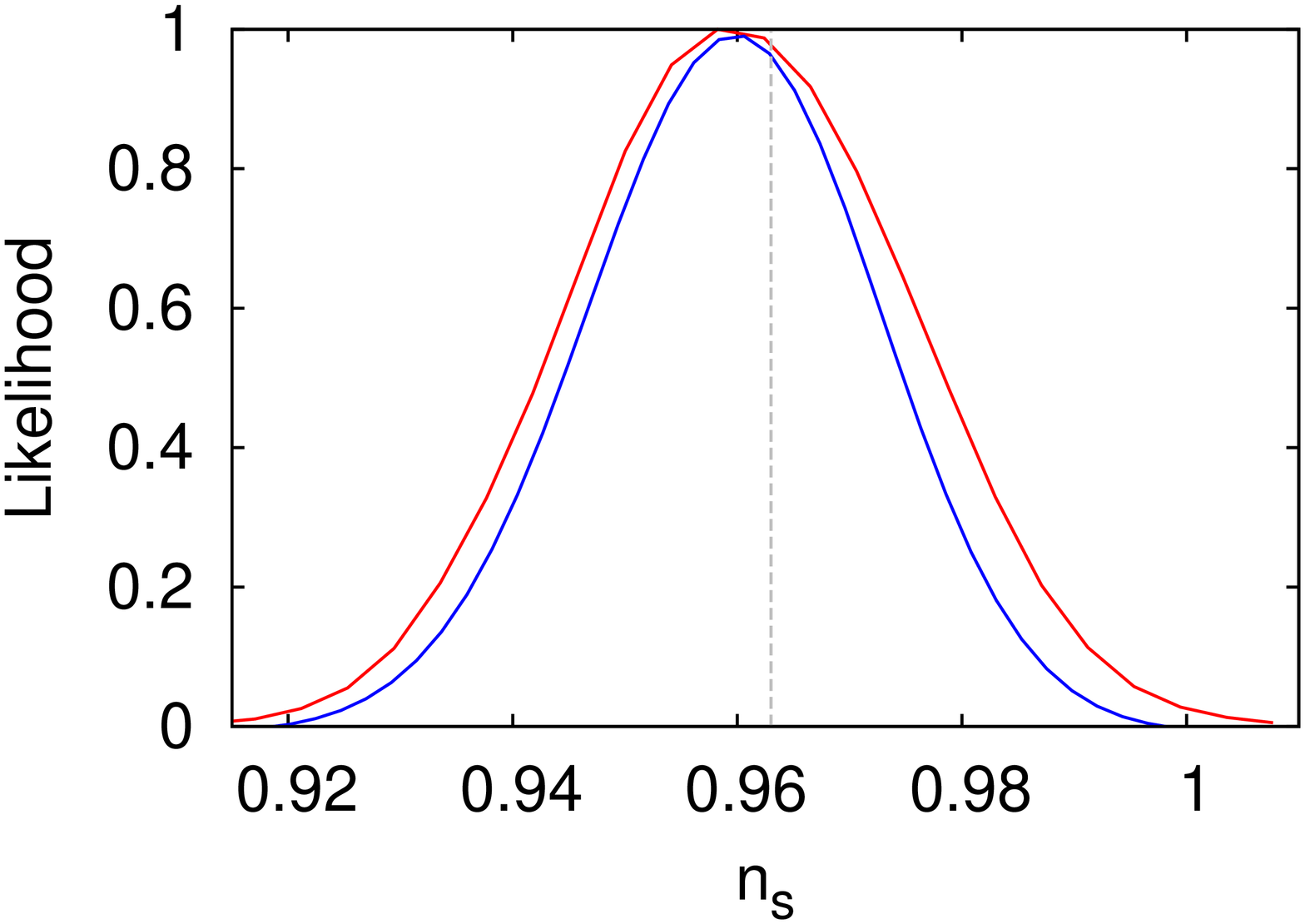}
  \includegraphics[width=0.22\textwidth, angle=-90]{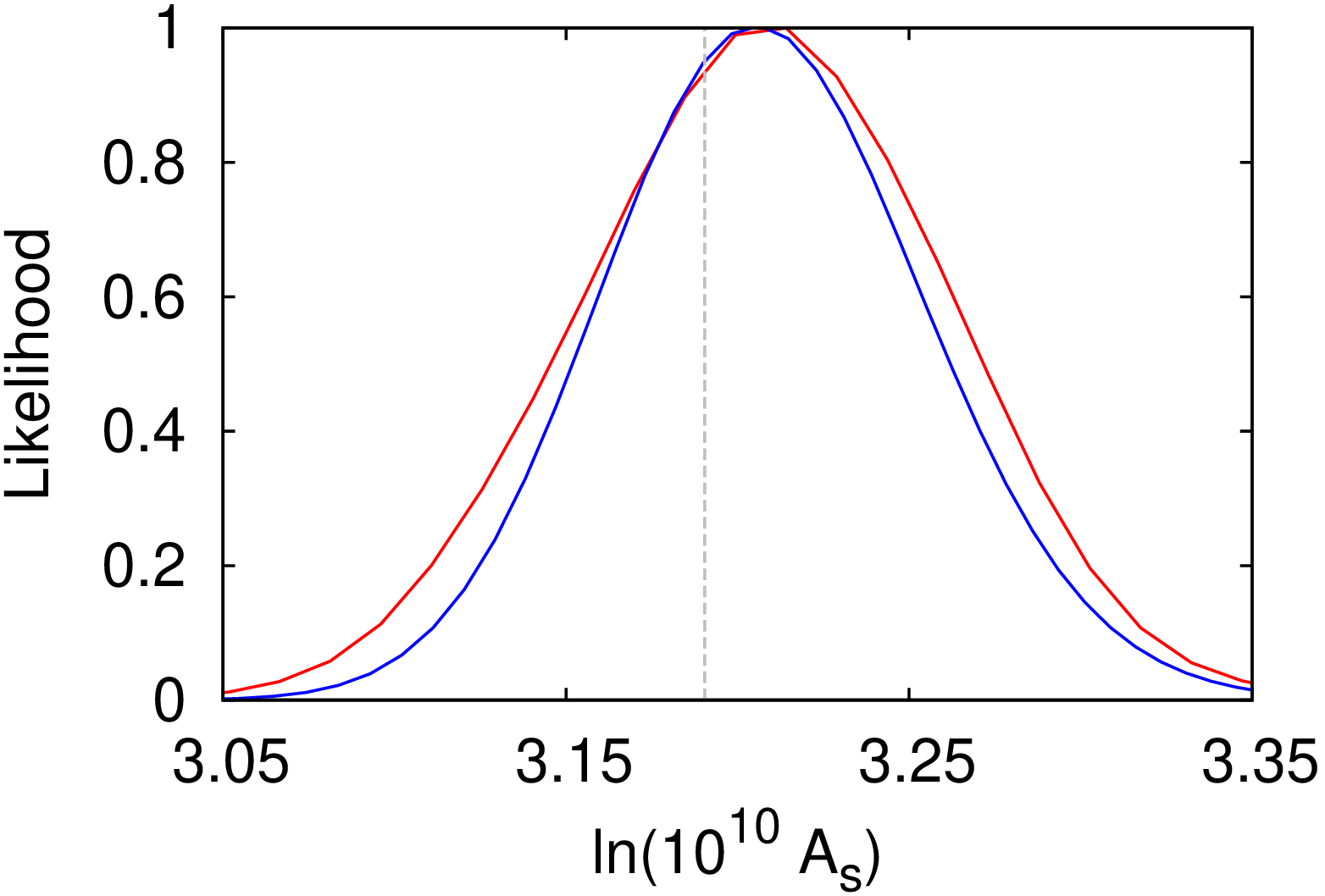}
     \vspace*{6mm}
  \caption{\label{fig:WMAPMCMCmarglike_TT_RN_5D} Same as Fig.~\ref{fig:WMAPMCMCmarglike_TT_th_5D} but for a WMAP-like experiment where the
  CMB power spectrum data set is generated from a random realization of a fiducial cosmology. In Appendix \ref{RAND_generate} we show how we generate our
  random data set. In the case above, we do not expect the posterior distributions (solid blue or solid red lines) to peak at the fiducial parameter input values
  shown with dashed gray lines.}
  \end{figure*}

 In Fig. \ref{fig:wmap7_TT_6_alphas} we show the compression vectors for all the parameters. Due to cosmic variance the data at lower multipoles
 is given a low weight, while for $l>900$ the amplitude of the vectors tends to zero due to the experimental noise. For a WMAP-like experiment, therefore, the vectors all
 peak in the range $l\approx 330-440$, with WMAP being cosmic variance limited up to around $l\sim550$. The jump at $l\sim600$ is due to a discontinuity in the 
 WMAP noise.

 We have already seen that the mode that captures the amplitude $A_s$ is as shown in the middle bottom panel: uniformly positive, but weighing
 the higher signal to noise modes most heavily. The mode that captures the baryon density differences the heights of the first and second peaks, as expected. The sound
 horizon angle is captured by its alternating effect on peaks and troughs. The dark matter density leaves its imprint on the first peak, normalized
 to the most constrained value at $l\sim400$. The optical depth is essentially the inverse vector
 of the amplitude because it enters the temperature spectrum via $A_se^{-2\tau}$, while the mode that captures the spectral index $n_s$ is sensitive to the
 decrease in $C_l$ amplitude as the spectral index increases, up to the first peak.

 In Fig. \ref{fig:wmap7_TT_6d_alphas} we show the marginalized vectors with other parameters removed and compare them to vectors from Fig.
 \ref{fig:wmap7_TT_6_alphas}. We find that many of the qualitative features remain, but the vectors are reduced in amplitude. This is
 the cost of removing the information about other parameters: information degenerate with those parameters about the parameter
 of interest is also removed. 

These vectors can be a useful tool to determine the relative importance of obtaining cosmic variance limited measurement of the power spectrum versus 
a higher sensitivity measurement $C_l$ at smaller scales. A recent example is the apparent need for a precise measurement of the reionization bump
 in order to break parameter degeneracies and obtain the best constraints on the sum of neutrino masses from a stage 4 CMB
 experiment \cite{2015arXiv150907471A}.
 
\subsection{Sensitivity to fiducial choice}
 
The vectors shown in the previous section are solutions to an eigenvalue problem that minimizes the error on each parameter, and leaves the Fisher matrix locally unchanged.
At the fiducial point, at which the derivatives and the covariance are computed, we expect the errors from the compressed Fisher matrix using the extreme compression to equal
those from the full Fisher matrix. But how well can we recover the parameters if the coefficients $\alpha^{i}_l$ are chosen away from the fiducial point, and how much does the error bar increase?

In Fig.~\ref{fig:wmap7_TT_6_error}, we show the ratio of the error from the Fisher matrix obtained with the extreme compression $\text{F}^{y}_{ii}$ to the error
obtained using the full data set from Eq.~(\ref{9c}). Even over a wide range of parameter space (roughly the same as the expected width of the marginalized posteriors from
WMAP) we find that $\Delta\theta_i$ changes by less than 0.2\% for parameters $n_s$, $A_s$ and $\tau$, while the errors increase by at most 2\% for $\omega_b$ 
and $\theta_s$. For the physical cold dark matter density $\omega_c$, the error change is less than 8\%. At the 
fiducial point, the compression is locally lossless.

Another important question that we address is whether the fiducial cosmology used in the compression affects the results. To test whether the choice
of the fiducial point matters, we created a new set of compression vectors $\alpha^{i}_l$ computed at a different cosmology, denoted as EC 2, with the
following values for the cosmological parameters: $\omega_c=0.12$, $\omega_b=0.0235$, $100*\theta_s=1.0485995$, $n_s=0.98$, ln$(10^{10} A_s) = 3.258$, and
$\tau=0.085$. We then marginalized over all other parameters and used the new marginalized vectors to compress an exact theory $C_l$ data set
with WMAP-like noise.

In Fig.~\ref{clsthrn22ccc}, we plot the recovered likelihoods when compressing the data with our fiducial cosmology denoted as EC 1 (solid blue lines), and
the new cosmology as EC 2 (solid black lines). Figure \ref{clsthrn22ccc} shows that no matter what the fiducial point we choose, we still get back the correct answer.
 
\subsection{How does extreme compression compare to a MCMC calculation?}

Once we form the compression vectors, it is easy and very fast to compute the likelihood for each parameter, with a typical time of less than a 
minute. To test the method and to see how well we can recover the posterior probabilities, we first analyze a mock data set, where the observed data set is
the set of theory $C_l$'s. In this case, we expect an unbiased estimate of the input cosmology from both our method and the MCMC.

Since the spherical harmonic coefficients $a_{lm}$ are Gaussian random variates and are statistically isotropic, the likelihood function for the temperature 
power spectrum is a
Wishart distribution with $\mathcal{P}(\hat{\mC_{l}}|\mC_{l}) \propto \mathcal{L}(\mC_{l}|\hat{\mC_{l}})$ and
\begin{eqnarray}
\chi_{\text{eff}}^{2} = && -2\text{ln}\mathcal{L}(\mC_{l}|\hat{\mC_{l}}) = \nonumber\\
 && \sum^{lmax}_{l=2}(2l+1)\left( \frac{\hat{\mC_{l}}}{\mC^{th}_{l}} + \ln\left(\frac{\mC^{th}_{l}}{\hat{\mC_{l}}}\right) - 1\right), \nonumber\\
\label{eqn:cmblike}
\end{eqnarray}

where $\hat{\mC_{l}}$ is the observed data set \cite{2000ApJ...533...19B,2008PhRvD..77j3013H,2009PhRvD..79h3012H}.
The above likelihood is a general case for an experiment with no noise and a full-sky coverage. In practice, experiments have noise and observe
only a fraction of the sky. We modify Eq.~(\ref{eqn:cmblike}) by replacing $\mC^{th}_l$ with $\mC^{th}_{l}+\mN_l$, and by decreasing the number of modes on the sky
from $(2l+1)$ to $(2l+1)\fsky$. Both $\mC^{th}_l$ and $\mN_l$ appear in the likelihood because they are both Gaussian random fields. Note that the
likelihood above is normalized such that $\chi_{\text{eff}}^{2} = 0$, when $\hat{\mC_{l}}=\mC^{th}_{l}$. 

In our WMAP mock MCMC likelihood calculations we assume that the fraction of the remaining sky after applying the WMAP mask KQ85y7 is 78.3\% \cite{2011ApJS..192...16L}.
When analyzing the WMAP seven-year data however, we use the sky fraction contained in the WMAP likelihood code, which varies with the multipole $l$. 

We show our results in Fig.~\ref{fig:WMAPMCMCmarglike_TT_th}, where we plot the MCMC posteriors in solid red and the result using our compressed vectors
in solid blue. Because the Thomson scattering optical depth due to
reionization is not well constrained by the temperature spectrum alone, the MCMC posterior has a wide, non-Gaussian distribution and the 95\% C.L. upper
limit for $\tau$ is 0.36. The extreme compression formalism implicitly assumes Gaussian distributions for the parameters, so the $\tau$ distribution
offers a nice test of the impact of the breakdown of this assumption on the full analysis. Figure \ref{fig:WMAPMCMCmarglike_TT_th} shows that the impact
falls mainly on the parameter $A_s$ with which $\tau$ is degenerate (recall that the amplitude of the perturbations is roughly $A_s e^{-2\tau}$). The
ensuing bias on $A_s$ is small: relative to the mean $\mu$ from MCMC, the value of $\text{ln}(10^{10}A_s)$ is biased low by $0.88\sigma$, where the error on $\text{ln}(10^{10}A_s)$ is
$\sigma = 0.0814$. Note that in general the ensuing biases are smaller when the maximum likelihood is used, as opposed to the mean likelihood. In 
Table \ref{tab:table1}, we show the bias on $\text{ln}(10^{10}A_s)$ for exact theory $C_l$ and a random catalog.

If we fix the optical depth to its fiducial value of $\tau=0.088$, we obtain the results shown in Fig.~\ref{fig:WMAPMCMCmarglike_TT_th_5D}, and then
the likelihood results from the MCMC and EC are in very good agreement. In this case the MCMC means and the estimates from EC coincide
with the input cosmology.

Figure \ref{fig:WMAPMCMCmarglike_TT_th_5D} also illustrates that the EC method is insensitive to the choice of fiducial parameters. The dashed black curves show the likelihoods when the coefficients
$\alpha^{i}_{l}$ are chosen assuming the nonfiducial parameter set:  $\omega_c=0.12$, $\omega_b=0.0235$, $100*\theta_s=1.0485995$, $n_s=0.98$, ln$(10^{10} A_s) = 3.258$ and $\tau=0.085$. The figure shows that shifts of this order leave no imprint on the final likelihood.

Before analyzing real data, we investigate how our method performs on a random mock. We create a realistic mock for a full-sky 
CMB experiment with WMAP noise. We discuss random mock generation in Appendix \ref{RAND_generate}. Figure \ref{fig:WMAPMCMCmarglike_TT_RN_5D} shows the posteriors in a $\Lambda$CDM model with $\tau$ fixed at its fiducial value. Again the two distributions agree very well.

In the next section we apply the methods discussed so far to the seven-year WMAP temperature spectrum, and compress the 
temperature spectrum to estimate the cosmological parameters with WMAP precision.
\begin{table}[t]
\caption{\label{tab:table1} Bias in the recovered values of $\text{ln}(10^{10}A_s)$ using extreme compression relative to both
the mean $\mu$ and the best fit (the maximum likelihood point $\theta_{\text{ML}}$) from MCMC.}
\begin{ruledtabular}
\begin{tabular}{C{1.9cm}C{1.9cm}C{1.9cm}C{1.5cm}}
 \head{0.5cm}{\text{Model \&} \text{Parameters}}  & \head{0.5cm}{\text{Bias relative} \text{to $\mu$}} & \head{0.5cm}{\text{Bias relative} \text{to $\theta_{\text{ML}}$}}  & \head{0.5cm}{\text{Standard} \text{Deviation $\sigma$}}
\\ [0.5ex]
\colrule
\rule{0pt}{2.5ex} TH $\tau$ free    &   $-0.88 \sigma$  &   $-0.43 \sigma$  &  0.0814  \\
\rule{0pt}{2.5ex} TH $\tau$ fixed   &   $-0.07 \sigma$  &   0.15$\sigma$  &  0.0492  \\
\rule{0pt}{2.5ex} RN $\tau$ fixed   &   $-0.05 \sigma$  &   0.20$\sigma$  &  0.0508  \\
%\rule{0pt}{2.5ex} Data $\tau$ free  &   -1.56$\sigma$  &   0.06$\sigma$  &  0.0858  \\
%\rule{0pt}{2.5ex} Data $\tau$ fixed &   -0.37$\sigma$  &   0.34$\sigma$  &  0.0470  \\
\end{tabular}
\end{ruledtabular}
\end{table}
\section{Results}
\label{results}
In the previous section we analyzed mock data to see how well we can recover the input cosmology, and we compared the results of the extreme
compression to the MCMC means and best-fit (maximum likelihood) MCMC results. In
this section we apply the methods to a real data set and as an example choose the seven-year WMAP temperature spectrum. Although this is not the most
up-to-date CMB data set, it is a useful test which will inform further development of the EC method. For this analysis, we formulate the 
vectors that compress the WMAP spectrum using the same WMAP noise and fraction of the sky observed as in the WMAP likelihood.
Since the WMAP likelihood is not a simple Gaussian, and consists of a number of components, we review the likelihood briefly in the next section. We
discuss how this will affect our results in Sec. \ref{data7}.
\subsection{WMAP likelihood}
\label{explain}
The full WMAP likelihood is made up of ten components, four of which form part of the temperature analysis. The analysis is split up into low-$l$ and high-$l$
components. For multipoles $l\le32$, there is a choice between a direct evaluation of the likelihood in pixel space and one using Gibbs
sampling (see \cite{2009ApJS..180..306D} and the references therein). The default is Gibbs sampling, where the spectrum is obtained using a
Blackwell-Rao estimator applied to a chain of Gibbs samples. For multipoles $l\ge33$, the likelihood uses the spectrum  derived from the MASTER
pseudo-$C_l$ quadratic estimator and a covariance matrix \cite{2002ApJ...567....2H,2007ApJS..170..288H}. In addition, there are terms in the likelihood
due to uncertainty in determining the
WMAP beam and the error in the extragalactic point source removal (for details see the appendix of \cite{2007ApJS..170..288H}). 

For a large $l$, Eq.~(\ref{eqn:cmblike}) can be approximated as Gaussian $\ln {\cal L}_{\rm Gauss}$, but since the likelihood function for the power
spectrum is slightly non-Gaussian, this gives a biased estimator. Although \cite{2000ApJ...533...19B} suggest using 
a log-normal distribution ${\cal L}_{\rm LN}$, both the Gaussian and the log-normal distributions are found to be biased estimators for
WMAP \cite{2003ApJS..148..195V}. The approximation for the $C_l$ likelihood used in the WMAP analysis, consists of a Gaussian and a log-normal distribution, where
\be
\ln {\cal L}=\frac{1}{3} \ln {\cal L}_{\rm Gauss}+\frac{2}{3}\ln {\cal L}'_{\rm LN}\;.
\label{eq:likelihoodform}
\ee

Clearly the likelihood 
in the real analysis is not trivial and since we do not account for such corrections, we expect that our results will differ from those obtained with
MCMC. An interesting question is by how much? How well does a simple method fare against the full, more complex likelihood? We explore these questions in the next section.
\subsection{Analyzing WMAP seven-year data}
\label{data7}
We analyze the WMAP seven-year temperature power spectrum, using the vectors shown in solid red, in Fig.~\ref{fig:wmap7_TT_6d_alphas}.
This analysis differs slightly from those in previous sections, in that here we use the sky fraction contained in the WMAP likelihood, which 
varies with $l$, rather than a fixed value of $f_{\text{sky}}=0.783$. The spectrum range included in the analysis is $2-1200$, and we neglect the effect of lensing on the CMB. 
We fix the Sunyaev-Zel'dovich (SZ) amplitude parameter in the MCMC, and we hold the helium fraction constant and equal to $Y_{\text{He}}=0.24$.
  \begin{figure*}[t!]
\includegraphics[width=0.22\textwidth, angle=-90]{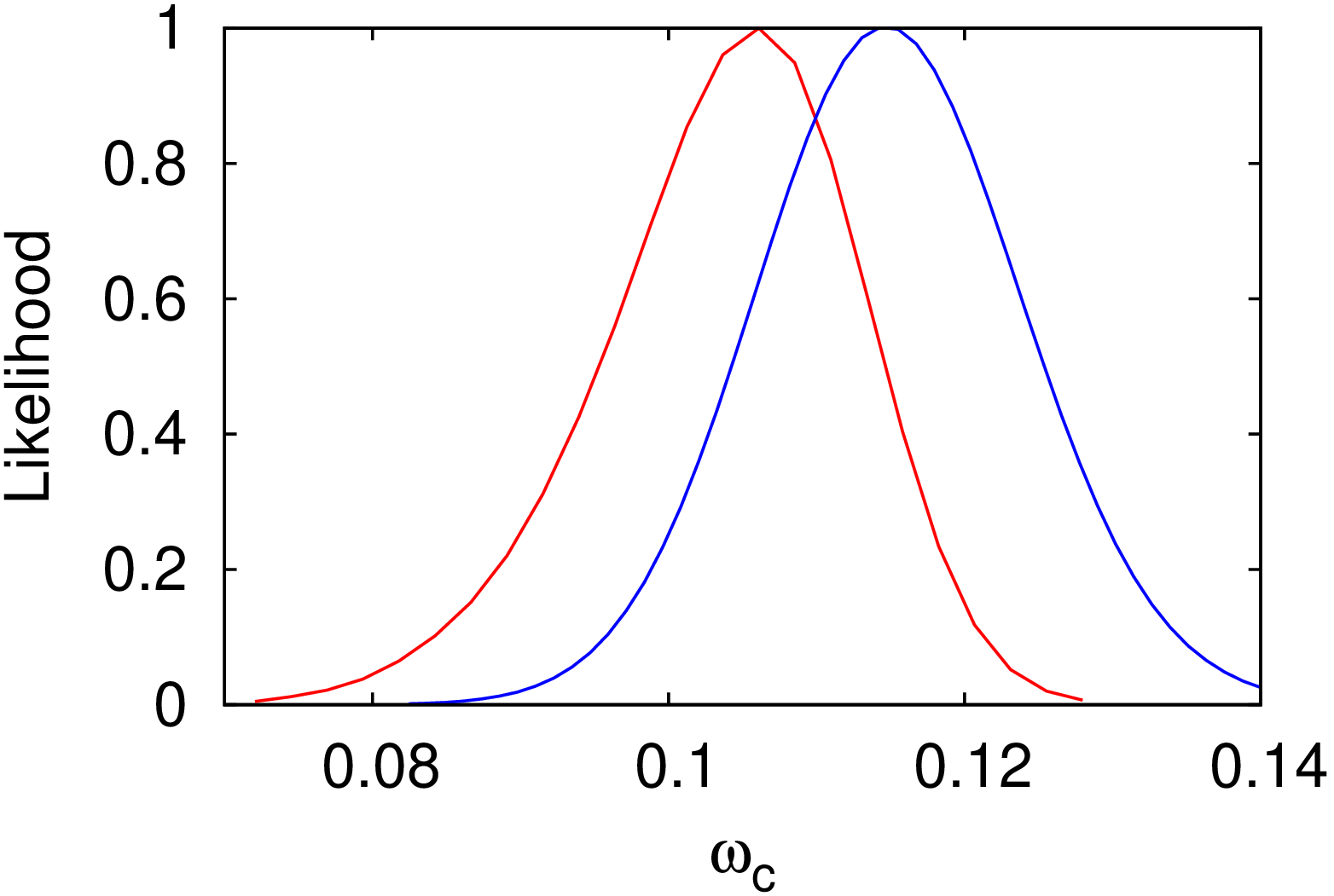}
   \vspace*{3mm}
\includegraphics[width=0.22\textwidth, angle=-90]{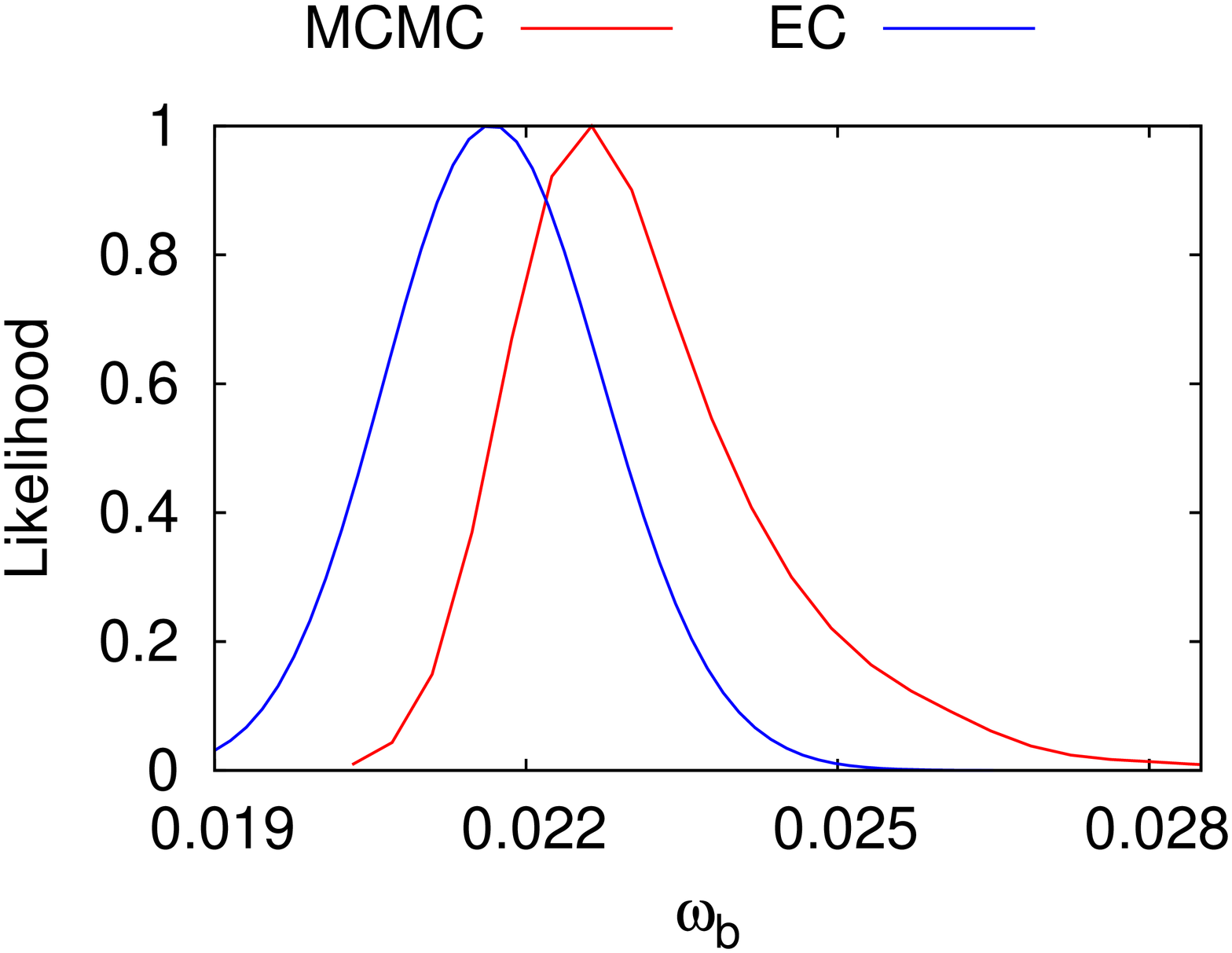}
\includegraphics[width=0.22\textwidth, angle=-90]{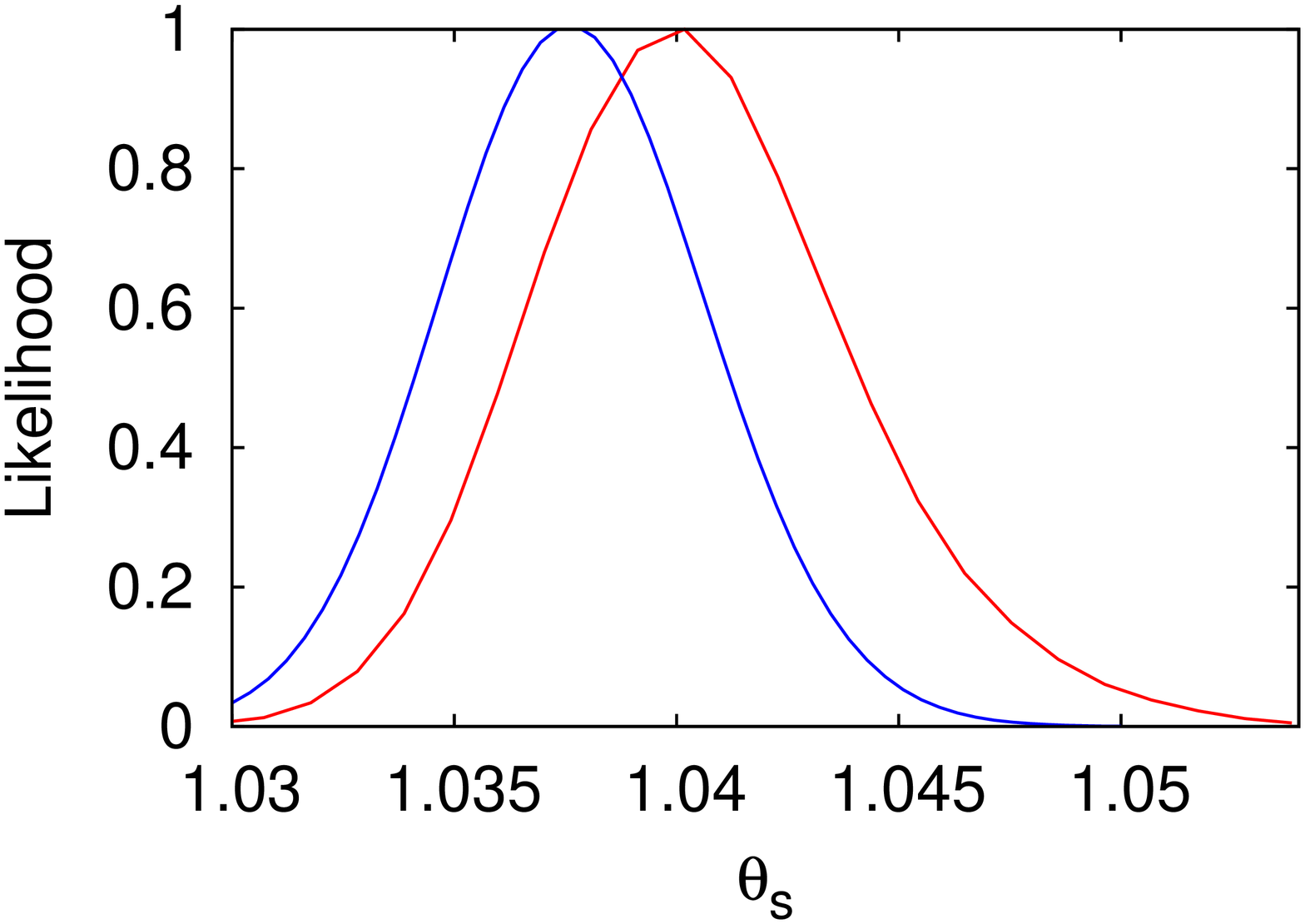}
\includegraphics[width=0.22\textwidth, angle=-90]{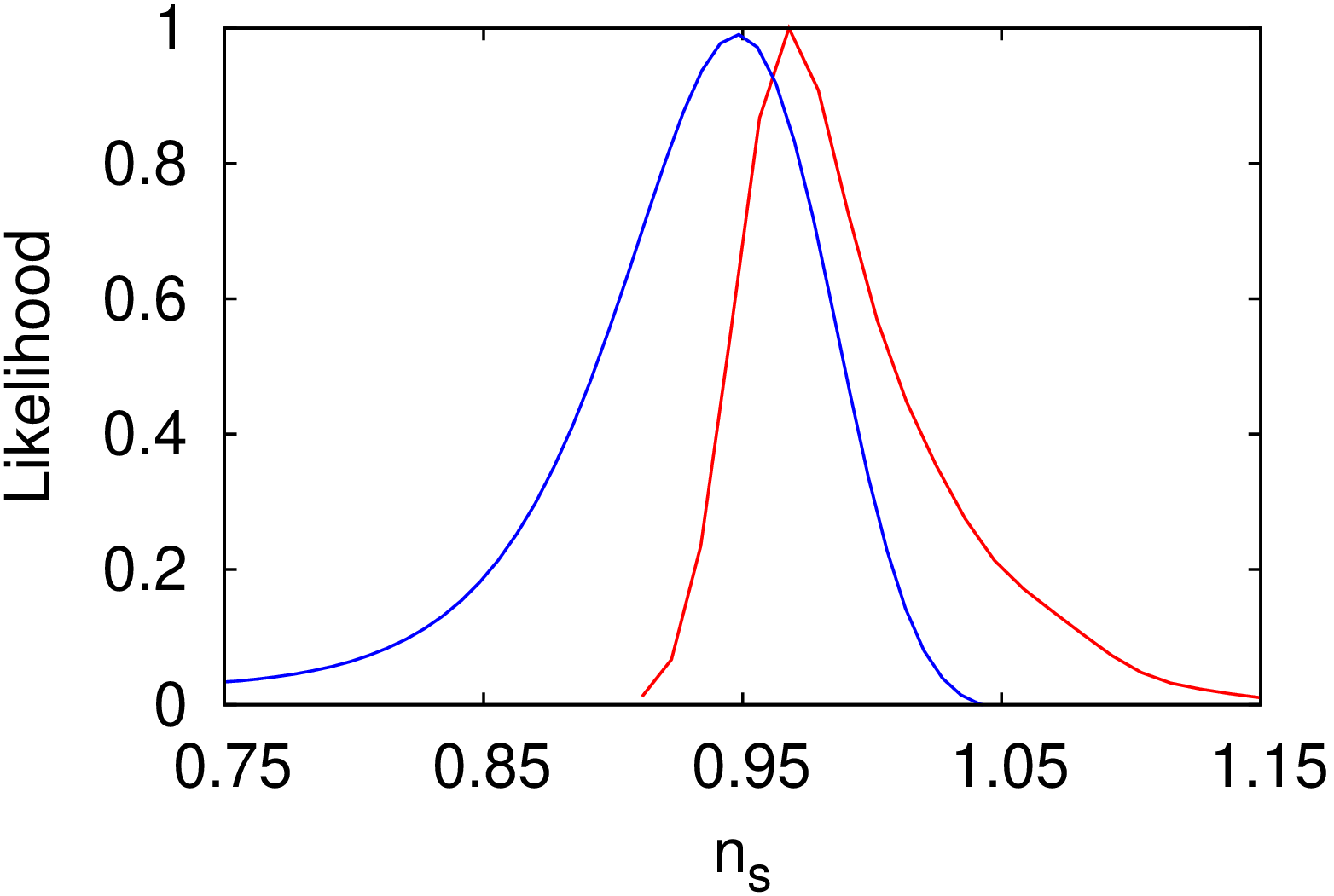}
\includegraphics[width=0.22\textwidth, angle=-90]{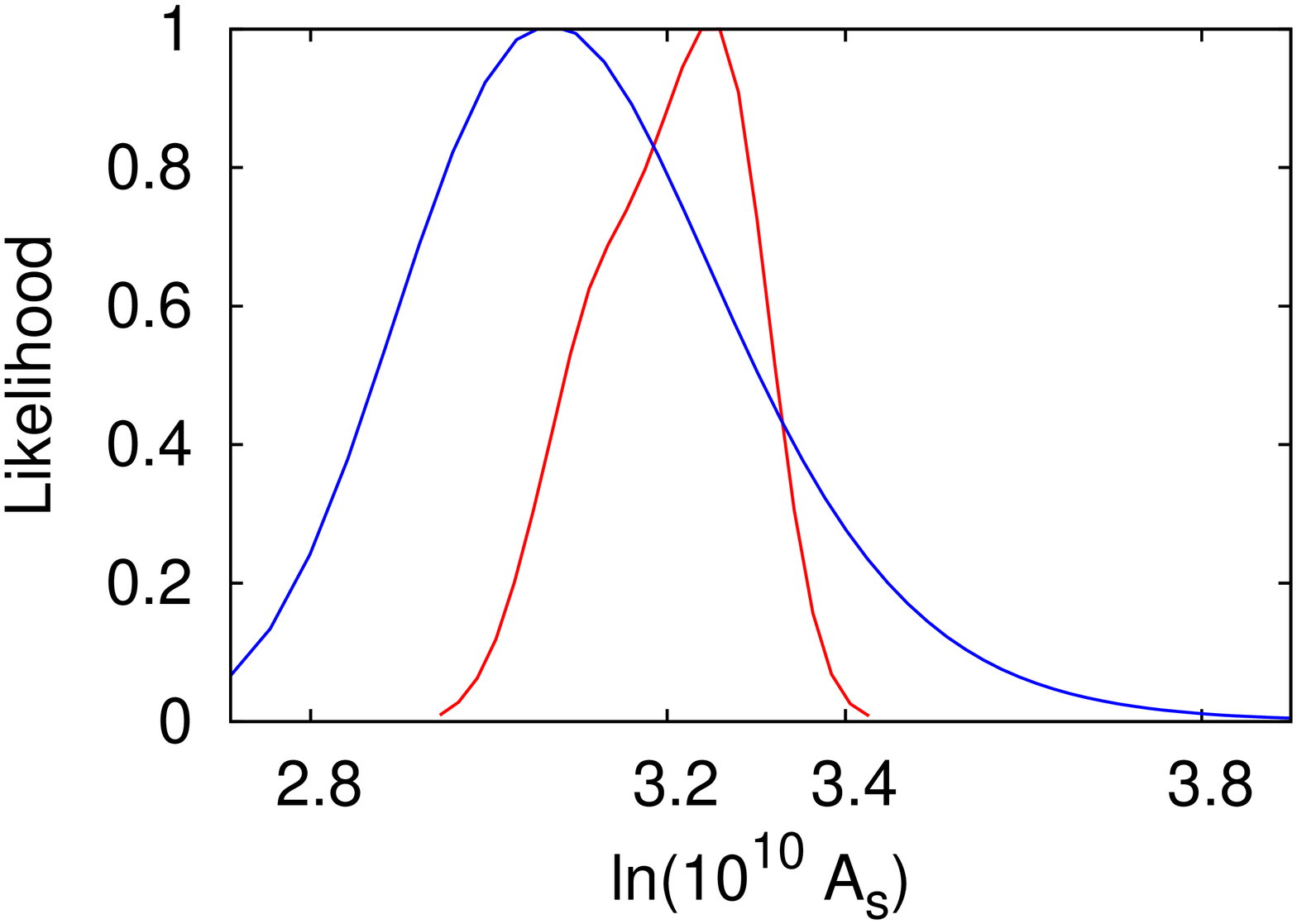}
\includegraphics[width=0.22\textwidth, angle=-90]{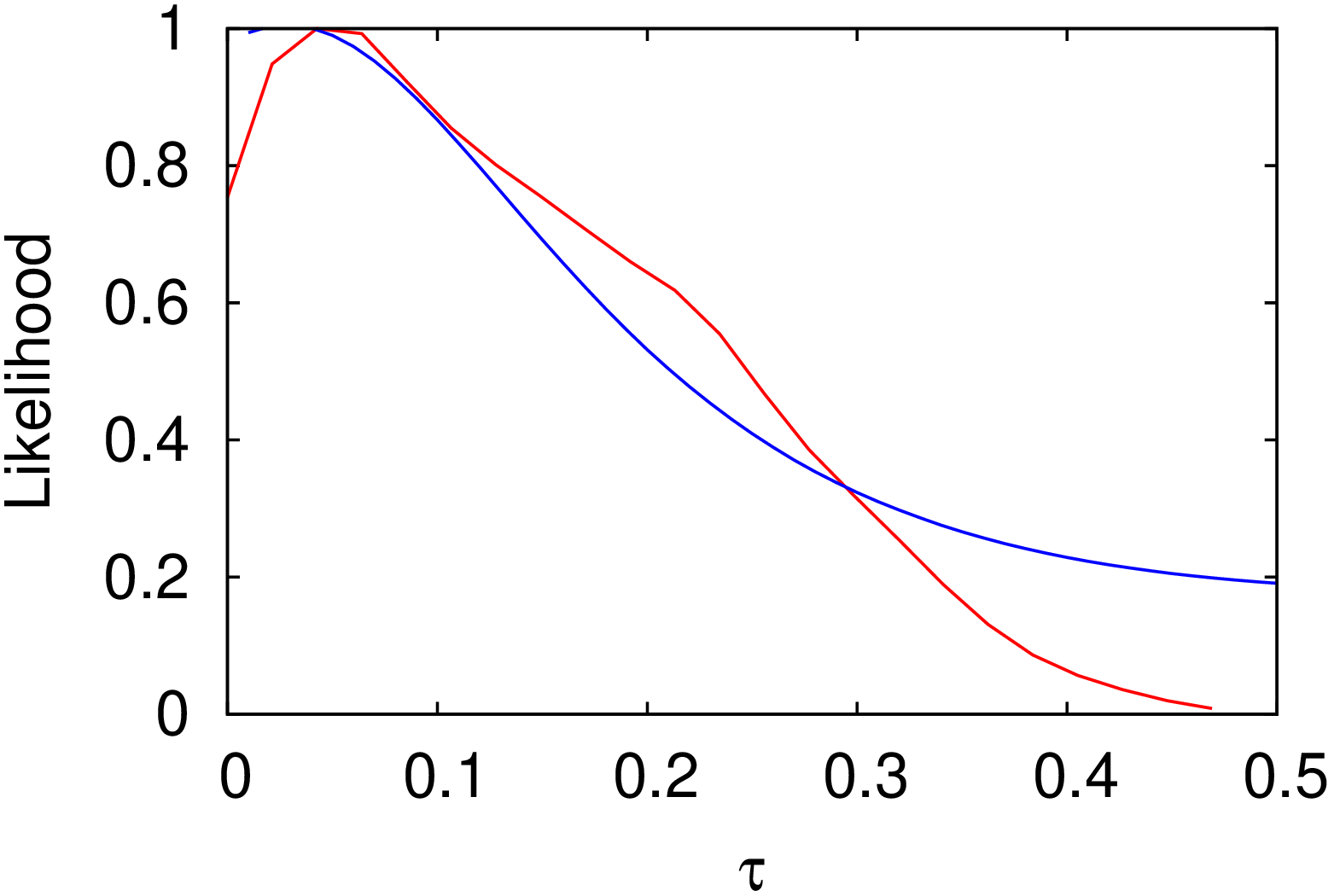}
   \vspace*{6mm}
\caption{\label{fig:WMAPMCMCmarglike_TT_data} Results of the EC analysis on the WMAP seven-year temperature data only (solid blue lines) and the full
likelihood evaluations with MCMC (solid red lines) in a $\Lambda$CDM model. Since the temperature data is not constraining enough to measure $\tau$, the posterior for the
optical depth is wide, resulting in a biased result for ln$(10^{10} A_s)$.}
\end{figure*}
\begin{figure*}[t!]
\includegraphics[width=0.22\textwidth, angle=-90]{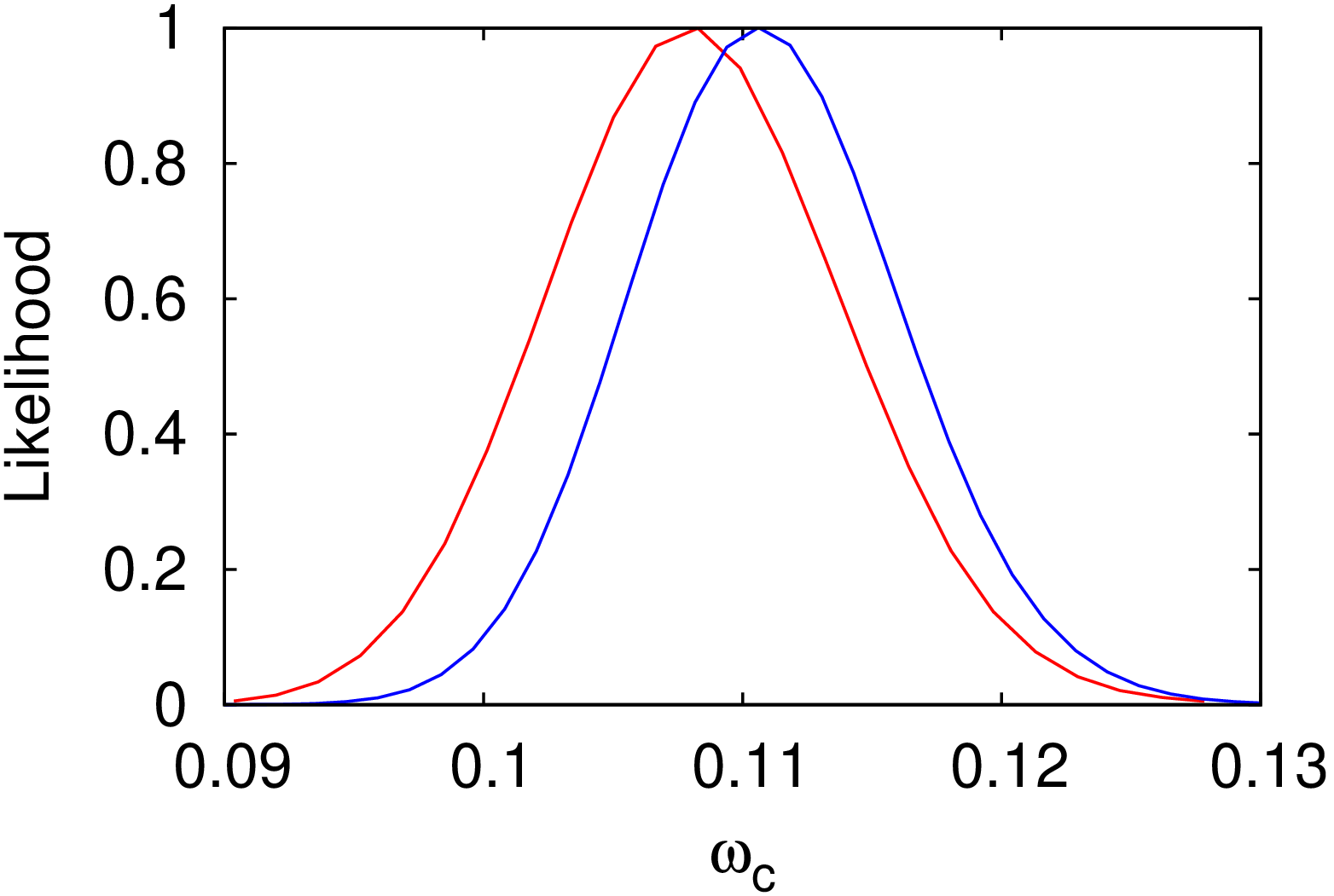}
\vspace*{3mm}
\includegraphics[width=0.22\textwidth, angle=-90]{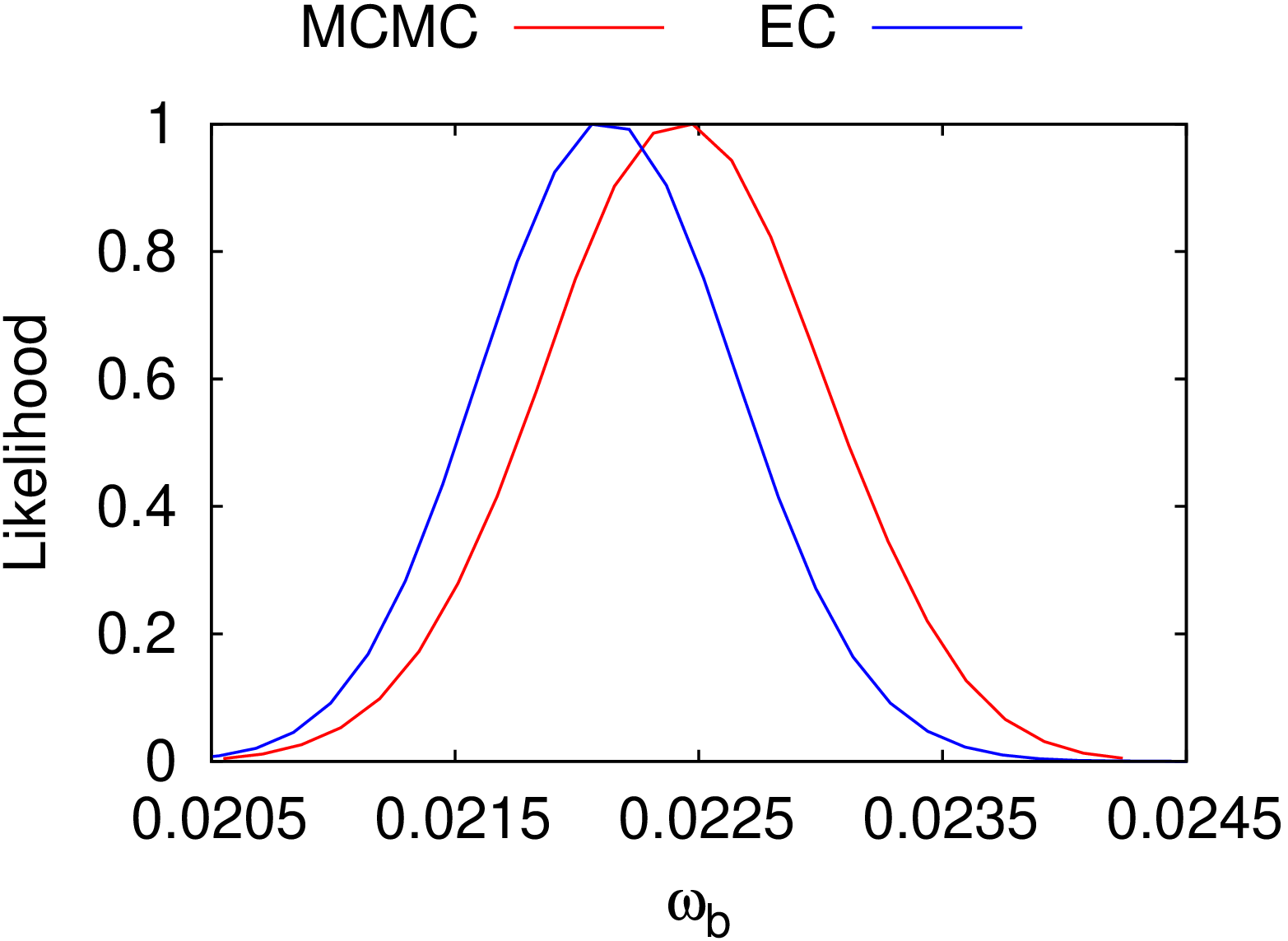}
\includegraphics[width=0.22\textwidth, angle=-90]{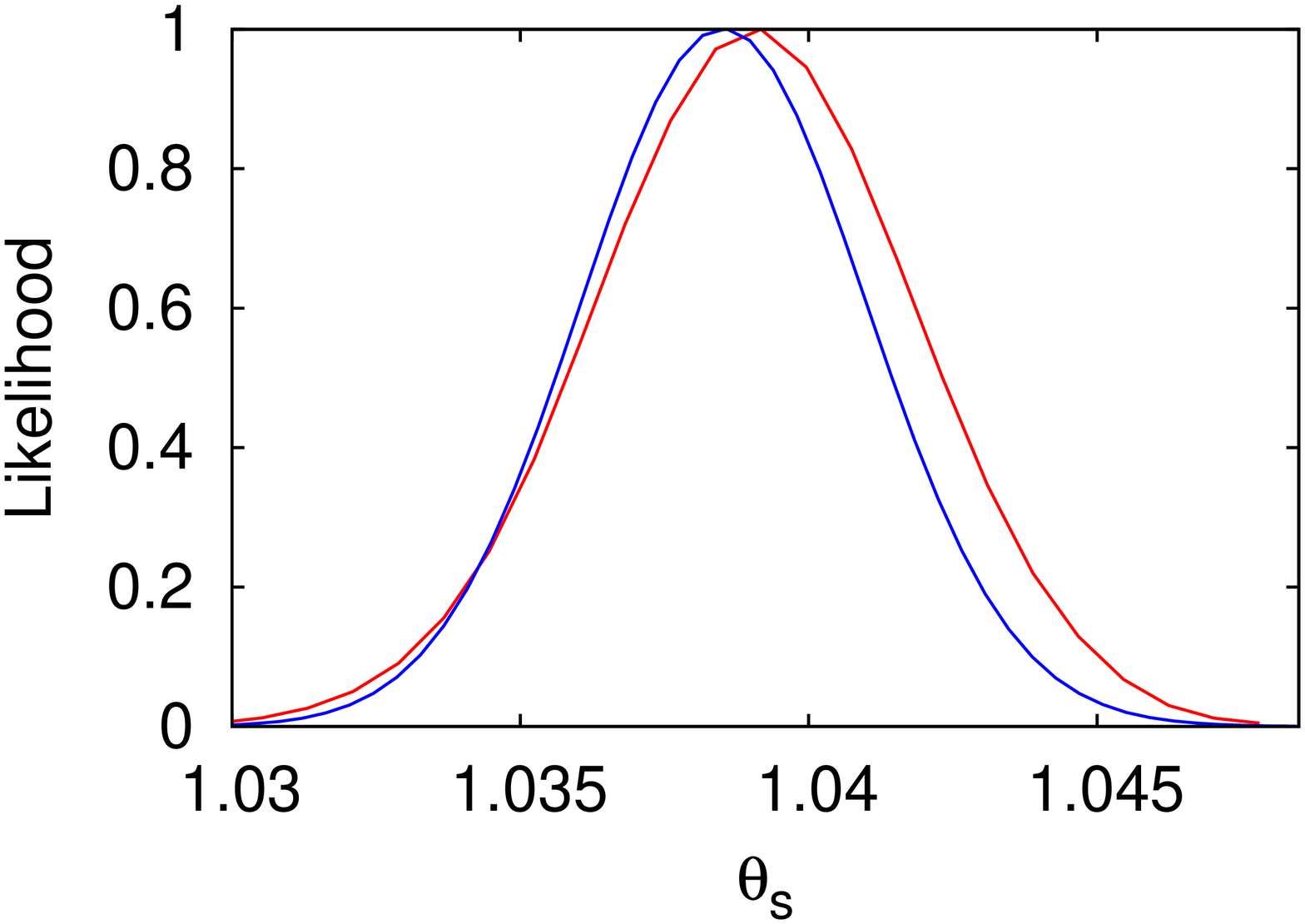}
\includegraphics[width=0.22\textwidth, angle=-90]{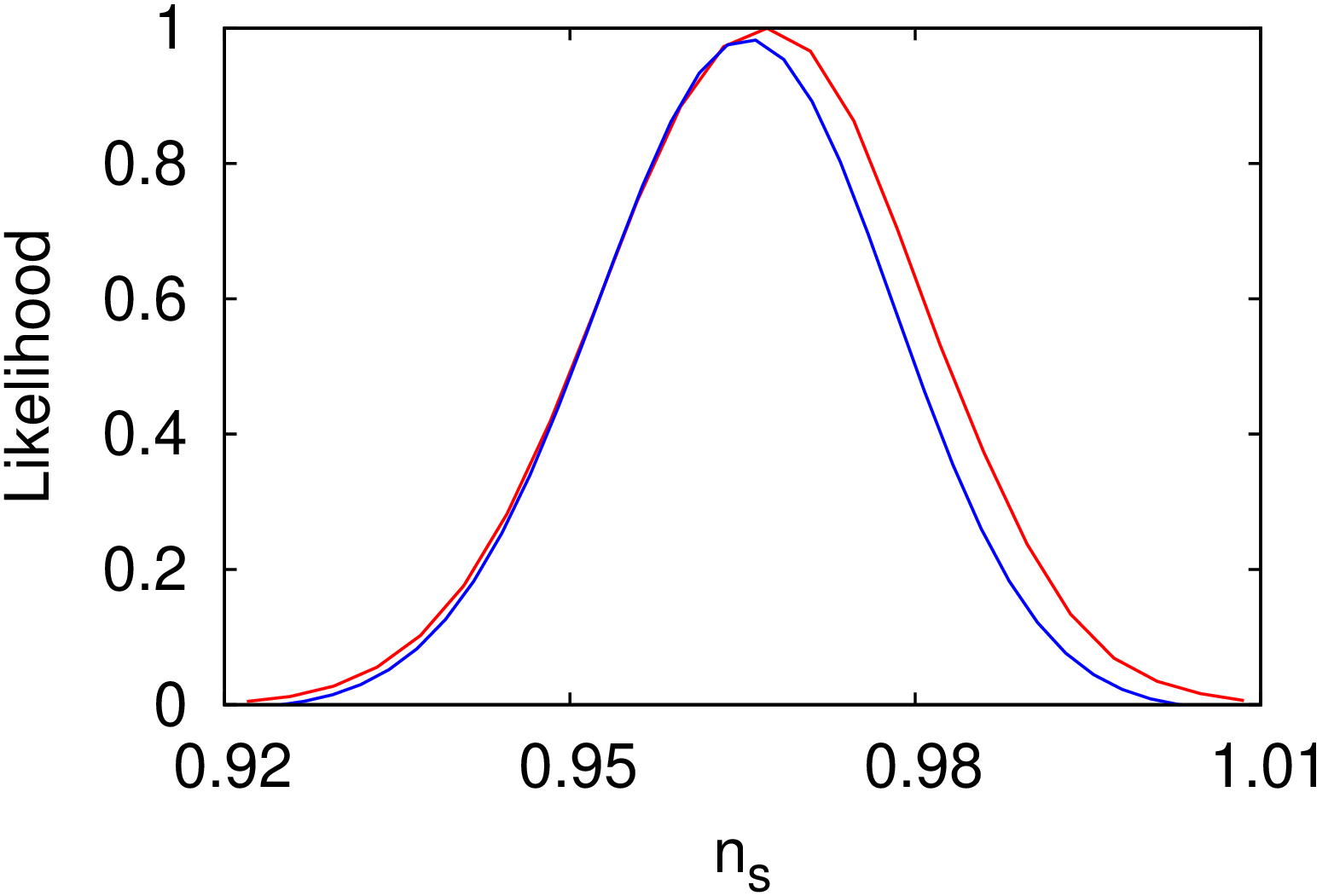}
\includegraphics[width=0.22\textwidth, angle=-90]{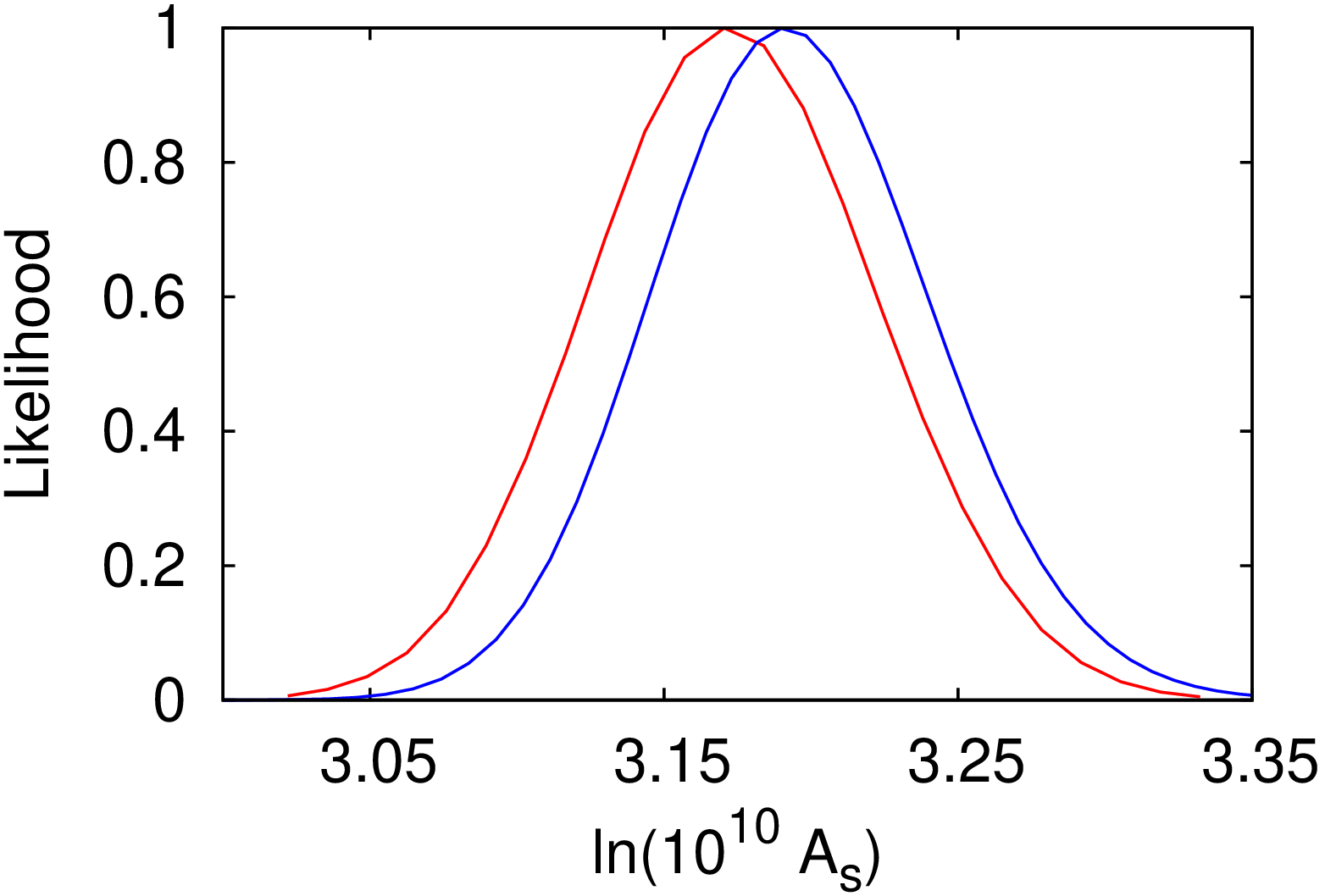}
\vspace*{6mm}
\caption{\label{fig:WMAPMCMCmarglike_TT_data_5D} Same as Fig.~\ref{fig:WMAPMCMCmarglike_TT_data} but with $\tau$ held fixed. Although the distributions
are a closer match, there is still some residual bias in the estimated parameters. We quote the bias on each parameter in Table \ref{tab:table2}.}
\end{figure*}

\subsubsection{$\Lambda$CDM including optical depth $\tau$}
In Fig.~\ref{fig:WMAPMCMCmarglike_TT_data}, we compare the results from extreme compression with MCMC assuming the
WMAP likelihood in Eq.~(\ref{eq:likelihoodform}). As we showed in Fig.~\ref{fig:WMAPMCMCmarglike_TT_th}, we do not expect that the posteriors from both methods
will agree exactly, in part because of the degeneracies due to poor constraints on the optical depth $\tau$. We also do not expect to obtain parameter
estimates equal to those of the base WMAP+SZ+LENS model, since we do not include polarization data. 
In this sense, we are
using compression vectors without assuming a ``correct'' fiducial model (as was done in Sec. \ref{wmaplikeexp}, 
Figs. \ref{fig:WMAPMCMCmarglike_TT_th} and \ref{fig:WMAPMCMCmarglike_TT_th_5D}). Further, we saw when analyzing mock data that the non-Gaussianity 
of the $\tau$ likelihood leads to a bias in $A_s$ in the EC method. Nonetheless, the biases shown in Fig.~\ref{fig:WMAPMCMCmarglike_TT_data} are 
still relatively small, with those estimated from the maximum of the likelihood significantly less than the statistical error. We show the bias between the
EC method and the MCMC results in Table \ref{tab:table2}, where we calculate the difference between the peak in the EC likelihood and the MCMC mean $\mu$ and the best-fit
($\theta_{\text{ML}}$) point, relative to the standard deviation $\sigma $ from MCMC.
\begin{table}[b!]
\caption{\label{tab:table2} Bias in the recovered parameter values using extreme compression relative to the MCMC results. We compare the parameter
value at the peak of the likelihood obtained with the EC method, to both the mean $\mu$ and the best fit (the maximum likelihood point $\theta_{\text{ML}}$)
from MCMC. The fourth column, is the standard deviation $\sigma$ of the MCMC samples.}
\begin{ruledtabular}
\begin{tabular}{C{1.8cm}C{1.8cm}C{1.8cm}C{1.5cm}}
 \head{0.5cm}{\text{Model \&} \text{Parameters}}  & \head{0.5cm}{\text{Bias relative} \text{to $\mu$}} & \head{0.5cm}{\text{Bias relative} \text{to $\theta_{\text{ML}}$}}  & \head{0.5cm}{\text{Standard} \text{Deviation $\sigma$}}
\\ [0.5ex]
\colrule
\rule{0pt}{2.5ex} \textbf{Theory $\tau$ free} \\
$\omega_c$                   &    0.67$\sigma$  &   0.01$\sigma$  &  0.0095  \\
$\omega_b$                   &   $-0.66 \sigma$  &   0.09$\sigma$  &  0.0014  \\
$\theta_s$              &   $-0.48 \sigma$  &   0.03$\sigma$  &  0.0036  \\
$n_s$                   &   $-0.71 \sigma$  &   0.04$\sigma$  &  0.0460  \\
$\text{ln}(10^{10}A_s)$ &   $-0.88 \sigma$  &  $-0.43 \sigma$  &  0.0814  \\
$\tau$                  &   $-0.69 \sigma$  &   0.08$\sigma$  &  0.1033  \\
\rule{0pt}{2.5ex} \textbf{Theory $\tau$ fixed} \\
$\omega_c$                   &   $-0.02 \sigma$  &   0.10$\sigma$  &  0.0054  \\
$\omega_b$                   &    0.06$\sigma$  &  $-0.17 \sigma$  &  0.0006  \\
$\theta_s$              &    0.04$\sigma$  &  $-0.10 \sigma$  &  0.0027  \\
$n_s$                   &    0.08$\sigma$  &  $-0.12 \sigma$  &  0.0137  \\
$\text{ln}(10^{10}A_s)$ &   $-0.07 \sigma$  &   0.15$\sigma$  &  0.0492  \\
\rule{0pt}{2.5ex} \textbf{Random $\tau$ fixed} \\
$\omega_c$                   &   $-0.04 \sigma$  &   0.24$\sigma$  &  0.0054  \\
$\omega_b$                   &    0.00$\sigma$  &  $-0.18 \sigma$  &  0.0006  \\
$\theta_s$              &    0.01$\sigma$  &  $-0.15 \sigma$  &  0.0027  \\
$n_s$                   &   $-0.03 \sigma$  &  $-0.21 \sigma$  &  0.0142  \\
$\text{ln}(10^{10}A_s)$ &   $-0.05 \sigma$  &   0.20$\sigma$  &  0.0508  \\
\rule{0pt}{2.5ex} \textbf{Data $\tau$ free} \\
$\omega_c$                   &    1.29$\sigma$  &   0.35$\sigma$  &  0.0084  \\
$\omega_b$                   &   $-1.19 \sigma$  &  $-0.43 \sigma$  &  0.0012  \\
$\theta_s$              &   $-0.82 \sigma$  &  $-0.44 \sigma$  &  0.0035  \\
$n_s$                   &   $-1.09 \sigma$  &  $-0.25 \sigma$  &  0.0399  \\
$\text{ln}(10^{10}A_s)$ &   $-1.56 \sigma$  &   0.06$\sigma$  &  0.0858  \\
$\tau$                  &   $-1.21 \sigma$  &   0.16$\sigma$  &  0.0972  \\
\rule{0pt}{2.5ex} \textbf{Data $\tau$ fixed} \\
$\omega_c$                   &    0.45$\sigma$  &   0.40$\sigma$  &  0.0055  \\
$\omega_b$                   &   $-0.56 \sigma$  &  $-0.63 \sigma$  &  0.0006  \\
$\theta_s$              &   $-0.19 \sigma$  &  $-0.19 \sigma$  &  0.0027  \\
$n_s$                   &   $-0.08 \sigma$  &  $-0.09 \sigma$  &  0.0131  \\
$\text{ln}(10^{10}A_s)$ &   $-0.37 \sigma$  &   0.34$\sigma$  &  0.0470  \\
\end{tabular}
\end{ruledtabular}
\end{table}
\subsubsection{$\Lambda$CDM and fixed optical depth $\tau$}
In Fig.~\ref{fig:WMAPMCMCmarglike_TT_data_5D} we show constraints from the compressed data set and MCMC results using the entire WMAP CMB temperature anisotropy
power spectrum. The agreement is best for $n_s$ and $\theta_s$, with the other parameters experiencing a bias of less than $\sim 0.5\sigma$. We show the results
from the EC method and any bias in determining the posterior mean and the maximum likelihood (ML) point in Table \ref{tab:table2}. As pointed out
in Sec. \ref{explain}, the likelihood used in the full WMAP analysis is not a simple Gaussian. In addition, we do not take into account in our compression method
the intricacies involved with beam corrections and point source subtraction. Neither do we account for non-Gaussianity of the data at the lowest multipoles.
The fact that WMAP uses Gibbs sampling for the lowest multipoles also means that our results will not be the same. Crucially, if we modify the code to either model the
likelihood as a full Gaussian, by discarding log-normal part in Eq.~(\ref{eq:likelihoodform}) or do not use Gibbs sampling and restrict the analysis to modes with $l > 30$, the 
resulting shifts in each of the parameter posteriors cause much larger differences than the ones quoted above. So,  the biases introduced in the EC method are smaller than those that emanate from much milder assumptions about the likelihood.

\section{Conclusion}
\label{conclude}
We have shown that a locally lossless extreme compression of modern CMB data sets gains significant speedup in the computation of marginalized
likelihoods in $\Lambda$CDM models. By requiring that the Fisher information matrix is unchanged, we derived the weighting vectors for the CMB that can estimate
cosmological parameters in less than a minute, much faster than MCMC. The method requires computations of 
the likelihood for one parameter at a time, instead of having to explore the whole parameter space with MCMC. We therefore achieve extreme data 
compression by (i) compressing the 
entire data set into just a few numbers, and (ii) reducing the dimensionality of the parameter space that needs to be explored.

The compression vectors for the CMB are also very useful since their shape and amplitude provide an intuitive feel for the physics of the CMB, 
the sensitivity of the observed spectrum to cosmological parameters. They can also inform about the relative sensitivity of different
experiments to cosmological parameters.

We have tested our method on exact theory $C_l$ as well as on a WMAP-like CMB data set generated from a random realization
of a fiducial cosmology. By comparing our results to those from full likelihood analyses using CosmoMC, we have been able to show that the method performs
very well, and is able to recover the maximum likelihood estimates for parameters even if the posterior is not Gaussian. If the posterior is Gaussian,
then the extreme compression method can recover the posterior means to better than $0.1\sigma$.

We have applied the compression method to the temperature power spectrum from the WMAP seven-year data release, and have found that
even though the likelihood for WMAP is nontrivial and non-Gaussian, our method is in good agreement with the
posteriors from a full MCMC analysis. The biases in our estimates of cosmological parameters, compared to the mean 
are: $\omega_b$ bias is $-0.56 \sigma$, $\omega_c$ bias is 0.45$\sigma$, $\theta_s$ bias is  $-0.19 \sigma$, $n_s$ is $-0.08 \sigma$, $A_s$ is $-0.37 \sigma$. The biases
relative to the best fit (the maximum likelihood) are comparable.

Furthermore, given the nontrivial nature of the likelihood, it is possible that the method may also work well with newer data and a more complicated Bayesian analysis, e.g., the Planck
likelihood. We will address this in a future investigation. 

Additionally as a bonus, including polarization data and extending the parameter space is not going to increase the computational costs. The vectors
can be precomputed and stored, and the 
calculation of the likelihood is limited only by the speed of one call to CAMB, times the number of samples we wish to obtain. The increase in parameter space, can be
accommodated by running
each compression separately, one after another, or at the same time using $n$ nodes. In this case, the time for the likelihood computation for the entire parameter space is no longer than a computation for a single parameter, which takes less than a minute
 \begin{acknowledgments}
 The authors would like to thank the anonymous referee for the valuable comments on the manuscript. A.Z. would like to thank Wayne Hu for the useful discussions. This work was completed
 in part with resources provided by the University of Chicago Research Computing Center as well as the Joint Fermilab - KICP Supercomputing Cluster,
 supported by grants from Fermilab, the Kavli Institute for Cosmological Physics, and the University of Chicago. A.Z. acknowledges support
 from KICP, the Brinson Foundation and U.S. Department of Energy Contract No. DE-FG02-13ER41958. The work of S.D. is 
 supported by the U.S. Department of Energy, including Grant No. DE-FG02-95ER40896.
 \end{acknowledgments}
\appendix
\section{CHOOSING THE RIGHT PARAMETRIZATION IN A MODEL}
\label{bad_par}
\begin{figure}[t]
\centering
\includegraphics[width=0.40\textwidth]{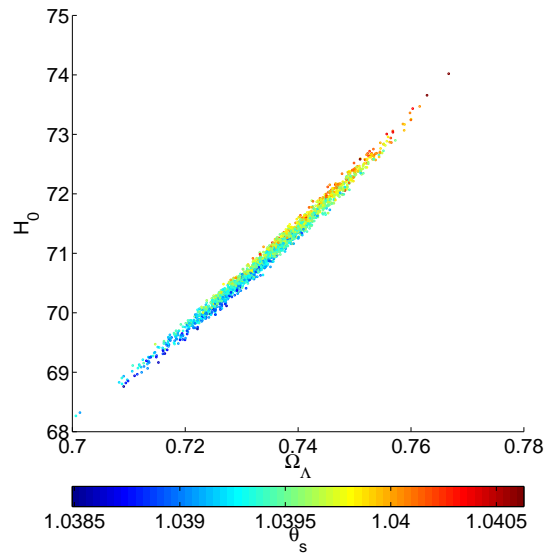}
 \caption{\label{fig:3D}The extent of the geometric degeneracy in MCMC samples between the cosmological constant density parameter $\Omega_{\Lambda}$ and
 the Hubble expansion rate $H_0$. The color scale shows the values of $100\theta_{\text{MC}}$, an approximation for $r_{s}(z_{\star}) / D_A(z_{\star})$, the
 angular scale of the sound horizon at last scattering. The data used in this case were exact theory $C_l$.}
  \end{figure}

If there exist known degeneracies in the data, e.g., the geometric degeneracy in the CMB, then the choice of parametrization will matter. For the CMB, we find that
a bad parametrization may have an adverse effect on the compression and therefore the recovered posterior distributions may be non-Gaussian and/or multimodal.
In the specific case of the CMB, we found that using $\Omega_{\Lambda}$ instead of $\theta_s$ results in a bimodal
distribution for  $\Omega_{\Lambda}$ with all other parameters not affected (that is, their posteriors were all correct).
The root of the problem can be seen in Fig.~\ref{fig:3D}, where we plot the geometric degeneracy between $H_0$ and $\Omega_{\Lambda}$. The color coding
shows various values of $100\, \theta_s$. The optimal parameter vector is $\Theta = \{\omega_c,\omega_b,100 \theta_s,n_s,A_s,\tau\}$.
\section{CMB DATA GENERATION}
\label{RAND_generate}
We generate two kinds of data sets using the Boltzmann code CAMB, computing the temperature power spectrum $C_l$ up to $l=1200$.
For the first data set (referred to as exact theory $C_l$), we assume white isotropic noise and Gaussian beams, and add the noise $N^{}_l$ 
given by Eq.~(\ref{14}) to $C_l$. In the MCMC analysis, we use the likelihood in Eq.~(\ref{eqn:cmblike}) to get parameter constraints. This is because
the likelihood is a function of $C_l+N_l$ and not just $C_l$ [see Eq.~(\ref{eqn:cmblike})]. The EC calculation assumes the data is exact theory $C_l$, with the
noise $N_l$ included in the covariance in Eq.~(\ref{eqn:cmbnoise2}).

The second data set that we use in our analysis makes use of a random realization of the underlying theory $C_l$. To create a random mock data set
we generate four sets of Gaussian random deviates $a$, $b$, $c$ and $d$, with $\mu=0$ and $\sigma^2 = 1$. We use these random deviates to create two complex
Gaussian fields, $g_{lm}=\frac{1}{\sqrt{2}} (a + \imath b)$ and $h_{lm}=\frac{1}{\sqrt{2}} (c + \imath d)$ with $\langle h_{lm}^{*} h_{lm}^{}\rangle = 1$ and
$\langle g_{lm}^{*} g_{lm}^{}\rangle = 1$.
For completeness, we include the generation of both $a_{lm}^{T}$ and $a_{lm}^{E}$, such that $ C^{TT}_{l} C^{EE}_{l}-(C^{TE}_{l})^2 > 0$ and $a_{l,m}^{*XX} = (-1)^m a_{l,-m}^{XX}$.
The spherical harmonic coefficients for temperature are
\begin{equation}
 a_{lm}^{T} = \sqrt{C_l^{TT} + N_l^{TT}} \ g_{lm}
\end{equation}
and the polarization coefficients are given by
\begin{eqnarray}
 a_{lm}^{E} &=& \frac{C_l^{TE}}{C_l^{TT} + N_l^{TT}} \sqrt{C_l^{TT} + N_l} \ g_{lm}\nonumber\\
 &+&h_{lm} \sqrt{ (C_l^{EE} + N_l^{EE}) - \frac{(C_l^{TE})^2}{(C_l^{TT} + N_l^{TT})}}.
\end{eqnarray}
The random mock can then be generated using the full-sky power spectra estimators for the temperature, the E-mode polarization, and the cross spectrum between
the temperature and the E-mode polarization given by 
\begin{equation}
\hat{C}_l^{TT} = \frac{1}{2l+1} \sum_{m=-l}^{m=l} \langle a_{lm}^{*TT} a_{lm}^{TT}\rangle
\end{equation}
\begin{equation}
\hat{C}_l^{EE} = \frac{1}{2l+1} \sum_{m=-l}^{m=l} \langle a_{lm}^{*EE} a_{lm}^{EE}\rangle
\end{equation}
\begin{equation}
\hat{C}_l^{TE} = \frac{1}{2l+1} \sum_{m=-l}^{m=l} \langle a_{lm}^{*TT} a_{lm}^{TE}\rangle.
\end{equation}
We have tested this prescription using MCMC, and find that on average seven out of ten times the
estimate of $\theta_i$ is within $1\sigma$ of the fiducial input value.
%\bibliography{../libs/library}

\begin{thebibliography}{25}%
\makeatletter
\providecommand \@ifxundefined [1]{%
 \@ifx{#1\undefined}
}%
\providecommand \@ifnum [1]{%
 \ifnum #1\expandafter \@firstoftwo
 \else \expandafter \@secondoftwo
 \fi
}%
\providecommand \@ifx [1]{%
 \ifx #1\expandafter \@firstoftwo
 \else \expandafter \@secondoftwo
 \fi
}%
\providecommand \natexlab [1]{#1}%
\providecommand \enquote  [1]{``#1''}%
\providecommand \bibnamefont  [1]{#1}%
\providecommand \bibfnamefont [1]{#1}%
\providecommand \citenamefont [1]{#1}%
\providecommand \href@noop [0]{\@secondoftwo}%
\providecommand \href [0]{\begingroup \@sanitize@url \@href}%
\providecommand \@href[1]{\@@startlink{#1}\@@href}%
\providecommand \@@href[1]{\endgroup#1\@@endlink}%
\providecommand \@sanitize@url [0]{\catcode `\\12\catcode `\$12\catcode
  `\&12\catcode `\#12\catcode `\^12\catcode `\_12\catcode `\%12\relax}%
\providecommand \@@startlink[1]{}%
\providecommand \@@endlink[0]{}%
\providecommand \url  [0]{\begingroup\@sanitize@url \@url }%
\providecommand \@url [1]{\endgroup\@href {#1}{\urlprefix }}%
\providecommand \urlprefix  [0]{URL }%
\providecommand \Eprint [0]{\href }%
\providecommand \doibase [0]{http://dx.doi.org/}%
\providecommand \selectlanguage [0]{\@gobble}%
\providecommand \bibinfo  [0]{\@secondoftwo}%
\providecommand \bibfield  [0]{\@secondoftwo}%
\providecommand \translation [1]{[#1]}%
\providecommand \BibitemOpen [0]{}%
\providecommand \bibitemStop [0]{}%
\providecommand \bibitemNoStop [0]{.\EOS\space}%
\providecommand \EOS [0]{\spacefactor3000\relax}%
\providecommand \BibitemShut  [1]{\csname bibitem#1\endcsname}%
\let\auto@bib@innerbib\@empty
%</preamble>
\bibitem [{\citenamefont {{Davis}}\ and\ \citenamefont
  {{Peebles}}(1983)}]{1983ApJ...267..465D}%
  \BibitemOpen
  \bibfield  {author} {\bibinfo {author} {\bibfnamefont {M.}~\bibnamefont
  {{Davis}}}\ and\ \bibinfo {author} {\bibfnamefont {P.~J.~E.}\ \bibnamefont
  {{Peebles}}},\ }\href {\doibase 10.1086/160884} {\bibfield  {journal}
  {\bibinfo  {journal} {\apj}\ }\textbf {\bibinfo {volume} {267}},\ \bibinfo
  {pages} {465} (\bibinfo {year} {1983})}\BibitemShut {NoStop}%
\bibitem [{\citenamefont {{Seljak}}\ and\ \citenamefont
  {{Bertschinger}}(1993)}]{1993ApJ...417L...9S}%
  \BibitemOpen
  \bibfield  {author} {\bibinfo {author} {\bibfnamefont {U.}~\bibnamefont
  {{Seljak}}}\ and\ \bibinfo {author} {\bibfnamefont {E.}~\bibnamefont
  {{Bertschinger}}},\ }\href {\doibase 10.1086/187081} {\bibfield  {journal}
  {\bibinfo  {journal} {\apjl}\ }\textbf {\bibinfo {volume} {417}},\ \bibinfo
  {pages} {L9} (\bibinfo {year} {1993})},\ \Eprint
  {http://arxiv.org/abs/astro-ph/9309003} {astro-ph/9309003} \BibitemShut
  {NoStop}%
\bibitem [{\citenamefont {{Gorski}}(1994)}]{1994ApJ...430L..85G}%
  \BibitemOpen
  \bibfield  {author} {\bibinfo {author} {\bibfnamefont {K.~M.}\ \bibnamefont
  {{Gorski}}},\ }\href {\doibase 10.1086/187444} {\bibfield  {journal}
  {\bibinfo  {journal} {\apjl}\ }\textbf {\bibinfo {volume} {430}},\ \bibinfo
  {pages} {L85} (\bibinfo {year} {1994})},\ \Eprint
  {http://arxiv.org/abs/astro-ph/9403066} {astro-ph/9403066} \BibitemShut
  {NoStop}%
\bibitem [{\citenamefont {{Bond}}\ \emph {et~al.}(1994)\citenamefont {{Bond}},
  \citenamefont {{Crittenden}}, \citenamefont {{Davis}}, \citenamefont
  {{Efstathiou}},\ and\ \citenamefont {{Steinhardt}}}]{1994PhRvL..72...13B}%
  \BibitemOpen
  \bibfield  {author} {\bibinfo {author} {\bibfnamefont {J.~R.}\ \bibnamefont
  {{Bond}}}, \bibinfo {author} {\bibfnamefont {R.}~\bibnamefont
  {{Crittenden}}}, \bibinfo {author} {\bibfnamefont {R.~L.}\ \bibnamefont
  {{Davis}}}, \bibinfo {author} {\bibfnamefont {G.}~\bibnamefont
  {{Efstathiou}}}, \ and\ \bibinfo {author} {\bibfnamefont {P.~J.}\
  \bibnamefont {{Steinhardt}}},\ }\href {\doibase 10.1103/PhysRevLett.72.13}
  {\bibfield  {journal} {\bibinfo  {journal} {Physical Review Letters}\
  }\textbf {\bibinfo {volume} {72}},\ \bibinfo {pages} {13} (\bibinfo {year}
  {1994})},\ \Eprint {http://arxiv.org/abs/astro-ph/9309041} {astro-ph/9309041}
  \BibitemShut {NoStop}%
\bibitem [{\citenamefont {{Vogeley}}\ and\ \citenamefont
  {{Szalay}}(1996)}]{1996ApJ...465...34V}%
  \BibitemOpen
  \bibfield  {author} {\bibinfo {author} {\bibfnamefont {M.~S.}\ \bibnamefont
  {{Vogeley}}}\ and\ \bibinfo {author} {\bibfnamefont {A.~S.}\ \bibnamefont
  {{Szalay}}},\ }\href {\doibase 10.1086/177399} {\bibfield  {journal}
  {\bibinfo  {journal} {\apj}\ }\textbf {\bibinfo {volume} {465}},\ \bibinfo
  {pages} {34} (\bibinfo {year} {1996})},\ \Eprint
  {http://arxiv.org/abs/astro-ph/9601185} {astro-ph/9601185} \BibitemShut
  {NoStop}%
\bibitem [{\citenamefont {{Tegmark}}\ \emph {et~al.}(1997)\citenamefont
  {{Tegmark}}, \citenamefont {{Taylor}},\ and\ \citenamefont
  {{Heavens}}}]{1997ApJ...480...22T}%
  \BibitemOpen
  \bibfield  {author} {\bibinfo {author} {\bibfnamefont {M.}~\bibnamefont
  {{Tegmark}}}, \bibinfo {author} {\bibfnamefont {A.~N.}\ \bibnamefont
  {{Taylor}}}, \ and\ \bibinfo {author} {\bibfnamefont {A.~F.}\ \bibnamefont
  {{Heavens}}},\ }\href {\doibase 10.1086/303939} {\bibfield  {journal}
  {\bibinfo  {journal} {\apj}\ }\textbf {\bibinfo {volume} {480}},\ \bibinfo
  {pages} {22} (\bibinfo {year} {1997})},\ \Eprint
  {http://arxiv.org/abs/astro-ph/9603021} {astro-ph/9603021} \BibitemShut
  {NoStop}%
\bibitem [{\citenamefont {{Heavens}}\ \emph {et~al.}(2000)\citenamefont
  {{Heavens}}, \citenamefont {{Jimenez}},\ and\ \citenamefont
  {{Lahav}}}]{2000MNRAS.317..965H}%
  \BibitemOpen
  \bibfield  {author} {\bibinfo {author} {\bibfnamefont {A.~F.}\ \bibnamefont
  {{Heavens}}}, \bibinfo {author} {\bibfnamefont {R.}~\bibnamefont
  {{Jimenez}}}, \ and\ \bibinfo {author} {\bibfnamefont {O.}~\bibnamefont
  {{Lahav}}},\ }\href {\doibase 10.1046/j.1365-8711.2000.03692.x} {\bibfield
  {journal} {\bibinfo  {journal} {\mnras}\ }\textbf {\bibinfo {volume} {317}},\
  \bibinfo {pages} {965} (\bibinfo {year} {2000})},\ \Eprint
  {http://arxiv.org/abs/astro-ph/9911102} {astro-ph/9911102} \BibitemShut
  {NoStop}%
\bibitem [{\citenamefont {{Reichardt}}\ \emph {et~al.}(2001)\citenamefont
  {{Reichardt}}, \citenamefont {{Jimenez}},\ and\ \citenamefont
  {{Heavens}}}]{2001MNRAS.327..849R}%
  \BibitemOpen
  \bibfield  {author} {\bibinfo {author} {\bibfnamefont {C.}~\bibnamefont
  {{Reichardt}}}, \bibinfo {author} {\bibfnamefont {R.}~\bibnamefont
  {{Jimenez}}}, \ and\ \bibinfo {author} {\bibfnamefont {A.~F.}\ \bibnamefont
  {{Heavens}}},\ }\href {\doibase 10.1046/j.1365-8711.2001.04768.x} {\bibfield
  {journal} {\bibinfo  {journal} {\mnras}\ }\textbf {\bibinfo {volume} {327}},\
  \bibinfo {pages} {849} (\bibinfo {year} {2001})},\ \Eprint
  {http://arxiv.org/abs/astro-ph/0101074} {astro-ph/0101074} \BibitemShut
  {NoStop}%
\bibitem [{\citenamefont {{Gupta}}\ and\ \citenamefont
  {{Heavens}}(2002)}]{2002MNRAS.334..167G}%
  \BibitemOpen
  \bibfield  {author} {\bibinfo {author} {\bibfnamefont {S.}~\bibnamefont
  {{Gupta}}}\ and\ \bibinfo {author} {\bibfnamefont {A.~F.}\ \bibnamefont
  {{Heavens}}},\ }\href {\doibase 10.1046/j.1365-8711.2002.05499.x} {\bibfield
  {journal} {\bibinfo  {journal} {\mnras}\ }\textbf {\bibinfo {volume} {334}},\
  \bibinfo {pages} {167} (\bibinfo {year} {2002})},\ \Eprint
  {http://arxiv.org/abs/astro-ph/0108315} {astro-ph/0108315} \BibitemShut
  {NoStop}%
\bibitem [{\citenamefont {{Protopapas}}\ \emph {et~al.}(2005)\citenamefont
  {{Protopapas}}, \citenamefont {{Jimenez}},\ and\ \citenamefont
  {{Alcock}}}]{2005MNRAS.362..460P}%
  \BibitemOpen
  \bibfield  {author} {\bibinfo {author} {\bibfnamefont {P.}~\bibnamefont
  {{Protopapas}}}, \bibinfo {author} {\bibfnamefont {R.}~\bibnamefont
  {{Jimenez}}}, \ and\ \bibinfo {author} {\bibfnamefont {C.}~\bibnamefont
  {{Alcock}}},\ }\href {\doibase 10.1111/j.1365-2966.2005.09305.x} {\bibfield
  {journal} {\bibinfo  {journal} {\mnras}\ }\textbf {\bibinfo {volume} {362}},\
  \bibinfo {pages} {460} (\bibinfo {year} {2005})},\ \Eprint
  {http://arxiv.org/abs/astro-ph/0502301} {astro-ph/0502301} \BibitemShut
  {NoStop}%
\bibitem [{\citenamefont {{Graff}}\ \emph {et~al.}(2011)\citenamefont
  {{Graff}}, \citenamefont {{Hobson}},\ and\ \citenamefont
  {{Lasenby}}}]{2011MNRAS.413L..66G}%
  \BibitemOpen
  \bibfield  {author} {\bibinfo {author} {\bibfnamefont {P.}~\bibnamefont
  {{Graff}}}, \bibinfo {author} {\bibfnamefont {M.~P.}\ \bibnamefont
  {{Hobson}}}, \ and\ \bibinfo {author} {\bibfnamefont {A.}~\bibnamefont
  {{Lasenby}}},\ }\href {\doibase 10.1111/j.1745-3933.2011.01034.x} {\bibfield
  {journal} {\bibinfo  {journal} {\mnras}\ }\textbf {\bibinfo {volume} {413}},\
  \bibinfo {pages} {L66} (\bibinfo {year} {2011})},\ \Eprint
  {http://arxiv.org/abs/1010.5907} {arXiv:1010.5907 [astro-ph.IM]} \BibitemShut
  {NoStop}%
\bibitem [{\citenamefont {{Gjerl{\o}w}}\ \emph {et~al.}(2015)\citenamefont
  {{Gjerl{\o}w}}, \citenamefont {{Colombo}}, \citenamefont {{Eriksen}},
  \citenamefont {{G{\'o}rski}}, \citenamefont {{Gruppuso}}, \citenamefont
  {{Jewell}}, \citenamefont {{Plaszczynski}},\ and\ \citenamefont
  {{Wehus}}}]{2015ApJS..221....5G}%
  \BibitemOpen
  \bibfield  {author} {\bibinfo {author} {\bibfnamefont {E.}~\bibnamefont
  {{Gjerl{\o}w}}}, \bibinfo {author} {\bibfnamefont {L.~P.~L.}\ \bibnamefont
  {{Colombo}}}, \bibinfo {author} {\bibfnamefont {H.~K.}\ \bibnamefont
  {{Eriksen}}}, \bibinfo {author} {\bibfnamefont {K.~M.}\ \bibnamefont
  {{G{\'o}rski}}}, \bibinfo {author} {\bibfnamefont {A.}~\bibnamefont
  {{Gruppuso}}}, \bibinfo {author} {\bibfnamefont {J.~B.}\ \bibnamefont
  {{Jewell}}}, \bibinfo {author} {\bibfnamefont {S.}~\bibnamefont
  {{Plaszczynski}}}, \ and\ \bibinfo {author} {\bibfnamefont {I.~K.}\
  \bibnamefont {{Wehus}}},\ }\href {\doibase 10.1088/0067-0049/221/1/5}
  {\bibfield  {journal} {\bibinfo  {journal} {\apjs}\ }\textbf {\bibinfo
  {volume} {221}},\ \bibinfo {eid} {5} (\bibinfo {year} {2015})},\ \Eprint
  {http://arxiv.org/abs/1506.04273} {arXiv:1506.04273 [astro-ph.IM]}
  \BibitemShut {NoStop}%
\bibitem [{\citenamefont {{Wang}}\ and\ \citenamefont
  {{Wang}}(2013)}]{2013PhRvD..88d3522W}%
  \BibitemOpen
  \bibfield  {author} {\bibinfo {author} {\bibfnamefont {Y.}~\bibnamefont
  {{Wang}}}\ and\ \bibinfo {author} {\bibfnamefont {S.}~\bibnamefont
  {{Wang}}},\ }\href {\doibase 10.1103/PhysRevD.88.043522} {\bibfield
  {journal} {\bibinfo  {journal} {\prd}\ }\textbf {\bibinfo {volume} {88}},\
  \bibinfo {eid} {043522} (\bibinfo {year} {2013})},\ \Eprint
  {http://arxiv.org/abs/1304.4514} {arXiv:1304.4514 [astro-ph.CO]} \BibitemShut
  {NoStop}%
\bibitem [{\citenamefont {{Chu}}\ \emph {et~al.}(2003)\citenamefont {{Chu}},
  \citenamefont {{Kaplinghat}},\ and\ \citenamefont
  {{Knox}}}]{2003ApJ...596..725C}%
  \BibitemOpen
  \bibfield  {author} {\bibinfo {author} {\bibfnamefont {M.}~\bibnamefont
  {{Chu}}}, \bibinfo {author} {\bibfnamefont {M.}~\bibnamefont {{Kaplinghat}}},
  \ and\ \bibinfo {author} {\bibfnamefont {L.}~\bibnamefont {{Knox}}},\ }\href
  {\doibase 10.1086/378039} {\bibfield  {journal} {\bibinfo  {journal} {\apj}\
  }\textbf {\bibinfo {volume} {596}},\ \bibinfo {pages} {725} (\bibinfo {year}
  {2003})},\ \Eprint {http://arxiv.org/abs/astro-ph/0212466} {astro-ph/0212466}
  \BibitemShut {NoStop}%
\bibitem [{\citenamefont {Lewis}\ \emph {et~al.}(2000)\citenamefont {Lewis},
  \citenamefont {Challinor},\ and\ \citenamefont {Lasenby}}]{Lewis:1999bs}%
  \BibitemOpen
  \bibfield  {author} {\bibinfo {author} {\bibfnamefont {A.}~\bibnamefont
  {Lewis}}, \bibinfo {author} {\bibfnamefont {A.}~\bibnamefont {Challinor}}, \
  and\ \bibinfo {author} {\bibfnamefont {A.}~\bibnamefont {Lasenby}},\
  }\href@noop {} {\bibfield  {journal} {\bibinfo  {journal} {Astrophys. J.}\
  }\textbf {\bibinfo {volume} {538}},\ \bibinfo {pages} {473} (\bibinfo {year}
  {2000})},\ \Eprint {http://arxiv.org/abs/astro-ph/9911177} {astro-ph/9911177}
  \BibitemShut {NoStop}%
%%CITATION = ASTRO-PH 9911177;%%
\bibitem [{\citenamefont {{Knox}}(1995)}]{1995PhRvD..52.4307K}%
  \BibitemOpen
  \bibfield  {author} {\bibinfo {author} {\bibfnamefont {L.}~\bibnamefont
  {{Knox}}},\ }\href {\doibase 10.1103/PhysRevD.52.4307} {\bibfield  {journal}
  {\bibinfo  {journal} {\prd}\ }\textbf {\bibinfo {volume} {52}},\ \bibinfo
  {pages} {4307} (\bibinfo {year} {1995})},\ \Eprint
  {http://arxiv.org/abs/astro-ph/9504054} {astro-ph/9504054} \BibitemShut
  {NoStop}%
\bibitem [{\citenamefont {{Allison}}\ \emph {et~al.}(2015)\citenamefont
  {{Allison}}, \citenamefont {{Caucal}}, \citenamefont {{Calabrese}},
  \citenamefont {{Dunkley}},\ and\ \citenamefont
  {{Louis}}}]{2015arXiv150907471A}%
  \BibitemOpen
  \bibfield  {author} {\bibinfo {author} {\bibfnamefont {R.}~\bibnamefont
  {{Allison}}}, \bibinfo {author} {\bibfnamefont {P.}~\bibnamefont {{Caucal}}},
  \bibinfo {author} {\bibfnamefont {E.}~\bibnamefont {{Calabrese}}}, \bibinfo
  {author} {\bibfnamefont {J.}~\bibnamefont {{Dunkley}}}, \ and\ \bibinfo
  {author} {\bibfnamefont {T.}~\bibnamefont {{Louis}}},\ }\href@noop {}
  {\bibfield  {journal} {\bibinfo  {journal} {ArXiv e-prints}\ } (\bibinfo
  {year} {2015})},\ \Eprint {http://arxiv.org/abs/1509.07471}
  {arXiv:1509.07471} \BibitemShut {NoStop}%
\bibitem [{\citenamefont {{Bond}}\ \emph {et~al.}(2000)\citenamefont {{Bond}},
  \citenamefont {{Jaffe}},\ and\ \citenamefont {{Knox}}}]{2000ApJ...533...19B}%
  \BibitemOpen
  \bibfield  {author} {\bibinfo {author} {\bibfnamefont {J.~R.}\ \bibnamefont
  {{Bond}}}, \bibinfo {author} {\bibfnamefont {A.~H.}\ \bibnamefont {{Jaffe}}},
  \ and\ \bibinfo {author} {\bibfnamefont {L.}~\bibnamefont {{Knox}}},\ }\href
  {\doibase 10.1086/308625} {\bibfield  {journal} {\bibinfo  {journal} {\apj}\
  }\textbf {\bibinfo {volume} {533}},\ \bibinfo {pages} {19} (\bibinfo {year}
  {2000})},\ \Eprint {http://arxiv.org/abs/astro-ph/9808264} {astro-ph/9808264}
  \BibitemShut {NoStop}%
\bibitem [{\citenamefont {{Hamimeche}}\ and\ \citenamefont
  {{Lewis}}(2008)}]{2008PhRvD..77j3013H}%
  \BibitemOpen
  \bibfield  {author} {\bibinfo {author} {\bibfnamefont {S.}~\bibnamefont
  {{Hamimeche}}}\ and\ \bibinfo {author} {\bibfnamefont {A.}~\bibnamefont
  {{Lewis}}},\ }\href {\doibase 10.1103/PhysRevD.77.103013} {\bibfield
  {journal} {\bibinfo  {journal} {\prd}\ }\textbf {\bibinfo {volume} {77}},\
  \bibinfo {eid} {103013} (\bibinfo {year} {2008})},\ \Eprint
  {http://arxiv.org/abs/0801.0554} {arXiv:0801.0554} \BibitemShut {NoStop}%
\bibitem [{\citenamefont {{Hamimeche}}\ and\ \citenamefont
  {{Lewis}}(2009)}]{2009PhRvD..79h3012H}%
  \BibitemOpen
  \bibfield  {author} {\bibinfo {author} {\bibfnamefont {S.}~\bibnamefont
  {{Hamimeche}}}\ and\ \bibinfo {author} {\bibfnamefont {A.}~\bibnamefont
  {{Lewis}}},\ }\href {\doibase 10.1103/PhysRevD.79.083012} {\bibfield
  {journal} {\bibinfo  {journal} {\prd}\ }\textbf {\bibinfo {volume} {79}},\
  \bibinfo {eid} {083012} (\bibinfo {year} {2009})},\ \Eprint
  {http://arxiv.org/abs/0902.0674} {arXiv:0902.0674 [astro-ph.CO]} \BibitemShut
  {NoStop}%
\bibitem [{\citenamefont {{Larson}}\ \emph {et~al.}(2011)\citenamefont
  {{Larson}}, \citenamefont {{Dunkley}}, \citenamefont {{Hinshaw}},
  \citenamefont {{Komatsu}}, \citenamefont {{Nolta}}, \citenamefont
  {{Bennett}}, \citenamefont {{Gold}}, \citenamefont {{Halpern}}, \citenamefont
  {{Hill}}, \citenamefont {{Jarosik}}, \citenamefont {{Kogut}}, \citenamefont
  {{Limon}}, \citenamefont {{Meyer}}, \citenamefont {{Odegard}}, \citenamefont
  {{Page}}, \citenamefont {{Smith}}, \citenamefont {{Spergel}}, \citenamefont
  {{Tucker}}, \citenamefont {{Weiland}}, \citenamefont {{Wollack}},\ and\
  \citenamefont {{Wright}}}]{2011ApJS..192...16L}%
  \BibitemOpen
  \bibfield  {author} {\bibinfo {author} {\bibfnamefont {D.}~\bibnamefont
  {{Larson}}}, \bibinfo {author} {\bibfnamefont {J.}~\bibnamefont {{Dunkley}}},
  \bibinfo {author} {\bibfnamefont {G.}~\bibnamefont {{Hinshaw}}}, \bibinfo
  {author} {\bibfnamefont {E.}~\bibnamefont {{Komatsu}}}, \bibinfo {author}
  {\bibfnamefont {M.~R.}\ \bibnamefont {{Nolta}}}, \bibinfo {author}
  {\bibfnamefont {C.~L.}\ \bibnamefont {{Bennett}}}, \bibinfo {author}
  {\bibfnamefont {B.}~\bibnamefont {{Gold}}}, \bibinfo {author} {\bibfnamefont
  {M.}~\bibnamefont {{Halpern}}}, \bibinfo {author} {\bibfnamefont {R.~S.}\
  \bibnamefont {{Hill}}}, \bibinfo {author} {\bibfnamefont {N.}~\bibnamefont
  {{Jarosik}}}, \bibinfo {author} {\bibfnamefont {A.}~\bibnamefont {{Kogut}}},
  \bibinfo {author} {\bibfnamefont {M.}~\bibnamefont {{Limon}}}, \bibinfo
  {author} {\bibfnamefont {S.~S.}\ \bibnamefont {{Meyer}}}, \bibinfo {author}
  {\bibfnamefont {N.}~\bibnamefont {{Odegard}}}, \bibinfo {author}
  {\bibfnamefont {L.}~\bibnamefont {{Page}}}, \bibinfo {author} {\bibfnamefont
  {K.~M.}\ \bibnamefont {{Smith}}}, \bibinfo {author} {\bibfnamefont {D.~N.}\
  \bibnamefont {{Spergel}}}, \bibinfo {author} {\bibfnamefont {G.~S.}\
  \bibnamefont {{Tucker}}}, \bibinfo {author} {\bibfnamefont {J.~L.}\
  \bibnamefont {{Weiland}}}, \bibinfo {author} {\bibfnamefont {E.}~\bibnamefont
  {{Wollack}}}, \ and\ \bibinfo {author} {\bibfnamefont {E.~L.}\ \bibnamefont
  {{Wright}}},\ }\href {\doibase 10.1088/0067-0049/192/2/16} {\bibfield
  {journal} {\bibinfo  {journal} {\apjs}\ }\textbf {\bibinfo {volume} {192}},\
  \bibinfo {eid} {16} (\bibinfo {year} {2011})},\ \Eprint
  {http://arxiv.org/abs/1001.4635} {arXiv:1001.4635 [astro-ph.CO]} \BibitemShut
  {NoStop}%
\bibitem [{\citenamefont {{Dunkley}}\ \emph {et~al.}(2009)\citenamefont
  {{Dunkley}}, \citenamefont {{Komatsu}}, \citenamefont {{Nolta}},
  \citenamefont {{Spergel}}, \citenamefont {{Larson}}, \citenamefont
  {{Hinshaw}}, \citenamefont {{Page}}, \citenamefont {{Bennett}}, \citenamefont
  {{Gold}}, \citenamefont {{Jarosik}}, \citenamefont {{Weiland}}, \citenamefont
  {{Halpern}}, \citenamefont {{Hill}}, \citenamefont {{Kogut}}, \citenamefont
  {{Limon}}, \citenamefont {{Meyer}}, \citenamefont {{Tucker}}, \citenamefont
  {{Wollack}},\ and\ \citenamefont {{Wright}}}]{2009ApJS..180..306D}%
  \BibitemOpen
  \bibfield  {author} {\bibinfo {author} {\bibfnamefont {J.}~\bibnamefont
  {{Dunkley}}}, \bibinfo {author} {\bibfnamefont {E.}~\bibnamefont
  {{Komatsu}}}, \bibinfo {author} {\bibfnamefont {M.~R.}\ \bibnamefont
  {{Nolta}}}, \bibinfo {author} {\bibfnamefont {D.~N.}\ \bibnamefont
  {{Spergel}}}, \bibinfo {author} {\bibfnamefont {D.}~\bibnamefont {{Larson}}},
  \bibinfo {author} {\bibfnamefont {G.}~\bibnamefont {{Hinshaw}}}, \bibinfo
  {author} {\bibfnamefont {L.}~\bibnamefont {{Page}}}, \bibinfo {author}
  {\bibfnamefont {C.~L.}\ \bibnamefont {{Bennett}}}, \bibinfo {author}
  {\bibfnamefont {B.}~\bibnamefont {{Gold}}}, \bibinfo {author} {\bibfnamefont
  {N.}~\bibnamefont {{Jarosik}}}, \bibinfo {author} {\bibfnamefont {J.~L.}\
  \bibnamefont {{Weiland}}}, \bibinfo {author} {\bibfnamefont {M.}~\bibnamefont
  {{Halpern}}}, \bibinfo {author} {\bibfnamefont {R.~S.}\ \bibnamefont
  {{Hill}}}, \bibinfo {author} {\bibfnamefont {A.}~\bibnamefont {{Kogut}}},
  \bibinfo {author} {\bibfnamefont {M.}~\bibnamefont {{Limon}}}, \bibinfo
  {author} {\bibfnamefont {S.~S.}\ \bibnamefont {{Meyer}}}, \bibinfo {author}
  {\bibfnamefont {G.~S.}\ \bibnamefont {{Tucker}}}, \bibinfo {author}
  {\bibfnamefont {E.}~\bibnamefont {{Wollack}}}, \ and\ \bibinfo {author}
  {\bibfnamefont {E.~L.}\ \bibnamefont {{Wright}}},\ }\href {\doibase
  10.1088/0067-0049/180/2/306} {\bibfield  {journal} {\bibinfo  {journal}
  {\apjs}\ }\textbf {\bibinfo {volume} {180}},\ \bibinfo {pages} {306}
  (\bibinfo {year} {2009})},\ \Eprint {http://arxiv.org/abs/0803.0586}
  {arXiv:0803.0586} \BibitemShut {NoStop}%
\bibitem [{\citenamefont {{Hivon}}\ \emph {et~al.}(2002)\citenamefont
  {{Hivon}}, \citenamefont {{G{\'o}rski}}, \citenamefont {{Netterfield}},
  \citenamefont {{Crill}}, \citenamefont {{Prunet}},\ and\ \citenamefont
  {{Hansen}}}]{2002ApJ...567....2H}%
  \BibitemOpen
  \bibfield  {author} {\bibinfo {author} {\bibfnamefont {E.}~\bibnamefont
  {{Hivon}}}, \bibinfo {author} {\bibfnamefont {K.~M.}\ \bibnamefont
  {{G{\'o}rski}}}, \bibinfo {author} {\bibfnamefont {C.~B.}\ \bibnamefont
  {{Netterfield}}}, \bibinfo {author} {\bibfnamefont {B.~P.}\ \bibnamefont
  {{Crill}}}, \bibinfo {author} {\bibfnamefont {S.}~\bibnamefont {{Prunet}}}, \
  and\ \bibinfo {author} {\bibfnamefont {F.}~\bibnamefont {{Hansen}}},\ }\href
  {\doibase 10.1086/338126} {\bibfield  {journal} {\bibinfo  {journal} {\apj}\
  }\textbf {\bibinfo {volume} {567}},\ \bibinfo {pages} {2} (\bibinfo {year}
  {2002})},\ \Eprint {http://arxiv.org/abs/astro-ph/0105302} {astro-ph/0105302}
  \BibitemShut {NoStop}%
\bibitem [{\citenamefont {{Hinshaw}}\ \emph {et~al.}(2007)\citenamefont
  {{Hinshaw}}, \citenamefont {{Nolta}}, \citenamefont {{Bennett}},
  \citenamefont {{Bean}}, \citenamefont {{Dor{\'e}}}, \citenamefont
  {{Greason}}, \citenamefont {{Halpern}}, \citenamefont {{Hill}}, \citenamefont
  {{Jarosik}}, \citenamefont {{Kogut}}, \citenamefont {{Komatsu}},
  \citenamefont {{Limon}}, \citenamefont {{Odegard}}, \citenamefont {{Meyer}},
  \citenamefont {{Page}}, \citenamefont {{Peiris}}, \citenamefont {{Spergel}},
  \citenamefont {{Tucker}}, \citenamefont {{Verde}}, \citenamefont {{Weiland}},
  \citenamefont {{Wollack}},\ and\ \citenamefont
  {{Wright}}}]{2007ApJS..170..288H}%
  \BibitemOpen
  \bibfield  {author} {\bibinfo {author} {\bibfnamefont {G.}~\bibnamefont
  {{Hinshaw}}}, \bibinfo {author} {\bibfnamefont {M.~R.}\ \bibnamefont
  {{Nolta}}}, \bibinfo {author} {\bibfnamefont {C.~L.}\ \bibnamefont
  {{Bennett}}}, \bibinfo {author} {\bibfnamefont {R.}~\bibnamefont {{Bean}}},
  \bibinfo {author} {\bibfnamefont {O.}~\bibnamefont {{Dor{\'e}}}}, \bibinfo
  {author} {\bibfnamefont {M.~R.}\ \bibnamefont {{Greason}}}, \bibinfo {author}
  {\bibfnamefont {M.}~\bibnamefont {{Halpern}}}, \bibinfo {author}
  {\bibfnamefont {R.~S.}\ \bibnamefont {{Hill}}}, \bibinfo {author}
  {\bibfnamefont {N.}~\bibnamefont {{Jarosik}}}, \bibinfo {author}
  {\bibfnamefont {A.}~\bibnamefont {{Kogut}}}, \bibinfo {author} {\bibfnamefont
  {E.}~\bibnamefont {{Komatsu}}}, \bibinfo {author} {\bibfnamefont
  {M.}~\bibnamefont {{Limon}}}, \bibinfo {author} {\bibfnamefont
  {N.}~\bibnamefont {{Odegard}}}, \bibinfo {author} {\bibfnamefont {S.~S.}\
  \bibnamefont {{Meyer}}}, \bibinfo {author} {\bibfnamefont {L.}~\bibnamefont
  {{Page}}}, \bibinfo {author} {\bibfnamefont {H.~V.}\ \bibnamefont
  {{Peiris}}}, \bibinfo {author} {\bibfnamefont {D.~N.}\ \bibnamefont
  {{Spergel}}}, \bibinfo {author} {\bibfnamefont {G.~S.}\ \bibnamefont
  {{Tucker}}}, \bibinfo {author} {\bibfnamefont {L.}~\bibnamefont {{Verde}}},
  \bibinfo {author} {\bibfnamefont {J.~L.}\ \bibnamefont {{Weiland}}}, \bibinfo
  {author} {\bibfnamefont {E.}~\bibnamefont {{Wollack}}}, \ and\ \bibinfo
  {author} {\bibfnamefont {E.~L.}\ \bibnamefont {{Wright}}},\ }\href {\doibase
  10.1086/513698} {\bibfield  {journal} {\bibinfo  {journal} {\apjs}\ }\textbf
  {\bibinfo {volume} {170}},\ \bibinfo {pages} {288} (\bibinfo {year}
  {2007})},\ \Eprint {http://arxiv.org/abs/astro-ph/0603451} {astro-ph/0603451}
  \BibitemShut {NoStop}%
\bibitem [{\citenamefont {{Verde}}\ \emph {et~al.}(2003)\citenamefont
  {{Verde}}, \citenamefont {{Peiris}}, \citenamefont {{Spergel}}, \citenamefont
  {{Nolta}}, \citenamefont {{Bennett}}, \citenamefont {{Halpern}},
  \citenamefont {{Hinshaw}}, \citenamefont {{Jarosik}}, \citenamefont
  {{Kogut}}, \citenamefont {{Limon}}, \citenamefont {{Meyer}}, \citenamefont
  {{Page}}, \citenamefont {{Tucker}}, \citenamefont {{Wollack}},\ and\
  \citenamefont {{Wright}}}]{2003ApJS..148..195V}%
  \BibitemOpen
  \bibfield  {author} {\bibinfo {author} {\bibfnamefont {L.}~\bibnamefont
  {{Verde}}}, \bibinfo {author} {\bibfnamefont {H.~V.}\ \bibnamefont
  {{Peiris}}}, \bibinfo {author} {\bibfnamefont {D.~N.}\ \bibnamefont
  {{Spergel}}}, \bibinfo {author} {\bibfnamefont {M.~R.}\ \bibnamefont
  {{Nolta}}}, \bibinfo {author} {\bibfnamefont {C.~L.}\ \bibnamefont
  {{Bennett}}}, \bibinfo {author} {\bibfnamefont {M.}~\bibnamefont
  {{Halpern}}}, \bibinfo {author} {\bibfnamefont {G.}~\bibnamefont
  {{Hinshaw}}}, \bibinfo {author} {\bibfnamefont {N.}~\bibnamefont
  {{Jarosik}}}, \bibinfo {author} {\bibfnamefont {A.}~\bibnamefont {{Kogut}}},
  \bibinfo {author} {\bibfnamefont {M.}~\bibnamefont {{Limon}}}, \bibinfo
  {author} {\bibfnamefont {S.~S.}\ \bibnamefont {{Meyer}}}, \bibinfo {author}
  {\bibfnamefont {L.}~\bibnamefont {{Page}}}, \bibinfo {author} {\bibfnamefont
  {G.~S.}\ \bibnamefont {{Tucker}}}, \bibinfo {author} {\bibfnamefont
  {E.}~\bibnamefont {{Wollack}}}, \ and\ \bibinfo {author} {\bibfnamefont
  {E.~L.}\ \bibnamefont {{Wright}}},\ }\href {\doibase 10.1086/377335}
  {\bibfield  {journal} {\bibinfo  {journal} {\apjs}\ }\textbf {\bibinfo
  {volume} {148}},\ \bibinfo {pages} {195} (\bibinfo {year} {2003})},\ \Eprint
  {http://arxiv.org/abs/astro-ph/0302218} {astro-ph/0302218} \BibitemShut
  {NoStop}%
\end{thebibliography}
%

\end{document}